\documentclass[usenatbib]{mnras}
\usepackage{graphicx,aas_macros,bm,amssymb,amsmath,times}
\usepackage{todonotes}
\usepackage{xfrac}
\numberwithin{equation}{section}
\usepackage{empheq}

\usepackage{xcolor}

\def\sma{\sum_a m_a}
\def\smb{\sum_b m_b}

\newcommand{\pder}[2]{\ensuremath{\frac{\partial #1}{\partial #2}}}
\newcommand{\evalat}[2]{\ensuremath{\left. #1 \right\rvert_{#2}}}
\def\d{\mathrm{d}}
\newcommand{\deriv}[2]{\ensuremath{\frac{\d #1}{\d #2}}}
\def\adgamma{{\gamma_{\mathrm{ad}}}}
\newcommand{\brkt}[1]{\ensuremath{\left(#1\right)}}

\allowdisplaybreaks

\defcitealias{chowmonaghan97}{CM97}

\title[SPH in the Kerr metric]{General relativistic smoothed particle hydrodynamics}

\author[Liptai and Price]{David Liptai$^{1}$\thanks{Contact e-mail: \href{mailto:david.liptai@monash.edu}{david.liptai@monash.edu}} and Daniel J. Price$^1$ \\
$^1$Monash Centre for Astrophysics (MoCA) and School of Physics and Astronomy, Monash University, Clayton Vic 3800, Australia
}

\pagerange{\pageref{firstpage}--\pageref{lastpage}} \pubyear{2019}

\begin{document}
\label{firstpage}
\bibliographystyle{mnras}
\maketitle
\begin{abstract}
   We present a method for general relativistic smoothed particle hydrodynamics (GRSPH), based on an entropy-conservative form of the general relativistic hydrodynamic equations for a perfect fluid. We aim to replace approximate treatments of general relativity in current SPH simulations of tidal disruption events and accretion discs. We develop an improved shock capturing formulation that distinguishes between shock viscosity and conductivity in relativity. We also describe a new Hamiltonian time integration algorithm for relativistic orbital dynamics and GRSPH. Our method correctly captures both Einstein and spin-induced precession around black holes. We benchmark our scheme in 1D and 3D against mildly and ultra relativistic shock tubes, exact solutions for epicyclic and vertical oscillation frequencies, and Bondi accretion. We assume fixed background metrics (Minkowski, Schwarzschild and Kerr in Cartesian Boyer-Lindquist coordinates) but the method lays the foundation for future direct coupling with numerical relativity.
\end{abstract}

%----------------------------------------------------------------------------------------------------------------
\section{Introduction}
Urgent motivation for relativistic hydrodynamics simulations arises from the coincident detection of an electromagnetic afterglow \citep{abbottabbottabbott17a} alongside the first detection of gravitational waves from a binary neutron star merger \citep{abbottabbottabbott17}. The coming decade should bring hundreds more such events as the Advanced Laser Interferometer Gravitational-Wave Observatory (LIGO) reaches design sensitivity \citep{abbottabbottabbott17}.

Smoothed particle hydrodynamics (SPH) \citep{gingoldmonaghan77,lucy77,rosswog09,price12} is perfectly suited for neutron star merger simulations because there is no preferred geometry, resolution follows mass, and it avoids the limitations imposed by a background density floor \citep[e.g.][]{oechslinrosswogthielemann02}. The main limitation of SPH in relativistic hydrodynamics to date is the approximate treatment of general relativity \citep{metzger17}.

Further motivation for a public GRSPH code comes from i) the race to find electromagnetic counterparts to binary black hole mergers detected by Advanced LIGO and the forthcoming Laser Interferometer Space Antenna \citep[e.g.][]{armitagenatarajan02,milosavljevicphinney05,ceriolilodatoprice16}; ii) disc formation in tidal disruption events \citep[e.g.][]{hayasakistoneloeb13,hayasakistoneloeb16,bonnerotrossilodato16,tejedagaftonrosswog17}; iii) the recently discovered phenomena of `tearing' in warped discs around spinning black holes \citep[e.g.][]{nixonkingprice12,nealonpricenixon15} and the possible relation to state transitions and quasi periodic oscillations \citep{nixonsalvesen14}; iv) forthcoming observations from the Event Horizon Telescope \citep{akiyamakuramochiikeda17}; and v) the need for relativistic SPH simulations of pulsar winds \citep{okazakinagatakinaito11}.

\citet{kheyfetsmillerzurek90} were the first to derive relativistic SPH equations, followed by \citet{mann91} and \citet{lagunamillerzurek93}. However, their formulations were not in conservative form and had difficulty handling shocks in even mildly relativistic flows. Conservative forms of the relativistic SPH equations were derived by \citet{chowmonaghan97} (hereafter \citetalias{chowmonaghan97}); \citet{sieglerriffert00,monaghanprice01,rosswog10a}~and~\citet{rosswog10} for both special and general relativity, but their applications to date have been limited to special relativity. Other recent studies have instead used post-Newtonian approximations within standard SPH codes to account for relativistic effects in fluid flows around black holes \citep[e.g.][]{tejedarosswog13,nealonpricenixon15,bonnerotrossilodato16,hayasakistoneloeb16}. SPH applications to fluid flows in strong self gravity assume conformal flatness \citep[e.g.][]{oechslinrosswogthielemann02,fabergrandclementrasio04,bausweinoechslinjanka10} but these papers do not include standardised tests in fixed metrics. Most recently, \citet{tejedagaftonrosswog17} applied GRSPH to tidal disruption events, but also showed only limited standardised tests of their method.

Our goal in this paper is to present a detailed description of a fully general relativistic SPH code, able to perform hydrodynamic simulations in any given fixed background metric, with the potential to be applied to dynamically evolving metrics. Our requirements for a modern GRSPH code include being able to capture relativistic shocks, a precise and accurate treatment of orbital dynamics, and the ability to work in Kerr geometry. To this end we derive an improved shock capturing method and a new Hamiltonian time integration algorithm suitable for GRSPH and relativistic orbital dynamics. We describe our complete method and its implementation in depth, before showing the results of various benchmark tests. We implement our method in \textsc{Phantom} \citep{pricewurstertricco18}, a publicly available SPH code.

\section{Relativistic hydrodynamics}
We follow \citet{monaghanprice01} who obtained the equations of general relativistic smoothed particle hydrodynamics from a Lagrangian. In the following, Greek indices are summed over ($0$,$1$,$2$,$3$) while Latin indices are summed over ($1$,$2$,$3$). The subscripts $a$ and $b$ are reserved for particle labels.
\subsection{Definitions}
The four velocity of a fluid at a given spacetime coordinate $x^\mu$ is given by
\begin{equation}
   U^{\mu} = \deriv{x^\mu}{\tau},
\end{equation}
where $\tau$ is the proper time, and $x^\mu \equiv (t,x^i)$. The coordinate velocity $v^{\mu}$ is defined with respect to the coordinate time $t$ such that
\begin{equation} \label{eq:coordinate-velocity}
   v^{\mu} \equiv \deriv{x^\mu}{t} = \frac{U^{\mu}}{U^{0}},
\end{equation}
where from the normalisation condition $U^\mu U_\mu = -1$ we have
\begin{equation} \label{eq:u0}
   U^{0} \equiv \deriv{t}{\tau} = \frac{1}{\sqrt{-g_{\mu\nu} v^{\mu} v^{\nu}}}.
\end{equation}
We assume standard relativistic units where $c=G=1$ and a metric signature ($-$,$+$,$+$,$+$).

\subsection{Conserved variables}
The conserved density, momentum and energy are respectively
\begin{align}
   \rho^* & = \sqrt{-g} \rho \, U^0,                                                      \label{eq:rhostar} \\
   p_i     & = U^0 w g_{i\mu} v^{\mu},                                                        \label{eq:pmom} \\
   e        & = U^0 \left[ w g_{i\mu} v^{\mu} v^{i}  - (1 + u) g_{\mu\nu} v^{\mu} v^{\nu} \right], \label{eq:etot}
\end{align}
where $g$ is the determinant of the covariant metric $g_{\mu\nu}$ and we use $w$ to denote the specific enthalpy
\begin{equation}
   w = 1 + u + \frac{P}{\rho}.
\end{equation}
The specific internal energy, pressure, and density in the rest frame of the fluid are denoted by $u, P$ and $\rho$, respectively.
Equivalently, given a $3+1$ decomposition of the metric \citep{arnowittdesermisner08} we can express the conserved quantities as
\begin{align}
   \rho^{*} & = \sqrt{\gamma} \, \Gamma \rho,              \label{eq:rhostar_3p1} \\
   p_i        & = w \Gamma V_i,                                     \label{eq:pmom_3p1}   \\
   e           & = p_i v^i + \frac{\alpha (1+u)}{\Gamma},  \label{eq:etot_3p1}
\end{align}
where $\Gamma = \brkt{1 - V^i V_i }^{-1/2}$ is the generalised Lorentz factor, and $V^i = \brkt{v^i + \beta^i}/ \alpha$ is the fluid velocity in the frame of a local Eulerian observer. The lapse function, shift vector, and spatial three-metric ($\alpha$, $\beta_i$, $\gamma_{ij}$) are related to the four-metric $ds^2 = g_{\mu \nu} \d x^\mu \d x^\nu$ via
\begin{align}
   ds^2 = -\alpha^2 \d t^2 + \gamma_{ij} (\d x^i + \beta^i \d t) ( \d x^j + \beta^j \d t).
\end{align}
Spatial vectors can then have their indices raised and lowered by the three-metric, e.g. $v^i v_i = \gamma_{ij} v^i v^j = \gamma^{ij} v_i v_j$. We note that $\gamma$ is the determinant of $\gamma_{ij}$, and that Equation~\ref{eq:etot_3p1} can be written alternatively in terms of the enthalpy and pressure as
\begin{equation}
   \label{eq:etot_3p1_alternate}
   e  = w \Gamma \brkt{ \alpha - V^i \beta_i } - \frac{\alpha P}{\Gamma \rho}.
\end{equation}

\subsection{Equations of relativistic hydrodynamics}
From \citet{monaghanprice01}, the equations of relativistic hydrodynamics in Lagrangian conservative form, assuming an ideal fluid, are given by
\begin{align}
   \deriv{\rho^*}{t} & = -\rho^* \pder{v^i}{x^i},                              \label{eq:relhydro1}\\
   \deriv{p_i}{t}    & = - \frac{1}{\rho^*} \pder{(\sqrt{-g} P)}{x^i} + f_i ,      \label{eq:relhydro2} \\
   \deriv{e}{t}      & = - \frac{1}{\rho^*} \pder{(\sqrt{-g} P v^i)}{x^i} + \Lambda, \label{eq:relhydro3}
\end{align}
where the source terms containing derivatives of the metric are defined according to
\begin{align}
   f_{i} \equiv & \frac{\sqrt{-g}}{2\rho^{*}} \brkt{ T^{\mu\nu} \pder{g_{\mu\nu}}{x^i} }, \\
   \Lambda \equiv & - \frac{\sqrt{-g}}{2\rho^{*}} \brkt{ T^{\mu\nu} \pder{g_{\mu\nu}}{t} }.
\end{align}
For an ideal fluid the stress-energy tensor is given by
\begin{align}
   T^{\mu\nu} & = (\rho + \rho u + P) U^{\mu} U^{\nu} + P g^{\mu\nu}, \\
   & = \rho w (U^{0})^{2} v^{\mu} v^{\nu} + P g^{\mu\nu}.
\end{align}
Thus, the term required to compute the source terms is
\begin{equation}
   \frac{\sqrt{-g}~T^{\mu\nu}}{2\rho^{*}} = \frac12 \left[ w U^{0} v^{\mu} v^{\nu} + \frac{P g^{\mu\nu}}{\rho ~ U^{0}} \right].
\end{equation}
We close the equation set with the ideal gas equation of state
\begin{equation}
   \label{eq:eos}
   P = \brkt{\adgamma - 1} \rho u,
\end{equation}
where $\adgamma$ is the adiabatic index. For our calculations we assume an ideal atomic gas, $\adgamma = 5/3$. The internal energy $u$ can be related to the gas temperature $T$ through the ideal gas law
\begin{equation}
   \label{eq:idealgaslaw}
   P = \frac{\rho k_\mathrm{B} T}{\mu m_\mathrm{H}},
\end{equation}
where $k_\mathrm{B}$ is Boltzmann's constant, $\mu$ is the mean molecular weight, and $m_\mathrm{H}$ is the mass of a Hydrogen atom.

\section{SPH Equations of relativistic hydrodynamics}
The discrete form of Equations \ref{eq:relhydro1}, \ref{eq:relhydro2}, and \ref{eq:relhydro3} in the absence of dissipation, are given respectively by \citep[e.g.][]{sieglerriffert00,monaghanprice01,rosswog10}
\begin{align}
   \deriv{\rho^*_a}{t} = & \frac{1}{\Omega_a} \smb (v^i_a - v^i_b) \pder{W_{ab}(h_a)}{x^i},                                                                                                           \label{eq:continuity-sph} \\
   \deriv{p_i^a}{t}     = & -\smb \left[\frac{\sqrt{-g_a} P_a}{\Omega_a \rho^{*2}_{a}} \pder{W_{ab} (h_a)}{x^i}  \right. \nonumber \\
                 & \qquad \qquad + \left. \frac{\sqrt{-g_b} P_b}{\Omega_b \rho^{*2}_{b}} \pder{W_{ab} (h_b)}{x^i} \right] + f_i^a,                                          \label{eq:momentum-sph} \\
   \deriv{e_a}{t}        = & -\smb \left[ \frac{\sqrt{-g_a} P_a v_{b}^{i}}{\Omega_a \rho^{*2}_{a}} \pder{W_{ab} (h_a)}{x^i} \right. \nonumber \\
                 & \qquad \qquad + \left. \frac{\sqrt{-g_b} P_b v_{a}^{i}}{\Omega_b \rho^{*2}_{b}} \pder{W_{ab} (h_b)}{x^i} \right] + \Lambda_a,                   \label{eq:energy-sph}
\end{align}
where $W_{ab}$ is the interpolating kernel, $h_a$ is the smoothing length, and $\Omega_a$ is a term related to the gradient of the smoothing length given by \citep{monaghan02,springelhernquist02}
\begin{align} \label{eq:omegaterm}
  \Omega_a = 1 - \pder{h_a}{\rho^*_a} \smb{\pder{W_{ab}(h_a)}{h_a}}.
\end{align}
The conserved density can also be computed directly via
\begin{equation}
   \label{eq:densitysum}
   \rho^{*}_{a} =  \smb W_{ab} (h_{a}).
\end{equation}
We choose to compute $\rho^*$ by summation, rather than by integration of Equation \ref{eq:continuity-sph}. Given equal mass particles, the smoothing length is adapted by simultaneously solving \ref{eq:densitysum} and
\begin{align} \label{eq:hrho}
  h_a = h_\mathrm{fac} \brkt{\frac{m_a}{\rho_a^*}}^{1/d},
\end{align}
where $d$ is the number of dimensions, and $h_\mathrm{fac}$ is a numerical parameter specifying the smoothing length in terms of the mean particle spacing. The term required in \ref{eq:omegaterm} is then
\begin{align}
  \pder{h_a}{\rho^*_a} = -\frac1d \frac{h_a}{\rho_a^*}.
\end{align}
For details on solving \ref{eq:densitysum} and \ref{eq:hrho} simultaneously we refer the reader to \citet{pricemonaghan07}.

\subsection{Interpolating kernel}
We use a kernel in the form
\begin{align}
   W_{ab}(h_a) = W(h_a,L_{ab}) \equiv \frac{C_{\mathrm{norm}}}{{h_a}^d} f(q),
\end{align}
where $L_{ab}$ is the inter-particle spacing, $C_{\mathrm{norm}}$ is a normalisation constant, and $q=L_{ab}/h_a$.
The standard choice for the dimensionless function $f(q)$ in SPH is the cubic spline. With $h_\mathrm{fac}=1.2$ this results in 57.9 neighbours per particle on average in 3D. In general, more accurate results can be obtained by using a smoother spline such as the quintic, given by
\begin{equation}
   f(q) = \left\{ \begin{array}{ll}
           (3-q)^5 - 6(2-q)^5 + 15(1-q)^5, & 0 \le q < 1, \\
           (3-q)^5 - 6(2-q)^5, & 1 \le q < 2, \\
           (3-q)^5, & 2 \le q < 3, \\
           0, & q \ge 3, \end{array} \right. \label{eq:quinticspline}
\end{equation}
for which $C_{\mathrm{norm}} = [1/120,1/(120\pi)]$ in 1D and 3D. This comes at the cost of a greater number of neighbours per particle; 113 on average in 3D when $h_\mathrm{fac}=1$.
For consistency, we choose to use the quintic spline everywhere, however the cubic spline gives satisfactory results also.
The neighbour finding procedure in more than one dimension is non trivial. We employ a $k$d-tree for this process, the details of which can be found in the \textsc{Phantom} paper, \citet{pricewurstertricco18} (see their section 2.1.7).

For the inter-particle spacing $L_{ab}$ we assume local flatness and thus use the Euclidean distance. We define this distance in Cartesian coordinates as
\begin{equation}
   L_{ab} \equiv \sqrt{\eta_{ij} r^i_{ab} r^j_{ab}},
\end{equation}
where $\eta_{ij}$ is the spatial part of the Minkowski metric, and $r^i_{ab} \equiv x^i_a - x^i_b$ is the line of sight vector pointing from particle $a$ to particle $b$ also in Cartesian coordinates.
Computing the kernel and its derivative is then identical to what is done in standard non-relativistic SPH. In Cartesian coordinates, since $\eta_{ij}$ is a constant, we can write the kernel derivative as
\begin{align}
   \pder{W_{ab}(h_a)}{x^i} = N_i \, F_{ab}(h_a),
\end{align}
where $F_{ab}\leq0$ is the scalar part of the kernel gradient. We define the line of sight projection operator,
\begin{equation}
   N _i \equiv \eta_{ij} \hat r^j_{ab},
\end{equation}
where,
\begin{align}
   \hat r^i_{ab}= \frac{r^i_{ab}}{L_{ab}},
\end{align}
is the line of sight unit vector.

We emphasise that the choice between using the Euclidean distance norm and proper lengths for neighbour distances is arbitrary. Using a different distance metric would simply require a different particle placement to obtain the same conserved density. So this choice reflects how one wishes to place particles to resolve features in the spacetime. We argue that the Euclidean norm is the most natural choice because the particle placement is independent of the metric. For example, a uniformly spaced lattice of particles produces a constant $\rho^*$. Using the proper length for neighbour distances would require stretching or squeezing the initial particle arrangement to compensate for the determinant of the metric in the volume element. It would also require computing a path integral in the given metric between all neighbouring particles.
 
One possible advantage to using the proper distance would be that smoothing spheres would never cross the event horizon. That is, neighbours close to the event horizon would become infinitely distant. However, it is no longer obvious how to set up the particles to achieve a particular $\rho^*$ in this case, and using different metrics would require different particle setups. We demonstrate in Section~\ref{sec:bondi} that using the Euclidean distance norm indeed produces the correct conserved densities close to the black hole so long as the horizon is sufficiently well resolved.
 
\section{Shock capturing} \label{sec:shock-capturing}
It is necessary to introduce dissipative terms in order to capture shocks. As in \citetalias{chowmonaghan97} and \citet{monaghan97} the construction of the dissipative terms is motivated by approximate Riemann solvers used by finite volume codes \citep{martibaezmiralles91}. To demonstrate this, consider the equation set written in conservative form and in one spatial dimension (for simplicity)
\begin{equation}
   \pder{\mathbf u}{t} + \pder{\mathbf F}{x} = 0,
\end{equation}
where $\mathbf{u}=(\rho^*,p_i,e)$ is the vector of conservative variables and $\mathbf{F}(\mathbf{u})$ is the flux. Finite volume schemes require computing the numerical flux between adjacent states $\mathbf{u}_\mathrm{L}, \mathbf{u}_\mathrm{R}$. The simplest approach utilises the `local Lax-Friedrichs' or `Rusanov' flux given by
\begin{equation}
   \mathbf{F}(\mathbf{u}_\mathrm{L}, \mathbf{u}_\mathrm{R}) = \frac12 \left[\mathbf{F}(\mathbf{u}_\mathrm{L}) + \mathbf{F}(\mathbf{u}_\mathrm{R})\right] - \frac{v_\mathrm{sig}}{2} \left( \mathbf{u}_\mathrm{R} - \mathbf{u}_\mathrm{L} \right)
\end{equation}
where $v_\mathrm{sig}$ is the maximum characteristic speed. Importantly, the dissipative term (proportional to $v_\mathrm{sig}$) acts on jumps in the conservative variables.

Although the exact relationship between the dissipation introduced in approximate Riemann solvers and physical dissipation terms is unclear in relativity, since the correct way to formulate viscosity and heat conduction in relativistic hydrodynamics is not well understood \cite[e.g.][]{anderssoncomer07}, the approach used above is standard in GRMHD codes \cite[e.g.][]{martibaezmiralles91,gammiemckinneytoth03,zhangmacfadyen06}.
 
Our SPH shock capturing formulation is based on approximate Riemann solvers, following \citetalias{chowmonaghan97}. If one considers adjacent particles in SPH as left and right states, an appropriate form of the dissipative terms requires a jump in the conservative variables and a maximum signal speed $v_\mathrm{sig}$ between the particles. \citetalias{chowmonaghan97} proposed corresponding terms in the momentum and energy equations in SPH given by
\begin{align}
   \left. \deriv{p_i^a}{t}\right|_\mathrm{diss} &= \alpha_\mathrm{AV} \sum_b \frac{m_b}{\bar{\rho}^*_{ab}} v_\mathrm{sig} \left(p^*_a - p^*_b \right) \pder{W_{ab}(h_a)}{x^i}, \\
      \left. \deriv{e_a}{t}\right|_\mathrm{diss} &= \alpha_\mathrm{AV} \sum_b \frac{m_b}{\bar{\rho}^*_{ab}} v_\mathrm{sig} \left(e_a^* - e_b^* \right) {F}_{ab}(h_a),
\end{align}
where $p^*$ and $e^*$ are computed using only velocities along the line of sight ($v^*$), $\bar{\rho}^*_{ab}=(\rho_a^*+\rho_b^*)/2$, and $\alpha_\mathrm{AV} \simeq 1$ is a numerical parameter to control the amount of dissipation.

The problem with the formulation given by \citetalias{chowmonaghan97} is that, while they proved that the entropy increase was positive definite for cold gas ($w=1$), we found this to no longer be the case when the enthalpy differs from unity. To remedy this, we adapted their formulation to \emph{not} use the full jump in momentum or total energy
 \begin{align}
    p_a^* - p_b^* &= w_a \Gamma_a^* V_a^* - w_b \Gamma_b^* V_b^*, \\
    e_a^*-e_b^* &= p_a^* v_a^* - p_b^* v_b^* + \frac{\alpha_a(1+u_a)}{\Gamma_a^*} - \frac{\alpha_b(1+u_b)}{\Gamma_b^*}, \label{eq:ejumpfull}
 \end{align}
where starred quantities again denote the use of the line of sight velocity $v^*$.
Instead, in order to have a positive definite change in entropy we use a jump only in the `kinetic terms', $ \Gamma_a^* V_a^* - \Gamma_b^* V_b^*$. For the momentum equation we multiply the corresponding term by an average enthalpy, and for the energy equation we multiply by an average of the term $wv$, in order to recover the correct dimensions. That is, we define
\begin{equation}
   p_a^* - p_b^* \equiv \overline{w}_{ab} \left(\Gamma_a^* V_a^* - \Gamma_b^* V_b^*\right),
\end{equation}
where $\overline{w}_{ab}$ denotes a suitable average of $w$ over particles $a$ and $b$.

A second problem with the \citetalias{chowmonaghan97} formulation was that it does not distinguish between viscosity and conductivity. This leads to over-smoothing of contact discontinuities (see Section~\ref{sec:mild-shock}). Showing how to split the dissipation into viscosity and conductivity in relativity is non-trivial. We demonstrate in Section~\ref{sec:mild-shock} and Appendix~\ref{sec:appendix-conductivity} that the relevant splitting can be achieved by defining the energy jump in two parts as follows
\begin{equation} \label{eq:ejump}
   e_a^* - e_b^* \equiv \underbrace{\overline{w}_{ab}\bar{v}^*_{ab} \, (\Gamma_a^*V_a^* - \Gamma_b^*V_b^*)}_\mathrm{viscosity} + \underbrace{\left(\frac{\alpha_a u_a}{\Gamma_a} - \frac{\alpha_b u_b}{\Gamma_b}\right)}_\mathrm{conductivity},
\end{equation}
where $\bar{v}^*_{ab}$ denotes an average of $v^*$ over particles $a$ and $b$ (see below).
This is equivalent to defining a relativistic thermal energy
\begin{equation}
\tilde u \equiv \frac{\alpha u}{\Gamma}.
\end{equation}

Our third change is more cosmetic. Rather than using separate averages for $\overline{w}_{ab}$ and $\bar{v}_{ab}^*$, we average them along with the density, signal speed and kernel derivative in the manner of \citet{pricefederrath10} following the non-relativistic terms in \textsc{Phantom} \citep{pricewurstertricco18}. That is, we write the viscosity as a $P + q$ term in the momentum and energy equations.

\subsection{Artificial viscosity}
Equations \ref{eq:momentum-sph} and \ref{eq:energy-sph} with our dissipation become
\begin{align}
   \deriv{p_i^a}{t}  = & -\smb \left[ \frac{(P_a + q_a)}{\rho^{*2}_a} D^a_i + \frac{(P_b + q_b)}{\rho^{*2}_b} D^b_i \right] + f_i^a,                                     \label{eq:momentum-sph-dissipation} \\
   \deriv{e_a}{t}       = & -\smb \left[ \frac{(P_a + q_a)}{\rho^{*2}_a} v_b^i D^a_i + \frac{(P_b + q_b)}{\rho^{*2}_b} v_a^i D^b_i \right] \nonumber \\
                         & \qquad \qquad +  \Pi^a_{\mathrm{cond}} + \Lambda_a,                                                                                                                                \label{eq:energy-sph-dissipation}
\end{align}
where $q_a$ and $q_b$ encapsulate the artificial viscosity, and $\Pi^a_{\mathrm{cond}}$ the artificial conductivity. For added simplicity in notation, we have defined
\begin{align}
   D^a_i \equiv \frac{\sqrt{-g_a}}{\Omega_a} \pder{W_{ab}(h_a)}{x^i}, \\
   D^b_i \equiv \frac{\sqrt{-g_b}}{\Omega_b} \pder{W_{ab}(h_b)}{x^i}.
\end{align}
Our artificial viscosity terms are
\begin{align}
   q_a &= -\frac12 \alpha_\mathrm{AV} \rho^*_a v_{\mathrm{sig},a} w_a \brkt{\Gamma^*_a V^*_a - \Gamma^*_b V^*_b}, \\
   q_b &= -\frac12 \alpha_\mathrm{AV} \rho^*_b v_{\mathrm{sig},b} w_b \brkt{\Gamma^*_a V^*_a - \Gamma^*_b V^*_b},
\end{align}
and involve a fluid velocity that is along the line of sight between particles, similar to \citetalias{chowmonaghan97}, i.e.
\begin{align}
   V^*_a &= N_i V^i_a, \\
   \Gamma^*_a &= \frac{1}{\sqrt{1-{V_a^*}^2}}.
\end{align}
If the particles are separating, $N_i (v_a^i - v_b^i) \geq 0$, we set $q_a=q_b=0$.

\subsection{Artificial conductivity}
Our artificial conductivity term is given by
\begin{align}
   \Pi_{\mathrm{cond}} &= \frac{\alpha_u}{2} \smb \brkt{\tilde{u}_a - \tilde{u}_b} \left(\frac{v^u_{\mathrm{sig},a} G_a}{\rho^*_a} + \frac{v^u_{\mathrm{sig},b} G_b}{\rho^*_b} \right),
\end{align}
where again for simplicity we define
\begin{align}
   G_a \equiv \frac{\sqrt{-g_a}}{\Omega_a} \pder{W_{ab}(h_a)}{x^i} \hat r^i_{ab} = D^a_i \hat r^i_{ab},
\end{align}
which in Cartesian coordinates is
\begin{align}
   G_a = \frac{\sqrt{-g_a}}{\Omega_a} F_{ab}(h_a).
\end{align}
Note that the terms $\alpha_\mathrm{AV}$ and $\alpha_u$ allow us to control the amount of artificial viscosity and conductivity, respectively. These should not be confused with $\alpha_a$ and $\alpha_b$ which are the lapse function at the positions of particles $a$ and $b$.

\subsection{Signal speed --- viscosity}
When constructing an approximation for the maximum signal speed in the artificial viscosity, $v_\mathrm{sig}$, we impose two requirements.
First, the signal speed must not exceed the speed of light.
Second, the signal speed should reduce to the form used in standard SPH codes in the non-relativistic limit, namely
\begin{equation}
   v_{\mathrm{sig},a} = \alpha_\mathrm{AV} c_{\mathrm{s},a} + \beta_\mathrm{AV} \left| \mathbf{v}_{ab} \cdot \mathbf{\hat r}_{ab} \right|,
\end{equation}
where $c_{\mathrm{s}}$ is the sound speed, $\mathbf{v}_{ab}=\mathbf{v}_{a}-\mathbf{v}_{b}$ is the relative velocity, and $\mathbf{\hat r}_{ab}$ is the unit vector along the line of sight. The numerical parameters $[\alpha_\mathrm{AV},\beta_\mathrm{AV}]$ control the amount of dissipation. Most importantly, the signal speed should involve a relative velocity between the particles to avoid excessive dissipation when there is a bulk velocity and individual particle speeds are high.

Because of these requirements, we do not use the signal speeds proposed by \citetalias{chowmonaghan97}, since in some of their formulations the maximum signal speed may exceed the speed of light. We also do not use the signal speed proposed by \citet{rosswog10}, based on the eigenvalues of the Euler equations. Their formulation, although limited by the speed of light, involves the absolute velocity of the SPH particles. This means their viscosity term is no longer Galilean invariant. We therefore require a signal speed that uses only the relative velocity between particles. To this end we propose a signal speed given by
\begin{align}
   v_{\mathrm{sig},a} = \frac{c_{\mathrm{s},a}+ \left| V_{ab}^* \right| }{1+c_{\mathrm{s},a} \left| V_{ab}^* \right|},
\end{align}
where the relative velocity between particles, considered along the line of sight, is defined as
\begin{equation}
   V_{ab}^* \equiv \frac{V_a^* -V_b^*}{1- V_a^* V_b^*},
\end{equation}
and the relativistic sound speed is given by
\begin{align}
   c_{\mathrm s} = \sqrt{\frac{P \adgamma}{\rho w}}.
\end{align}
We use the velocities in the frame of the local Eulerian observer $V^i$ rather than the coordinate velocities $v^i$, since only $V^i$ is a physical velocity that is limited by the speed of light, that is $| V_i V^i |\leq1$. The coordinate velocity on the other hand, is not strictly under the same restriction.

\subsection{Signal speed --- conductivity}
For simplicity one may assume the same signal speed for both viscosity and conductivity. That is, $v^u_{\mathrm{sig}} = v_{\mathrm{sig}}$. In general however, one should use different speeds since the conductivity is mainly required at contact discontinuities which travel at a different speed to the shock. \citet{price08} proposed using a signal speed based on the relative pressure jump, designed to eliminate the pressure blip at contact discontinuities. We generalise this to the relativistic case using
\begin{align} \label{eq:vsigu1}
   v^u_{\mathrm{sig}} = \min \left(1,\sqrt { \frac{|P_a-P_b|}{\overline{w \rho}_{ab}} }\right).
\end{align}
For cases where equilibrium pressure gradients exist due to other physics (e.g. gravity) it is better to simply use the relative speeds \citep{price08,wadsleyveeravallicouchman08}, giving
\begin{equation} \label{eq:vsigu2}
   v^u_{\mathrm{sig}} = |V_{ab}^*|,
\end{equation}
when generalised to the relativistic case. In the non-relativistic limit, both eq.~\ref{eq:vsigu1}~and~\ref{eq:vsigu2} reduce to the non-relativistic implementation in \textsc{Phantom} \citep{pricewurstertricco18}, where the choice is based on whether or not gravity and/or external forces are used.

 The main advantage of the above signal speeds is that conductivity becomes $\mathcal{O}(h^2)$ in the numerical code, rather than first order. For most of the tests in this paper, we simply use $v^u_{\mathrm{sig}} = v_{\mathrm{sig}}$ with a reduced value of $\alpha_u$. In section~\ref{sec:polytrope} we demonstrate the benefit of using Eq.~\ref{eq:vsigu2} in the case of a self-gravitating, oscillating polytrope.

\section{Evolving entropy}
Instead of the total specific energy $e$, we evolve an entropy variable (cf. \citealt{springelhernquist02}) in order to guarantee positivity of the thermal energy. We define
\begin{equation}
   \label{eq:polyk}
   K \equiv \frac{P}{\rho^{\adgamma}}.
\end{equation}
Taking the time derivative and using Equation~\ref{eq:eos} we find
\begin{align}
   \deriv{K}{t} &= \brkt{\adgamma - 1} \rho^{1 - \adgamma} \brkt{ \deriv{u}{t} - \frac{P}{\rho^2}\deriv{\rho}{t} }, \label{eq:polyk-dt-1} \\
   &= \frac{K}{u} \brkt{ \deriv{u}{t} - \frac{P}{\rho^2}\deriv{\rho}{t} }.                                                \label{eq:polyk-dt-2}
\end{align}
Comparing this to the second law of thermodynamics
\begin{align}
   T \deriv{s}{t} &= \deriv{u}{t} - \frac{P}{\rho^2} \deriv{\rho}{t},
\end{align}
we see that in the absence of dissipation
\begin{equation}
   \deriv{K}{t} = 0.
\end{equation}
Furthermore, using Equation~\ref{eq:idealgaslaw}, $K$ can be related to the specific entropy $s$
\begin{equation}
   \frac{1}{K} \deriv{K}{t} = \brkt{ \adgamma -1 } \frac{\mu m_\mathrm{H} }{k_\mathrm{B}} \deriv{s}{t}.
\end{equation}
More generally, we can rewrite Equation \ref{eq:polyk-dt-2} in terms of time derivatives of the conserved variables ($\rho^*$, $p_i$, $e$), giving
\begin{align}
   \label{eq:polyk-dt-3}
   \deriv{K}{t} = \frac{U^0 K}{u} \left[  \brkt{ \deriv{e}{t} - \Lambda }  - v^i \brkt{\deriv{p_i}{t} - f_i} - \frac{P \sqrt{-g}}{{\rho^*}^2} \deriv{\rho^*}{t} \right],
\end{align}
which, after substituting Equations~\ref{eq:continuity-sph},~\ref{eq:momentum-sph-dissipation}, and~\ref{eq:energy-sph-dissipation}, gives
\begin{equation} \label{eq:entropy-sph}
   \deriv{K_a}{t} = \frac{U^0_a K_a}{u_a} \brkt{ \Pi^a_\mathrm{cond} + \smb \frac{q_a \brkt{v_a^i - v_b^i}}{{\rho_a^*}^2} D_i^a }.
\end{equation}
Thus, the evolution of $K$ is governed purely by the dissipative terms. An important distinction between SPH and finite-volume schemes is that when evolving $K$, energy remains exactly conserved (to the accuracy of the time integration algorithm), since Eq~\ref{eq:polyk-dt-3} holds to round-off error. In Appendix~\ref{sec:appendix-pos_def} we demonstrate that Eq.~\ref{eq:entropy-sph} is always positive (for special relativity) thus ensuring that entropy always increases.

\section{Time Integration} \label{sec:timestepping}
Relativistic hydrodynamics, in the absence of dissipation, forms a Hamiltonian system where the canonical variables are the positions $x^{i}$ and momenta $p_{i}$. Ideally therefore one should use a geometric integrator that preserves the Hamiltonian properties. The main complication is that the Hamiltonian is non-separable --- it cannot be separated into kinetic, thermal, and potential energy terms.

\subsection{Generalised leapfrog method} \label{sec:generalised-leapfrog}
\citet{leimkuhlerreich05} suggest the `generalised leapfrog method' for the symplectic integration of Hamilton's equations with a non-separable Hamiltonian. With $p_{i}$ and $x^{i}$ as the conjugate variables, their scheme is given by
\begin{align}
   p_{i}^{n+\frac12} & = p_{i}^{n} + \frac{\Delta t}{2} \deriv{p_i}{t} (p^{n+\frac12}_{i}, x^{i, n}), \\
   x^{i, n+1} & = x^{i, n} + \frac{\Delta t}{2} \left[ \deriv{x^i}{t}  (p_{i}^{n+\frac12}, x^{i, n}) + \deriv{x^i}{t} (p_{i}^{n+\frac12}, x^{i, n+1}) \right],\\
   p_{i}^{n+1} & = p_{i}^{n+\frac12} + \frac{\Delta t}{2} \deriv{p_i}{t} (p^{n+\frac12}_{i}, x^{i, n+1}),
\end{align}
where the superscripts $n+m$ refer to quantities evaluated at times $t^{n+m} = t^{n} + m\Delta t$.
It may be seen that this reduces to the standard Kick-Drift-Kick leapfrog scheme when $v^{i} \equiv p_{i}$ and when the acceleration ${\d}p_{i}/{\d}t$ depends only on position. It is also time-reversible and symplectic. However, a general property of geometric integrators for non-separable Hamiltonians is that they are implicit in one or more of the steps \citep[e.g.][]{yoshida93}. In the above integrator this occurs in the first `Kick' step, where the momentum at the half step appears on both sides of the equation, and in the `Drift' step where the position at the full step appears on both sides.

\subsection{Alternative generalised leapfrog method} \label{sec:alternative-leapfrog}
For GRSPH, evaluating ${\d}p_{i}/{\d}t (p_{i}, x^{i})$ means performing a sum over particles, while evaluation of $ {{\d}x^{i}}/{{\d}t} (p_{i}, x^{i})$ means performing a primitive variable solve (see Section \ref{sec:cons2prim}) which can be performed independently for each particle. Hence the previous integrator would require two calls to the summation routines, since ${\d}p_{i}/{\d}t$ used in the second Kick step is different from the one required for the subsequent first Kick step at the start of the next timestep. For this reason we propose a similar integrator where the symmetrisation occurs in the drift step. Our alternative integrator is given by
\begin{align}
   p_{i}^{n+\frac12} & = p_{i}^{n} + \frac{\Delta t}{2} \deriv{p_i}{t} (p^{n}_{i}, x^{i, n}), \\
   x^{i, n+1} & = x^{i, n} + \frac{\Delta t}{2} \left[ \deriv{x^i}{t}  (p_{i}^{n+\frac12}, x^{i, n}) +   \deriv{x^i}{t} (p_{i}^{n+\frac12}, x^{i, n+1}) \right],\\
   p_{i}^{n+1} & = p_{i}^{n+\frac12} + \frac{\Delta t}{2} \deriv{p_i}{t} (p^{n+1}_{i}, x^{i, n+1}).
\end{align}
This integrator retains the time symmetry of the previous integrator, however it does not exactly conserve angular momentum. The difference is that the implicit steps are now in the Drift step, as previously, and in the second Kick step. Importantly, the derivative of the momentum used in the second Kick step is the same as that used at the beginning of the subsequent step, meaning that it can be re-used. The integrator is also easier to start, because it merely requires an evaluation of the momentum derivative involving position and momentum from the initial conditions.

The implicit solve in the Drift step (in the position derivative) involves only the primitive variable solver, so is independent for each particle. The implicit nature of the second Kick step (in the momentum derivative) in principle involves multiple loops over neighbouring particles, but this is similar to our existing approach to dealing with viscosity and other velocity-dependent terms in non-relativistic SPH (see below). With a good prediction of $p^{n+1}_{i}$, only one call to the summation routines is required.

\subsection{Hybrid leapfrog, and implementation in GRSPH} \label{sec:hybrid-leapfrog}
Since our modified integrator does not exactly conserve angular momentum, we implement a hybrid of the two algorithms in order to retain the benefits of both. Taking an operator splitting approach similar to that of the reversible reference system propagator algorithm (RESPA) derived by \citet{tuckermanbernemartyna92} we split the acceleration into `long range' and `short range' forces (see \citealt{pricewurstertricco18} for an implementation in SPH). In our case the `long range' forces are the SPH forces, while the `short range' forces are the external forces arising from the curvature of the spacetime. These correspond to the pressure gradient term and $f_i$ in the momentum equation, respectively. i.e.
\begin{empheq}{align}
   f_i^{\, \mathrm{sph}} \equiv & - \frac{1}{\rho^*} \pder{(\sqrt{-g} P)}{x^i}, \\
   f_i^{\, \mathrm{ext}}  \equiv & \quad f_i.                                                   % \frac{\sqrt{-g}}{2\rho^{*}} \brkt{ T^{\mu\nu} \pder{g_{\mu\nu}}{x^i} }
\end{empheq}
We use the `long range' forces $f_i^{\, \mathrm{sph}}$ to update the momentum of each particle and use the `short range' forces $f_i^{\, \mathrm{ext}}$ in updating their positions. Our complete hybrid algorithm is therefore
\begin{align}
   p_i & \rightarrow p_i + \frac{\Delta t_\mathrm{sph}}2 f_i^{\mathrm{sph}}(p_i,x^i),
\end{align}
\begin{empheq}[left=\textrm{substeps} \empheqlbrace]{align}
   \tilde p_i & = p_i + \frac{\Delta t_\mathrm{ext}}2 f_i^{\mathrm{ext}} (\tilde p_i,x^i), \label{eq:implicit-momentum} \\
   p_i & \rightarrow \tilde p_i,\\
   \tilde x^{i} & = x^i + \frac{\Delta t_\mathrm{ext}}2 \left[ v^i(p_i,x^i) + v^i(p_i,\tilde x^i) \right], \label{eq:implicit-position} \\
   x^i  & \rightarrow \tilde x^i,\\
   p_i & \rightarrow p_i + \frac{\Delta t_\mathrm{ext}}2 f_i^{\mathrm{ext}}(p_i,x^i),
\end{empheq}
\begin{align}
    \tilde p_i &= p_i + \frac{\Delta t_\mathrm{sph}}2 f_i^{\mathrm{sph}}(\tilde p_i,x^i), \label{eq:implicit-momentum-sph} \\
   p_i & \rightarrow \tilde p_i,
\end{align}
where the position update is done over $m$ substeps (indicated by the equations in braces), such that $m = \mathrm{int}(\Delta t_\mathrm{ext} / \Delta t_\mathrm{sph} + 1)$ and is the minimum over all particles. We denote the updating of a variable by an arrow, some of which first require solving and implicit equation, namely Equations~\ref{eq:implicit-momentum}, \ref{eq:implicit-position}, and \ref{eq:implicit-momentum-sph}.

The combination of the two algorithms ensures that one of the SPH force calculations can be reused between steps, reducing computational cost, and that angular momentum is conserved in the case where the SPH forces can be neglected. More importantly it ensures that our method is not restricted by the external force timestep, since computing $f_i^{\, \mathrm{ext}}$ is  cheap whereas $f_i^{\, \mathrm{sph}}$ is expensive.

It is useful to point out that in the absence of metric gradients the hybrid method reduces to our proposed scheme in section \ref{sec:alternative-leapfrog}, while for pressureless fluid it reduces to the method of \citet{leimkuhlerreich05} in section \ref{sec:generalised-leapfrog}.

\subsubsection{Evaluating implicit steps}
To evaluate Equations~\ref{eq:implicit-momentum} and \ref{eq:implicit-position}, we employ two first-order predictions which we refer to as $p^{*}_{i}$ and $x^{i,*}$.
That is, for Eq~\ref{eq:implicit-momentum} we do
\begin{align}
   p_i^*       & = p_i + \frac{\Delta t_\mathrm{ext}}2 f_i^{\, \mathrm{ext}} (p_i, x^i), \\
   \tilde p_i & = p_i^*  + \frac{\Delta t_\mathrm{ext}}2 \left[ f_i^{\, \mathrm{ext}} (p^*_i, x^i)  -  f_i^{\, \mathrm{ext}} (p_i, x^i)\right],           \label{eq:pcorr}
\end{align}
and for evaluating Eq~\ref{eq:implicit-position} we do
\begin{align}
   x^{i, *}         & = x^i + \Delta t_\mathrm{ext}  v^i  (p_i, x^i),  \\
   \tilde x^i    & = x^{i, *} + \frac{\Delta t_\mathrm{ext}}{2} \left[ v^i (p_i, x^{i, *}) - v^i (p_i, x^i)\right]. \label{eq:xcorr}
\end{align}
To complete the implicit step we check that
\begin{align}
   \vert \tilde p_i - p_i^* \vert &< \epsilon_p, \\
   \vert \tilde x^i - x^{i,*} \vert &< \epsilon_x,
\end{align}
after Equations~\ref{eq:pcorr} and \ref{eq:xcorr}, respectively, where $\epsilon_{x}$ and $\epsilon_{p}$ are small tolerances.
If these criteria are violated we iterate Equation~\ref{eq:pcorr} (or Eq.~\ref{eq:xcorr}) by a fresh evaluation of the corresponding derivative with the updated value, that is, simply replacing $p_i^*$ with $\tilde p_i$, and $p_i$ with $p_i^*$ on the right hand side of Eq.~\ref{eq:pcorr} (and similarly with $x^{i,*}$ and $x^i$ in Eq.~\ref{eq:xcorr}).

Equation~\ref{eq:implicit-momentum-sph} is evaluated in the same manner as \ref{eq:implicit-momentum} but with $f_i^{\, \mathrm{sph}}$ and $\Delta t_\mathrm{sph}$ instead of $f_i^{\, \mathrm{ext}}$ and $\Delta t_\mathrm{ext}$. We use the error criterion
\begin{align} \label{eq:}
   e = \frac{\max (|\Delta p|^2)}{\mathrm{rms}(|p|)} < \epsilon_p^\mathrm{sph},
\end{align}
where we define $|p| \equiv \sqrt{p_i p_j \eta^{ij}}$ and $\Delta p_i \equiv \tilde p_i - p_i^*$, consistent with the implementation of the equivalent implicit step in \textsc{Phantom} \cite[cf.][]{pricewurstertricco18}. The maximum is taken over all the particles in the numerator, while the denominator evaluates the root mean square, also over all the particles.

In general, iterations of Equations~\ref{eq:implicit-momentum} and \ref{eq:implicit-position} are cheap, while iterations of Equation~\ref{eq:implicit-momentum-sph} are expensive, so $\epsilon_p$ and $\epsilon_x$ can be set arbitrarily small while $\epsilon_p^\mathrm{sph}$ must allow for an efficient code. By default the tolerance is set to $\epsilon_p^\mathrm{sph} = 10^{-2}$ in \textsc{Phantom}. To try to prevent iterations from occurring, we constrain the timestep by
\begin{equation} \label{eq:tolv}
   \Delta t = \min \left( \Delta t, \frac{\Delta t}{\sqrt{\max(e) / \epsilon_p^\mathrm{sph}}} \right),
\end{equation}
since an iteration is as expensive as halving the timestep.

In terms of storage, we require only one position and momenta for each particle, but storage of two derivatives (current and previous) of both $v^i$ and $f_i$.

\subsection{Timestep constraint}
We adopt the usual stability conditions, namely
\begin{align} \label{eq:dt}
   \Delta t^a_\mathrm{sph} = \min \left(\frac{C_\mathrm{C} \, h_a}{\max (v_{\mathrm{sig},a})}, C_\mathrm{f} \sqrt{ \frac{h_a}{\left| f_a^{\mathrm{sph}}\right|} } \right),
\end{align}
where $C_\mathrm{C}$ and $C_\mathrm{f}$ are the usual safety factors, set to 0.3 and 0.25, respectively by default in \textsc{Phantom}, the maximum $v_\mathrm{sig}$ is over neighbours to particle $a$, and $|f| \equiv \sqrt{f_i f_j \eta^{ij}}$.  In our tests we use global timestepping where the global timestep $\Delta t_\mathrm{sph}$ is set to the minimum value of $\Delta t_a^\mathrm{sph}$ on the particles, although our scheme is simple to generalise to the case where particles have individual timesteps.
The timestep constraint for external force substeps is similarly
\begin{align} \label{eq:dtf}
   \Delta t_a^\mathrm{ext} = C_\mathrm{f} \sqrt{ \frac{h_a}{\left| f_a^{\mathrm{ext}}\right|} },
\end{align}
and the global $\Delta t_\mathrm{ext}$ is the minimum of $\Delta t_a^\mathrm{ext}$ on the particles.

\begin{figure*}
   \begin{center}
      \includegraphics[width=0.95\textwidth]{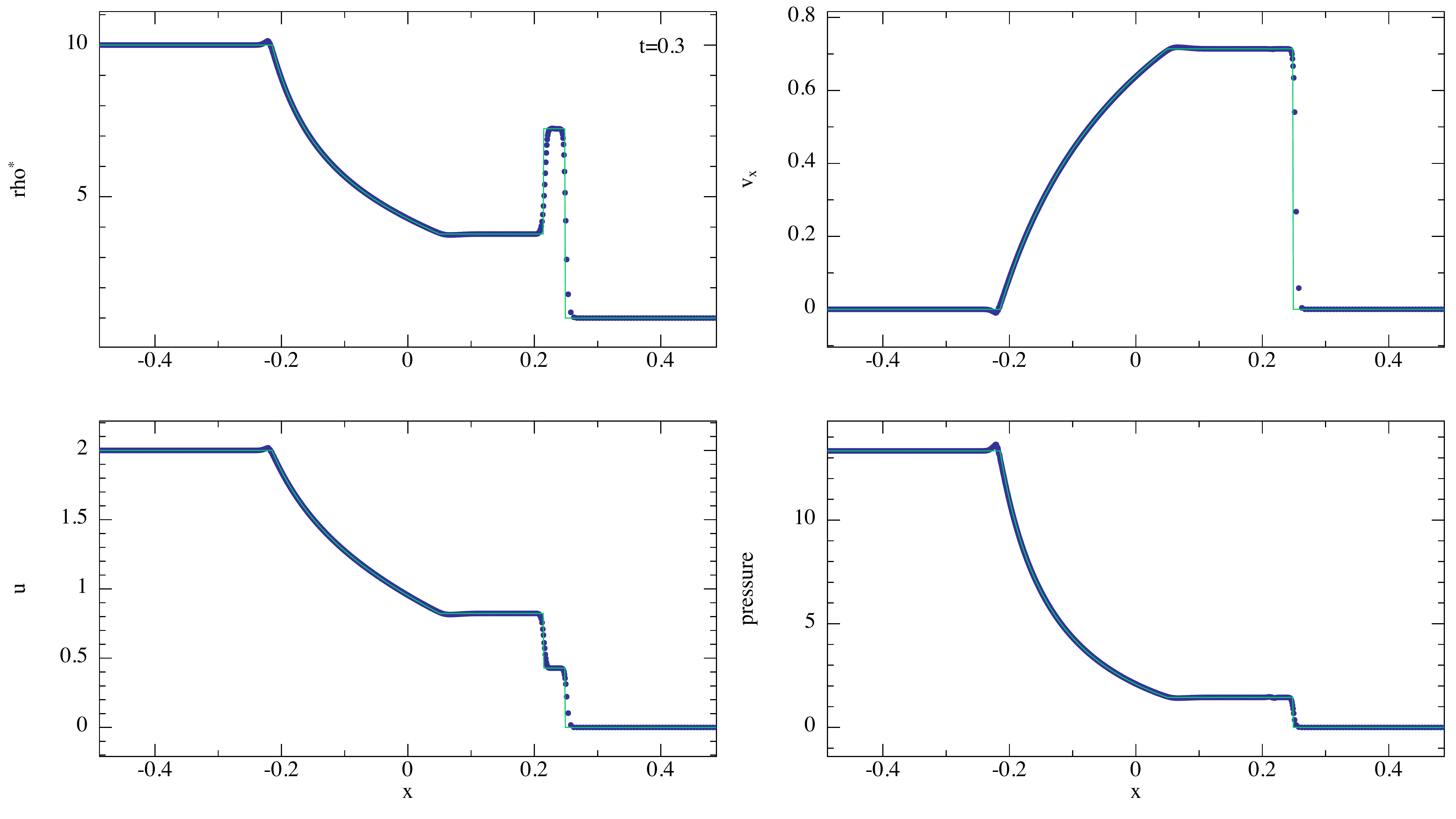}
      \caption{Mildly relativistic shock performed in 1D, with artificial conductivity. Blue circles show SPH particles, while green line shows the analytic solution from \citet{martimuller94}. The initial particle spacing to the left of the interface is 0.0005.}
      \label{fig:shocktube-mild-1D}
   \end{center}
   \vspace{1cm}
   \begin{center}
      \includegraphics[width=0.95\textwidth]{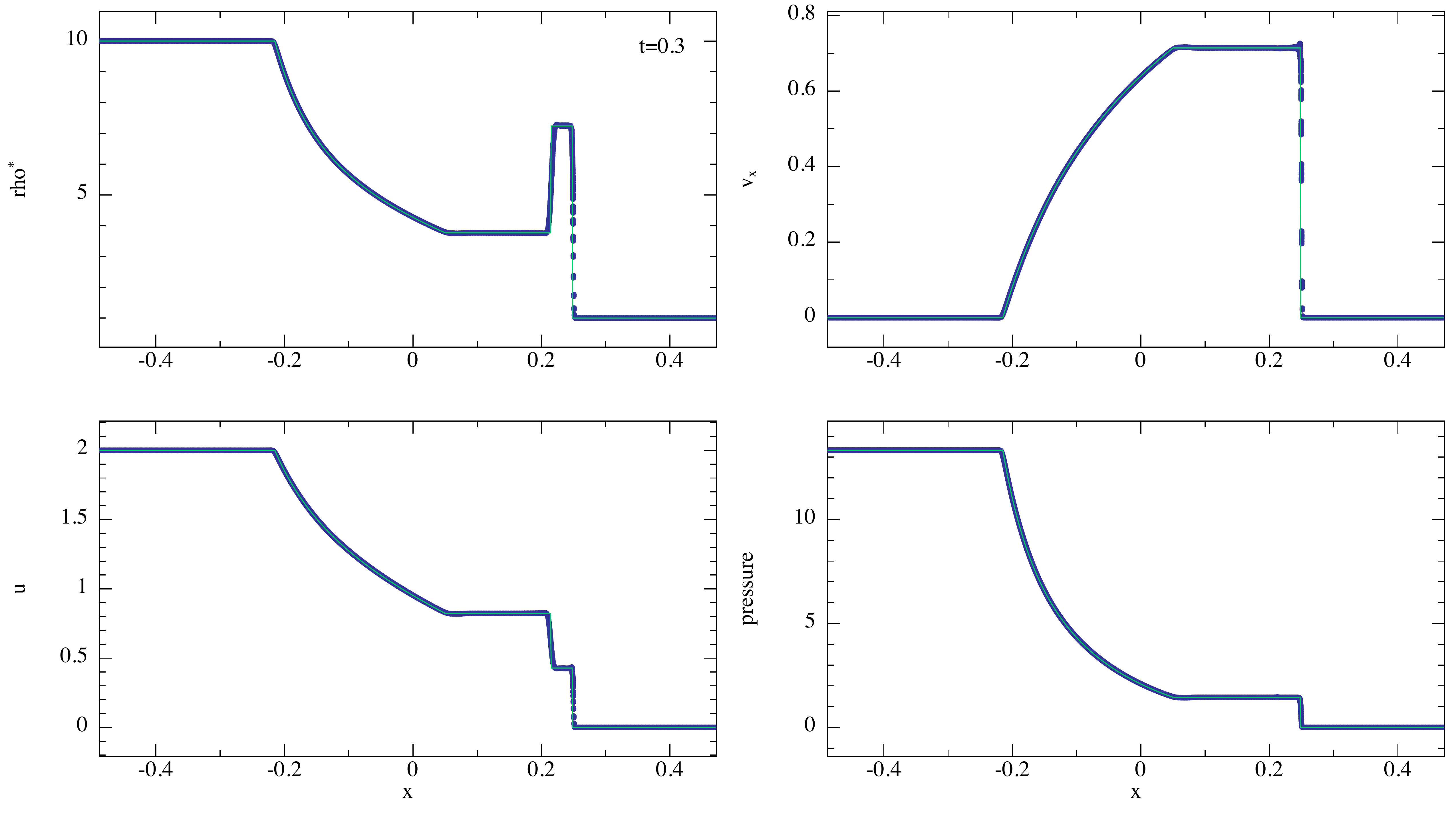}
      \caption{As in Figure~\ref{fig:shocktube-mild-1D} but in 3D. The initial particle spacing to the left of the interface is 0.0005. The solution is similar to that obtained with our 1D code.}
	\label{fig:shocktube-mild-3D}
   \end{center}
\end{figure*}

\begin{figure*}
   \begin{center}
      \includegraphics[width=0.95\textwidth]{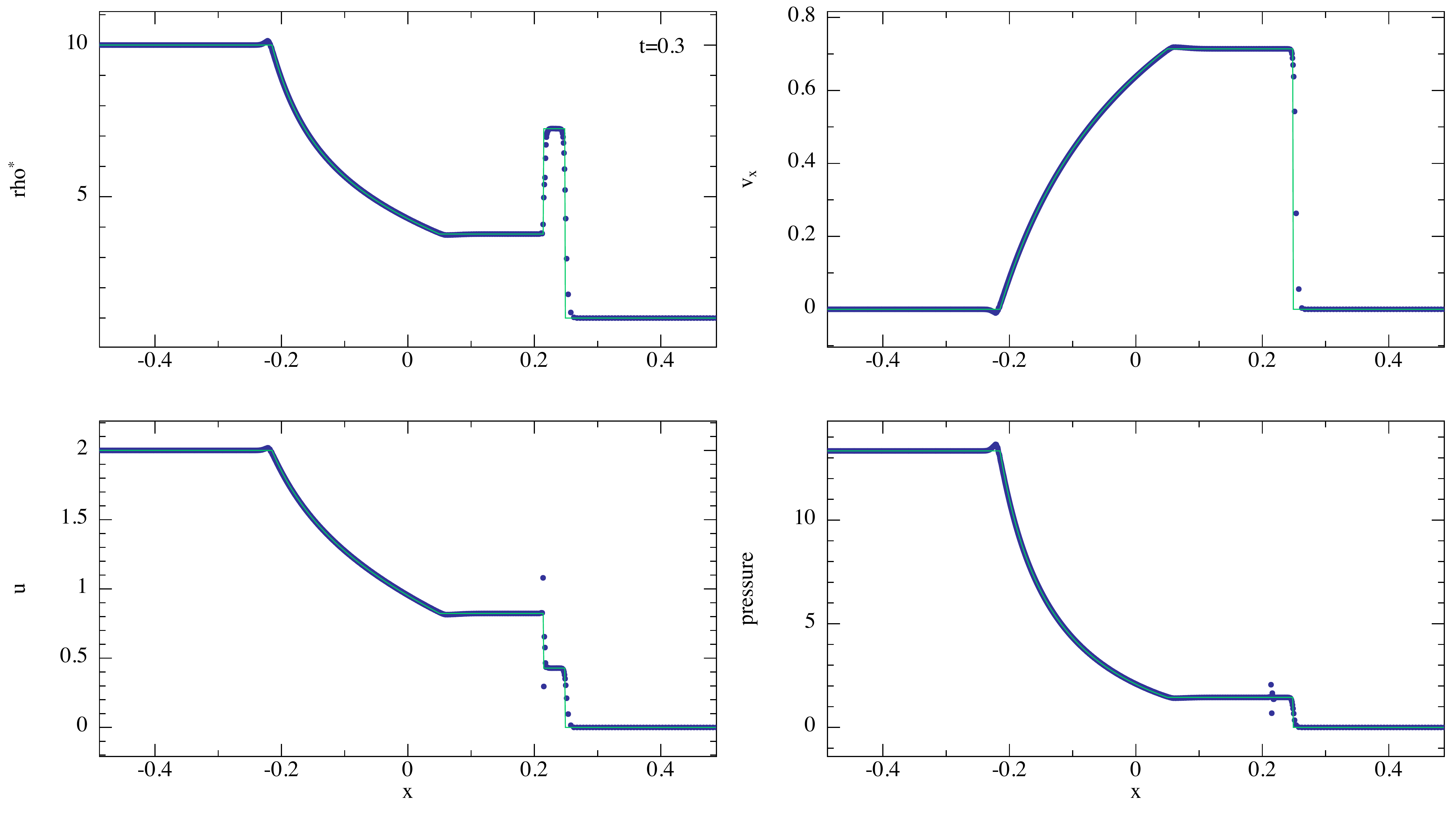}
      \caption{Mildly relativistic shock performed in 1D, without artificial conductivity ($\alpha_u = 0$). Blue circles show SPH particles, while green line shows the analytic solution from \citet{martimuller94}. The initial particle spacing to the left of the interface is 0.0005. A `blip' in the pressure and thermal energy is visible at the contact discontinuity.}
      \label{fig:shocktube-mild-1D-nocond}
   \end{center}
\end{figure*}

\section{Recovery of primitive variables} \label{sec:cons2prim}
Computing the right hand side of Equations~\ref{eq:continuity-sph},~\ref{eq:momentum-sph}, and~\ref{eq:energy-sph} explicitly requires the primitive variables $(\rho, u, v^i, P)$. We must therefore invert the set of equations describing the conserved variables (Eq.~\ref{eq:rhostar_3p1},~\ref{eq:pmom_3p1}, and~\ref{eq:etot_3p1}) at every timestep. To do this we use the recovery scheme given by \citet{tejeda12}.
\begin{enumerate}
   \item Use the equation of state to solve for the enthalpy as a function of density and pressure only,
   \begin{equation}
      \label{eq:cons2prim_enth}
      w = w(\rho,P).
   \end{equation}
   \item Express $\rho$, $P$ and $v^i$ as functions of only $w$ and the conserved variables,
   \begin{align}
      \rho(w) &= \frac{\rho^*}{\sqrt{\gamma} ~\Gamma(w)},                                                   \label{eq:cons2prim_rho}\\
      P(w)     &= \frac{\rho^*}{\alpha \sqrt{\gamma}} \brkt{ w ~\Gamma(w) \alpha - e - p_i \beta^i }, \label{eq:cons2prim_pre} \\
      v_i(w)   &= \frac{\alpha ~p_i}{w ~\Gamma(w)} - \beta_i,                                               \label{eq:cons2prim_v}
   \end{align}
   where $v_i = \gamma_{ij} ~v^j$, and $\Gamma(w)$ can be calculated from Eq.~\ref{eq:pmom_3p1} as,
   \begin{equation}
      \label{eq:cons2prim_gamma}
      \Gamma(w) = \sqrt{1 + \frac{p^i p_i}{w^2}}.
   \end{equation}
   \item Given Equations~\ref{eq:cons2prim_rho} and~\ref{eq:cons2prim_pre}, solve Equation~\ref{eq:cons2prim_enth} for the enthalpy using a standard root-finding algorithm. (See Appendix \ref{sec:appendix-cons2prim} for more detail). \label{step:rootfind}
   \item Compute $v_i$ using Equation~\ref{eq:cons2prim_v}, and raise the index using the three-metric $v^i = \gamma^{ij} v_j$.
\end{enumerate}

\section{Numerical tests}
We split the testing of our numerical method into three stages.
\begin{enumerate}
   \item First, we developed a dedicated 1D special relativistic SPH code to test the shock capturing terms in section~\ref{sec:shock-capturing}, and the conservative-to-primitive algorithm in section~\ref{sec:cons2prim}. For the time integration in this code we used the standard 2nd order Runge-Kutta method (RK2), with time steps constrained by Equation~\ref{eq:dt}). We use RK2 instead of the algorithm described in section~\ref{sec:timestepping} to show that the particular timestepping algorithm used is not important for shock tube tests.
   \item Next, we developed a 3D N-body code to test the force terms involving the metric derivatives $f_i$. This involved integrating Equation~\ref{eq:relhydro2} with $P=0$, i.e.
   \begin{align}
       \deriv{p_i}{t} = f_i \quad \mathrm{and} \quad \deriv{x^i}{t} = v^i.
    \end{align}
Although this is equivalent to solving the geodesic equation, it allowed us to test our time integration scheme (section~\ref{sec:hybrid-leapfrog}), as well as the conservative-to-primitive algorithm, again, but with more complicated metrics than the Minkowski metric. We implemented the Schwarzschild metric in both spherical and Cartesian coordinates, along with the Kerr metric in Boyer-Lindquist coordinates (both spherical and Cartesian). For this code we used a fixed time step.
   \item Finally, after testing individual aspects of our method in the previous two codes, we incorporated them all into \textsc{Phantom}, a fully 3-dimensional SPH code \citep{pricewurstertricco18}. In this code, we implemented the full, hybrid-Leapfrog method for the time integration, as described in section~\ref{sec:hybrid-leapfrog}. The timestep is constrained according to Equations~\ref{eq:tolv}, \ref{eq:dt}, \& \ref{eq:dtf}.
\end{enumerate}
For SPH simulations we always use equal mass particles, and assume an adiabatic gas with $\adgamma=\sfrac{5}{3}$. For the artificial dissipation we use $\alpha_\mathrm{AV}=1$ and $\alpha_u = 0.1$, unless otherwise stated.

We quantify the error in our simulations with the dimensionless $L_2$ error
\begin{equation} \label{eq:l2error}
   L_2 = \sqrt{\frac{1}{N \max(y_\mathrm{exact})} \sum_{i=1}^{N} |y_i - y_{\mathrm{exact}} | ^2 },
\end{equation}
where $N$ is the number of points and $y_{\mathrm{exact}}$ is the exact solution at point $i$. 

\begin{figure*}
   \begin{center}
      \includegraphics[width=0.95\textwidth]{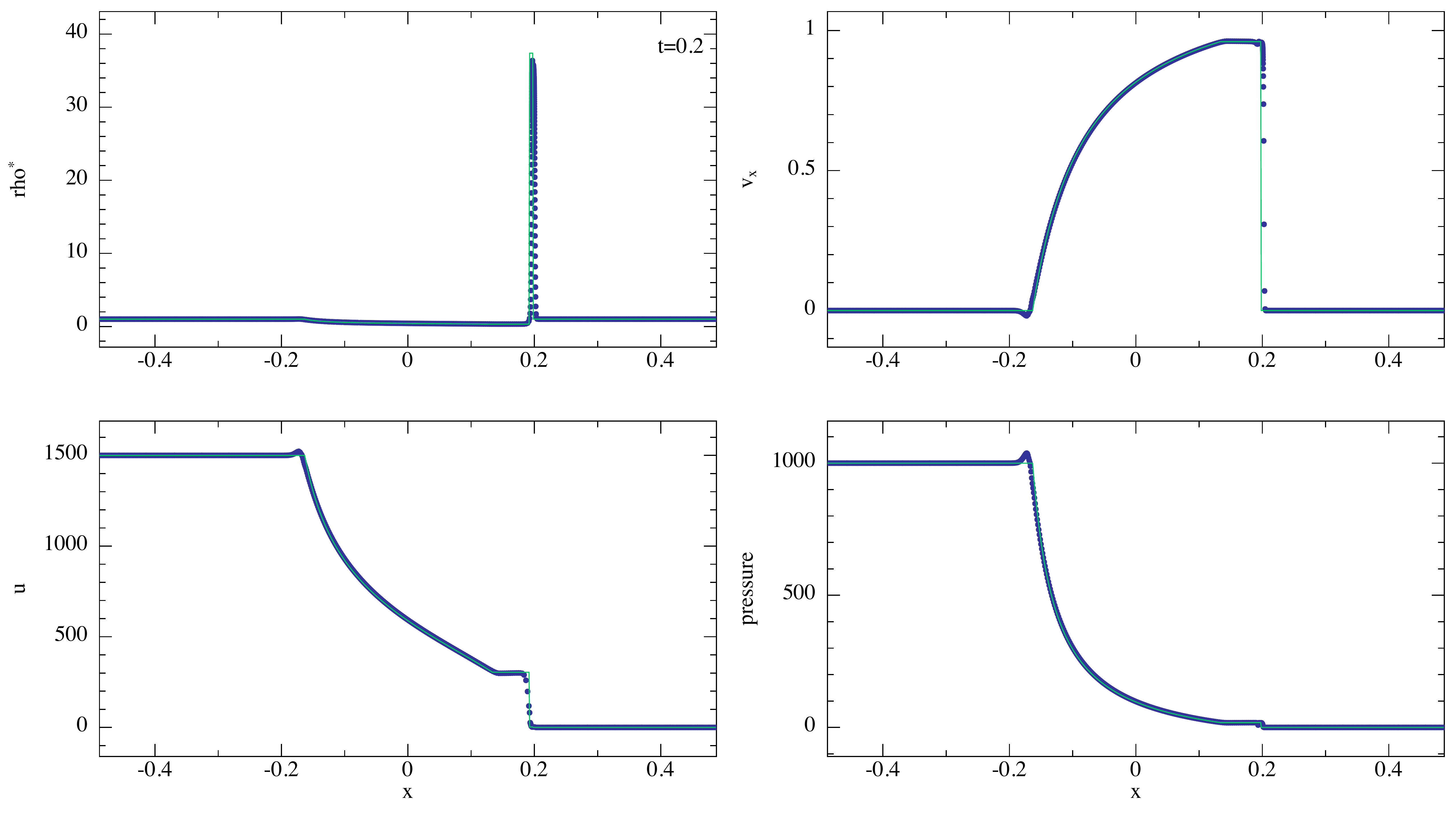}
      \caption{Results of the blast wave performed in 1D. Blue circles show SPH particles, while green lines show the analytic solution from \citet{martimuller94}. The initial particle spacing on both sides of the interface is 0.001.}
      \label{fig:blastwave-1D}
   \end{center}
   \vspace{1cm}
   \begin{center}
      \includegraphics[width=0.95\textwidth]{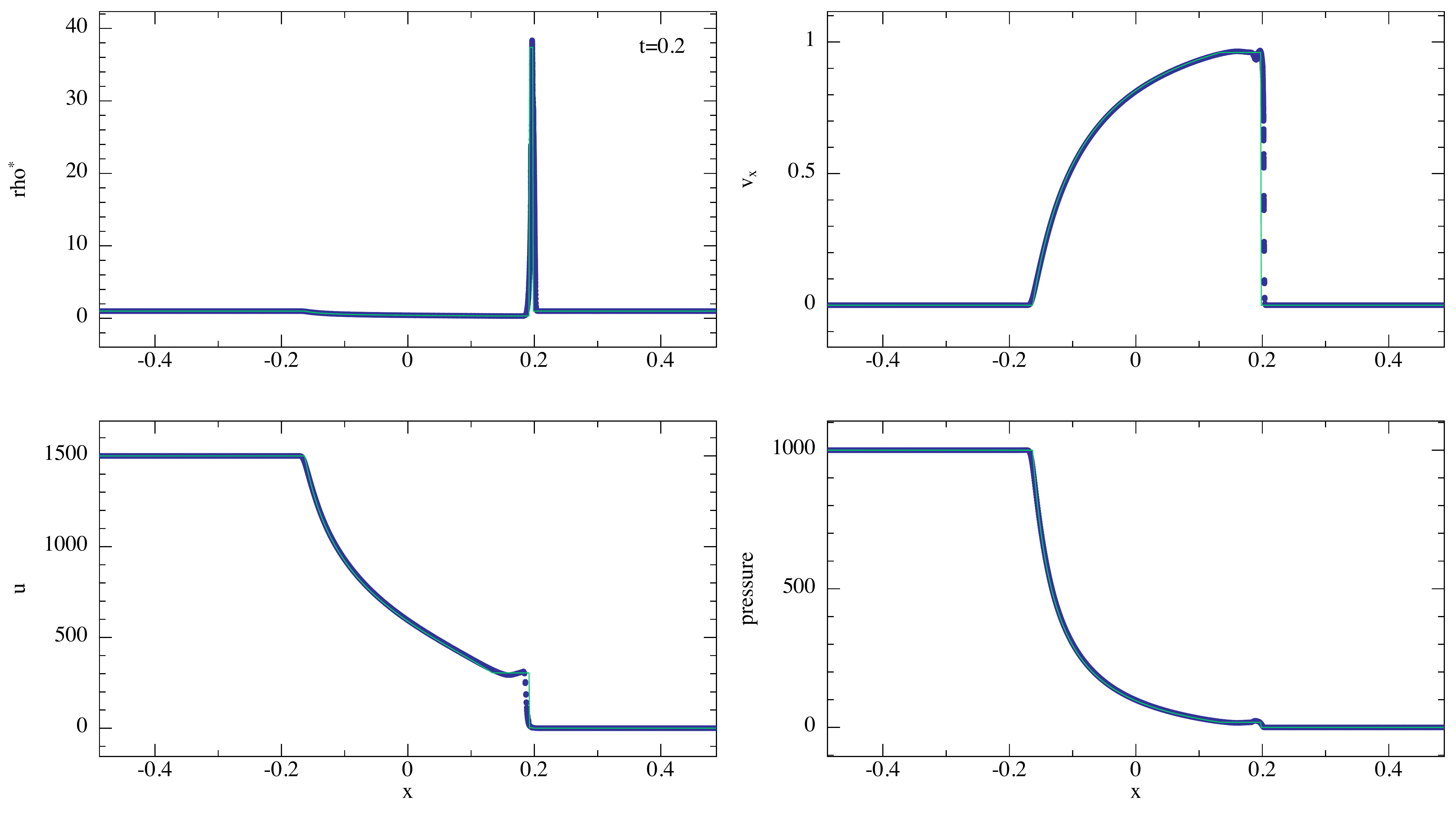}
      \caption{As in Figure~\ref{fig:blastwave-1D} but in 3D. Initial particle spacing is 0.001. According to \citet{martimuller03} this problem `is still a challenge for state-of-the-art codes today'. Our use of an entropy variable and positive definite dissipation means our code remains robust and accurate for this problem.}
      \label{fig:blastwave-3D}
   \end{center}
\end{figure*}

\begin{figure}
   \begin{center}
      \includegraphics[width=\columnwidth]{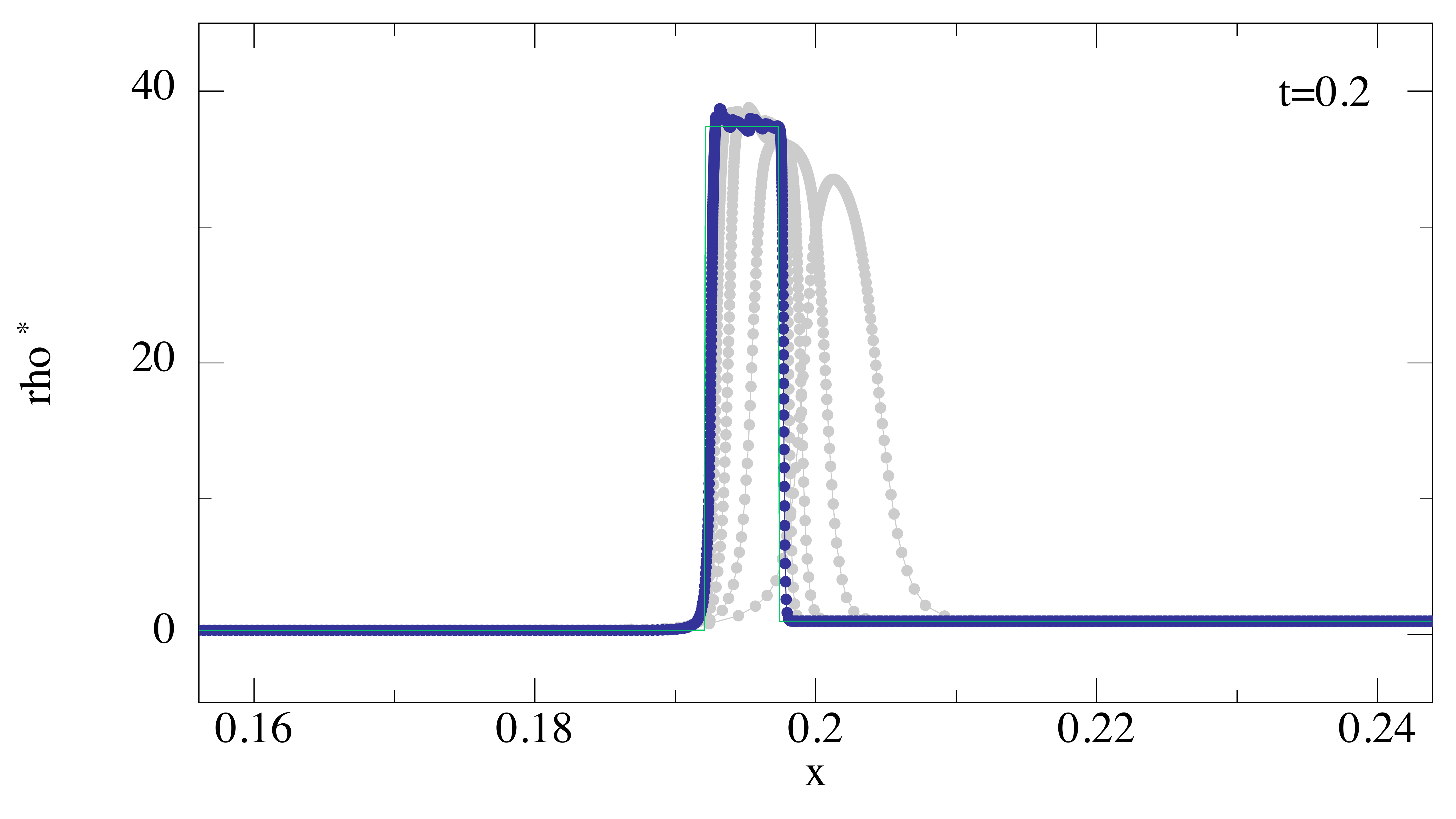}
      \caption{Details of the density spike at different resolutions for the blast wave problem in 1D. We show the results of 500, 1000, 2000, 4000, and 8000 particle simulations, going from right to left, beginning with the rightmost peak. This corresponds to initial particle spacings of $2\times10^{-3},10^{-3},5\times10^{-4},2.5\times10^{-4}$, and $1.25\times10^{-4}$, respectively.}
      \label{fig:spike-details}
   \end{center}
\end{figure}

\begin{figure*}
   \begin{center}
      \includegraphics[width=0.49\textwidth]{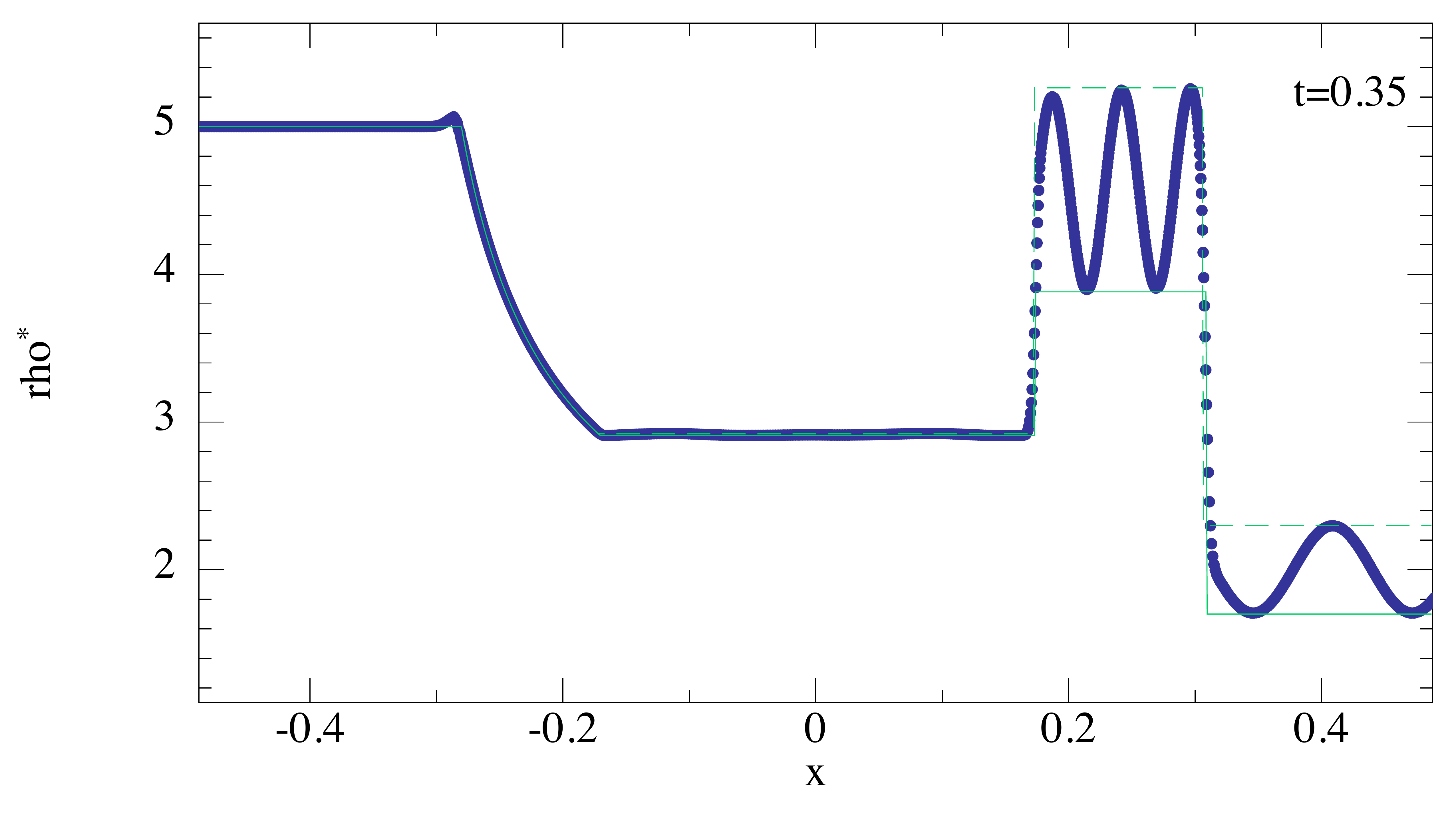}
      \includegraphics[width=0.49\textwidth]{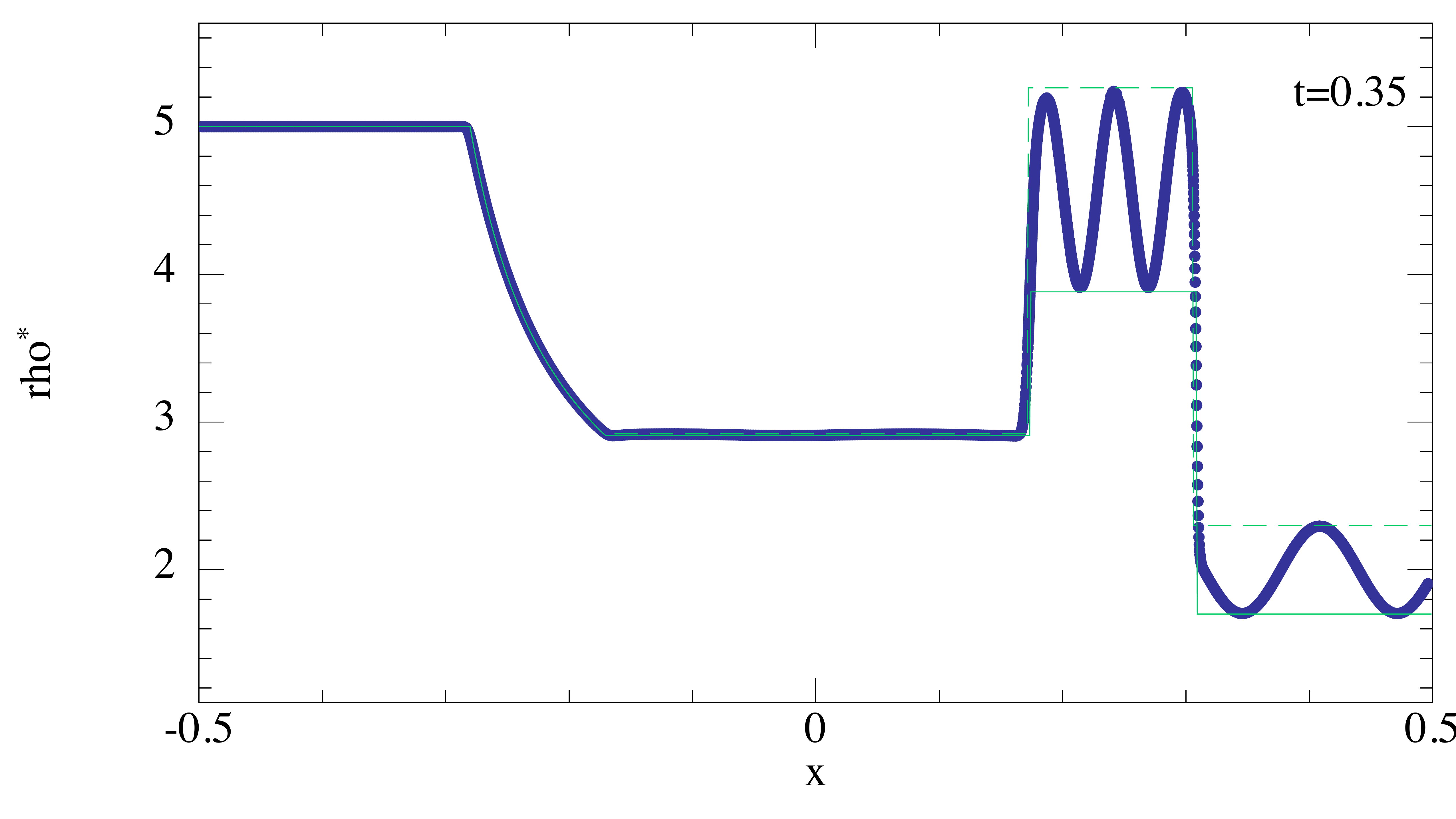}
      \caption{Results of the sine wave shocktube performed in 1D (left) and 3D (right). Blue circles show SPH particles, while green lines show the analytic solution for the two density extremes in the sine wave. The initial particle spacing to the left of the interface is 0.0004. The sine wave is accurately advected across the shock front, with no noticeable loss in accuracy.}
      \label{fig:sinewaves}
   \end{center}
\end{figure*}

\subsection{Special Relativity}
\subsubsection{Mildly relativistic shock} \label{sec:mild-shock}
In Figures~\ref{fig:shocktube-mild-1D}~\&~\ref{fig:shocktube-mild-3D} we show the results for a mildly relativistic shock in both 1D and 3D, respectively. This is `Problem 1' in \citet{martimuller03} with a maximum fluid Lorentz factor of $\Gamma=1.38$. For the initial conditions, we set
\begin{equation}
   [\rho,P] =
   \begin{cases}
      [10,\frac{40}{3}] &\text{for} \, x<0, \\
      [1,10^{-6}]         &\text{for} \, x>0. \\
   \end{cases}
\end{equation}
The particle spacing to the left of the interface is 0.0005, and 0.005 to the right. For the initial particle arrangement in 3D, we follow \citet{pricewurstertricco18}. We employ a uniform close-packed lattice of $1000 \times 26 \times 26$ particles for $x \in [-0.5,0]$ and $434 \times 12 \times 12$ particles for $x \in [0,0.5]$, with periodic boundaries in the $y$ and $z$ directions.
Of the four hydrodynamic variables plotted ($\rho^*$, $v^x$, $u$, $P$), the error is largest in $v^x$ with $L_2 = 4.0 \times 10^{-2}$ in 1D, and $L_2 = 3.2 \times 10^{-2}$ in 3D.
%1D Errors:
%rhostar : L2 err =  1.78414E-02 
%vel_x   : L2 err =  4.04524E-02 
%utherm  : L2 err =  8.23312E-03
%pressure: L2 err =  4.28148E-03 
%3D errors:
%rhostar : L2 err =  1.63998E-02
%vel_x   : L2 err =  3.18706E-02
%utherm  : L2 err =  7.14480E-03
%pressure: L2 err =  2.71544E-03

 The most noticeable difference between the 1D and 3D simulations is the slight overshoot at the left end of the rarefaction fan $x\approx-0.2$. This is caused by the lack of dissipation applied to the small discontinuity at the end of the rarefaction fan. It can be eliminated in 1D by applying the viscosity term for both approaching and receding particles. In 3D, even with viscosity applied only to approaching particles there is some dissipation applied at this location because the velocity differences are computed more isotropically. The small excess velocity at $x\approx 0.25$ seen in 3D is caused by particles rearranging behind the shock front.

Figure~\ref{fig:shocktube-mild-1D-nocond} shows the same shock but with $\alpha_u=0$. Without artificial conductivity we find a `blip' in the pressure profile at the contact discontinuity. The same feature is discussed by \citet{price08} and \citet{price12} in the context of non-relativistic SPH simulations. In both cases, the `blip' can be smoothed out by including artificial conductivity. Simulations by \citetalias{chowmonaghan97} do not contain such a `blip', but instead display excessive thermal diffusion at the contact discontinuity, since artificial conductivity is included implicitly within their dissipation terms (compare our results to figures 9 and 10 in their paper).

\subsubsection{Strong blast}
In Figures~\ref{fig:blastwave-1D}~\&~\ref{fig:blastwave-3D} we show the results for a strong blast wave in both 1D and 3D, respectively. This is `Problem 2' in \citet{martimuller03} with a maximum fluid Lorentz factor of $\Gamma=3.6$. We follow the same setup as for the mildly-relativistic shock but with differing initial conditions
\begin{equation}
   [\rho,P] =
   \begin{cases}
      [1,1000]     &\text{for} \, x<0, \\
      [1,10^{-2}]  &\text{for} \, x>0. \\
   \end{cases}
\end{equation}
The particle spacing is 0.001 on both sides of the interface, with $1000 \times 12 \times 12$ particles for the whole domain in 3D. Importantly, we use unsmoothed initial conditions, unlike in \citetalias{chowmonaghan97} and \citet{rosswog10}. We found this was only possible when we evolved the entropy rather than total energy as the conserved variable, otherwise negative pressures could result in the absence of smoothing. The error is again largest in $v^x$ for both cases, with $L_2 = 2.5 \times 10^{-1}$ in 1D and $L_2 = 3.1 \times 10^{-1}$ in 3D. The peak of the density spike is also within 2.7\% of the exact solution in both the 1D and 3D simulations. The same differences between 1D and 3D simulations occur as in the previous test, and for the same reasons.

%1D errors:
%rhostar : L2 err =  2.02594E-01
%vel_x   : L2 err =  2.52292E-01
%utherm  : L2 err =  1.73921E-02
%pressure: L2 err =  4.59166E-03
%3D errors  
%rhostar : L2 err =  2.76020E-01
%vel_x   : L2 err =  3.13155E-01
%utherm  : L2 err =  5.80450E-03
%pressure: L2 err =  8.01628E-03
%
% Spike peaks:
%1D: 36.4155731
%3D: 38.3855438
%Exact: 37.3865

Figure~\ref{fig:spike-details} shows the details of the density spike at different resolutions. At low resolutions the shock speed is overestimated, consistent with the results of \citetalias{chowmonaghan97}. With 500 particles the centre of the spike is ahead of the exact solution by approximately one spike width. As the resolution is increased, our simulations approach the exact solution.

\subsubsection{Sinusoidally perturbed shock}
In Figure~\ref{fig:sinewaves} we show the results in 1D and 3D for a shock tube with a sinusoidally perturbed right density state \citep[e.g.][]{del-zannabucciantini02,rosswog10} where the fluid reaches a maximum Lorentz factor of $\Gamma=1.15$. The initial conditions for this setup are
\begin{equation}
   [\rho,P] =
   \begin{cases}
      [5, \, 50]     &\text{for} \, x<0, \\
      [2 + 0.3 \sin (50x), \, 5]  &\text{for} \, x>0. \\
   \end{cases}
\end{equation}
The particle spacing is 0.0004 on the left side of the interface, and on the right such that the particle masses are the same as on the left. We find that the sine wave perturbation is easily transported across the shock front in both cases, with no noticeable loss in accuracy. The peaks of the propagated sine wave reach the correct levels of the two limiting solutions; to within 1.2\% in 1D, and within 1.3\% in 3D. The left end of the rarefaction fan in 1D displays an overshoot in the density, for the same reasons as discussed in the previous tests.
%-------------------------
% 1D peak errors
% peaks1: (exact = 5.262246)
% 5.20188236 -> 1.147% error
% 5.24737835 -> 0.283% error
% 5.25640869 -> 0.111% error
%
% troughs1: (exact = 3.8813093)
% 3.89690924 -> 0.402% error
% 3.90529227 -> 0.618% error
%
% peak2: (exact = 2.3)
% 2.29759383 -> 0.1046% error
%
% troughs2: (exact = 1.7)
% 1.70645452 -> 0.380%
% 1.70654988 -> 0.385%
%-------------------------
% 3D peak errors
% peaks1: (exact = 5.262246)
% 5.19591856 -> 1.260% error
% 5.23901844 -> 0.441% error
% 5.23325348 -> 0.551% error
% 
% troughs1: (exact = 3.8813093)
% 3.90979290 -> 0.734% error
% 3.91057062 -> 0.754% error
% 
% peak2: (exact = 2.3)
% 2.29499578 -> 0.218% error
% 
% troughs2: (exact = 1.7)
% 1.70386589 -> 0.227%
% 1.70387053 -> 0.228%
%-------------------------

\subsubsection{Shocks with transverse velocity}
We consider two problems where the shock contains a non-zero transverse velocity to the main flow. This is important in relativistic flows since the magnitude of the flow speed is limited by the speed of light, and the transverse velocity is coupled to the evolution of the other hydrodynamic variables through the Lorentz factor.

\begin{figure*}
   \begin{center}
      \includegraphics[width=\textwidth]{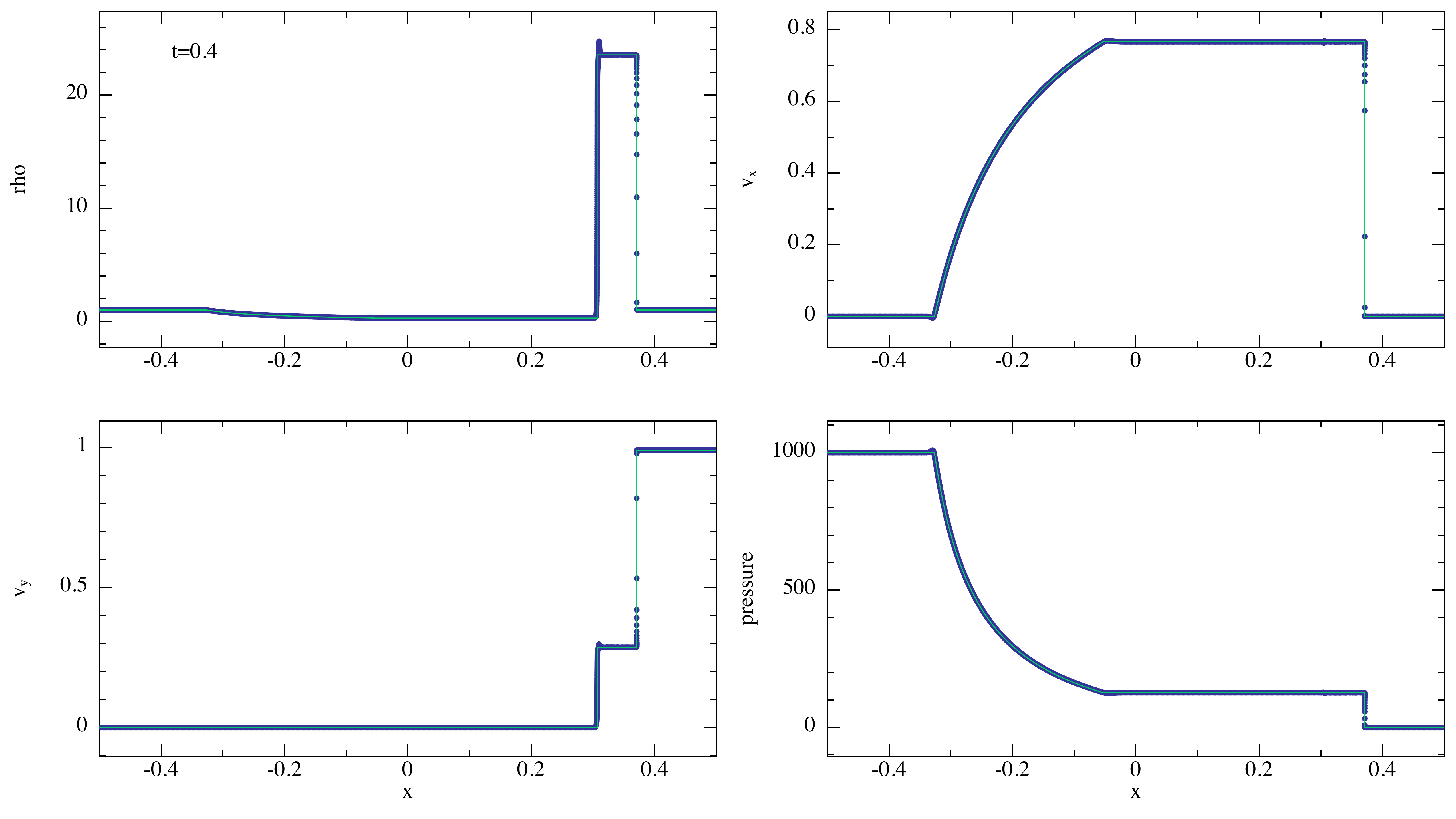}
      \caption{Results of the 1D shocktube with initial transverse velocity on the right side of the interface. The initial particle spacing is 0.0001 on the left side of the interface. We find a small overshoot in the density at the contact discontinuity.}
      \label{fig:easytest}
   \end{center}
   \begin{center}
      \includegraphics[width=\textwidth]{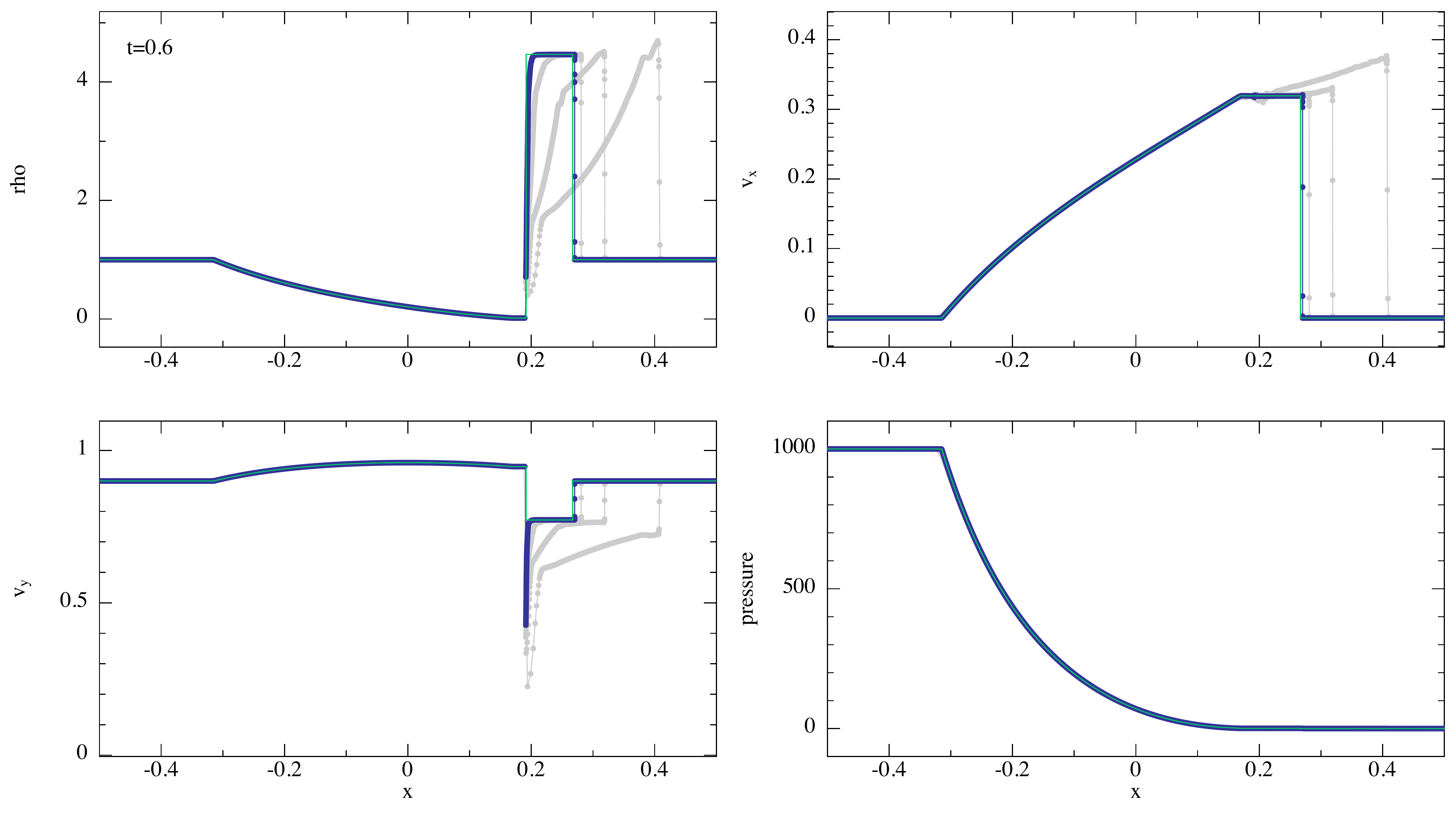}
      \caption{Results of the 1D shocktube with initial transverse velocity on both sides of the interface. Simulations with 1000, 4000, and 16000, particles are shown in grey for reference, and the highest resolution simulation of 64000 particles is shown in blue. These resolutions corresponds to initial particle spacings of $1\times10^{-3}, 2.5\times10^{-4}, 6.25\times10^{-5},$ and $1.5625\times10^{-5}$, respectively. The exact solution given by \citet{giacomazzorezzolla06,martimuller15} is shown in green. We find a large spike in $v^y$ at the contact discontinuity with even the highest resolution. }
      \label{fig:hardtest}
   \end{center}
\end{figure*}

Figure~\ref{fig:easytest} shows our results in 1D for a shocktube with initial conditions given by
\begin{align}
   [\rho,P,v^x,v^y] &=
   \begin{cases}
      [1,1000,0,0] &\text{for} \, x<0, \\
      [1,10^{-2},0,0.99]         &\text{for} \, x>0. \\
   \end{cases}
\end{align}
The initial particle spacing is 0.0001 on the left side of the interface.
Of the four hydrodynamic variables plotted ($\rho$, $v^x$, $v^y$, $P$), the error is greatest in $\rho$ with $L_2 = 3.7 \times 10^{-2}$, and a $5\%$ overshoot at the contact discontinuity. The fluid has a maximum Lorentz factor of $\Gamma=7.09$.
%overshoot: 24.8037205
%exact: 23.55644
%Errors:
%rho     : L2 err =  3.67026E-02
%vel_x   : L2 err =  3.20411E-02
%vel_y   : L2 err =  2.33018E-02
%pressure: L2 err =  3.87915E-03

A much more challenging test is the shocktube with initial conditions given by
\begin{align}
   [\rho,P,v^x,v^y] &=
   \begin{cases}
      [1,1000,0,0.9] &\text{for} \, x<0, \\
      [1,10^{-2},0,0.9]         &\text{for} \, x>0. \\
   \end{cases}
\end{align}
As pointed out by \citet{zhangmacfadyen06} this tests requires a `very high resolution to resolve the complicated structure of the transverse velocity'. Figure~\ref{fig:hardtest} shows the results for this problem in 1D. 
We find that this test is particularly challenging since the transverse velocity changes rapidly, requiring very high spatial resolution. The most obvious artefact is a spike in $v^y$ at $x\approx0.2$. This is also observed in \citet{zhangmacfadyen06} but is worse in SPH since this feature occurs in a low density part of the simulation, where the resolution is lowest. The largest error is in $\rho$ with $L_2 = 4.0 \times 10^{-1}$. The fluid reaches a maximum Lorentz factor of $\Gamma=44.5$.
%Errors:
%rho     : L2 err =  4.00847E-01
%vel_x   : L2 err =  3.41991E-02
%vel_y   : L2 err =  1.68738E-02
%pressure: L2 err =  1.37931E-01

\subsubsection{3D spherical blast wave}
\begin{figure*}
   \begin{center}
      \includegraphics[width=\textwidth]{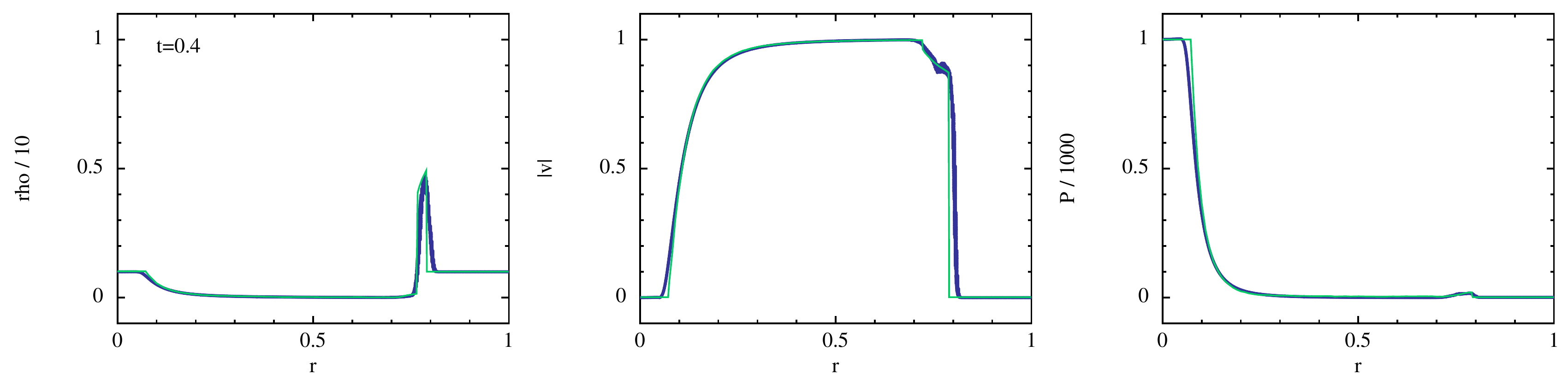}
      \caption{Results of the spherical blast wave in 3D with $640\times740\times784$ particles (blue circles). We compare to the high resolution 1D results of \citet{zhangmacfadyen06} (green line).}
      \label{fig:blastwave-ZhangMacFadyen06}
   \end{center}
\end{figure*}

In Figure~\ref{fig:blastwave-ZhangMacFadyen06} we show the results of a spherically symmetric relativistic blast in 3D, analogous to the non-relativistic Sedov-Taylor blast wave. This is the same problem as done by \citet{del-zannabucciantini02} and \citet{zhangmacfadyen06} (see their Figures 8 and 17, respectively).

For the initial conditions we set $640 \times 740 \times 784$ particles on a closepacked lattice within a uniform periodic box $x, y, z \in [-1, 1]$. The density is initially set to $\rho^* = \rho = 1$ everywhere. We inject a sphere of uniform pressure $P_\mathrm{in}=1000$ at the origin with radius $r_0=0.4$, and set the pressure outside the sphere to $P_\mathrm{out}=0$. Our setup and resolution is equivalent to \citet{zhangmacfadyen06}, however we simulate the whole box $x, y, z \in [-1, 1]$, instead of just the quadrant $x, y, z \in [0, 1]$ which is only possible in spherical symmetry.

Following \citet{chowmonaghan97} and \citet{rosswog10} we slightly smoothed the initial discontinuity in the pressure using
\begin{equation}
   P(r) = \frac{P_\mathrm{in} - P_\mathrm{out}}{1+\exp(\frac{r - r_0}{\Delta r})} - P_\mathrm{out},
\end{equation}
where for the characteristic transition length $\Delta r$ we used half the initial particle spacing.

\begin{figure}
   \begin{center}
      \includegraphics[width=0.8\columnwidth]{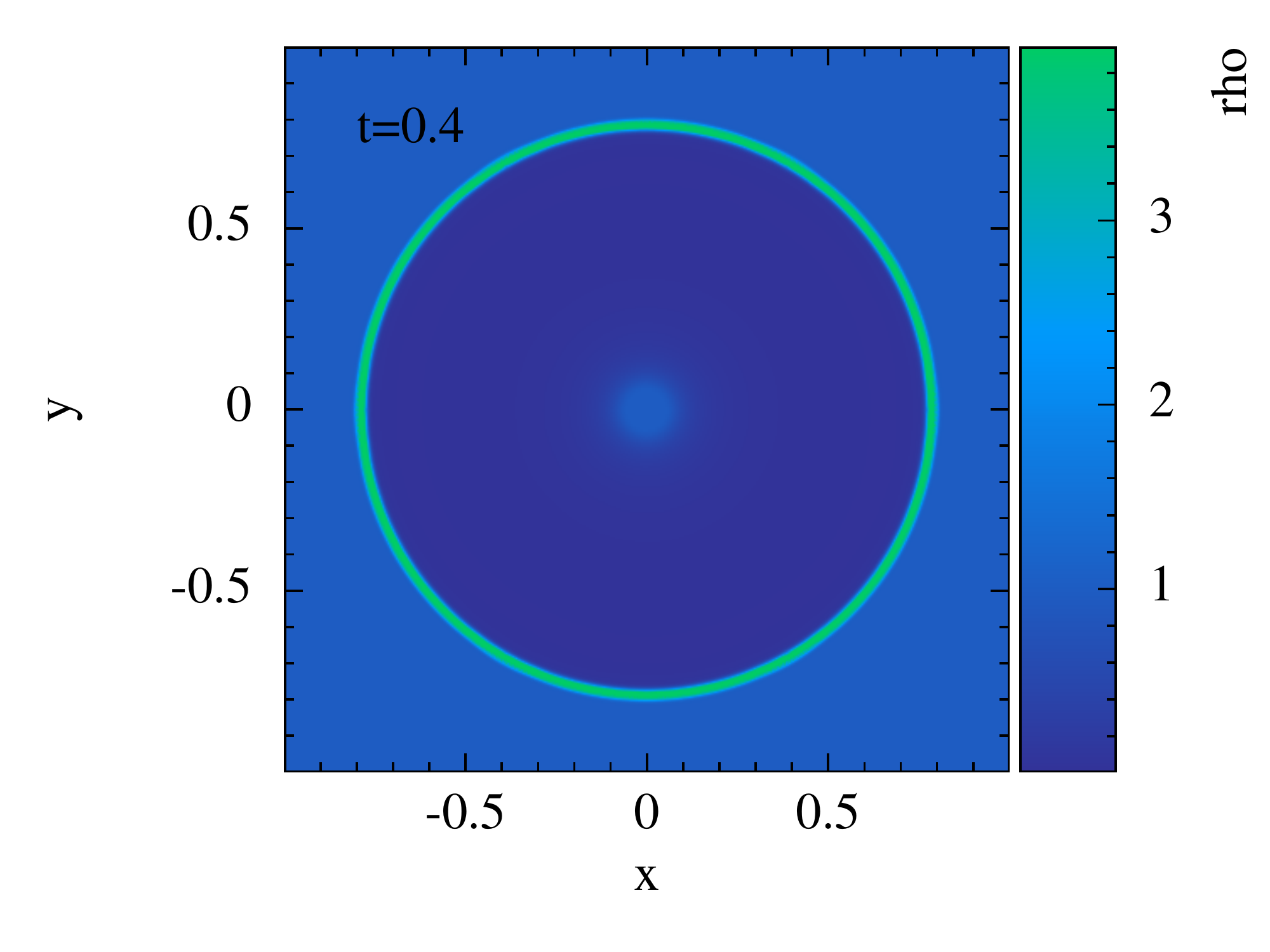}
      \caption{Slice through the spherical blast wave in 3D showing the primitive density $\rho$.}
      \label{fig:blastwave-densityslice}
   \end{center}
\end{figure}

We compare our results to the 1D high resolution results of \citet{zhangmacfadyen06}. Our solution shows qualitative agreement with theirs, with the most notable difference being an $\sim8\%$ increase in the speed of our shock front. There is also a small amount of smoothing at the left end of the rarefaction fan ($x\approx0.05$), which is present in the 3D simulations of \citet{del-zannabucciantini02} and \citet{zhangmacfadyen06}.
Figure~\ref{fig:blastwave-densityslice} shows the (primitive) density slice through the $z=0$ plane, demonstrating our code's ability to preserve radial symmetry and resolve the structure of the thin shell. The fluid reaches a maximum Lorentz factor of $\Gamma=19.56$.

\subsubsection{Oscillating polytrope} \label{sec:polytrope}
\begin{figure*}
   \begin{center}
      \includegraphics[width=0.9\textwidth]{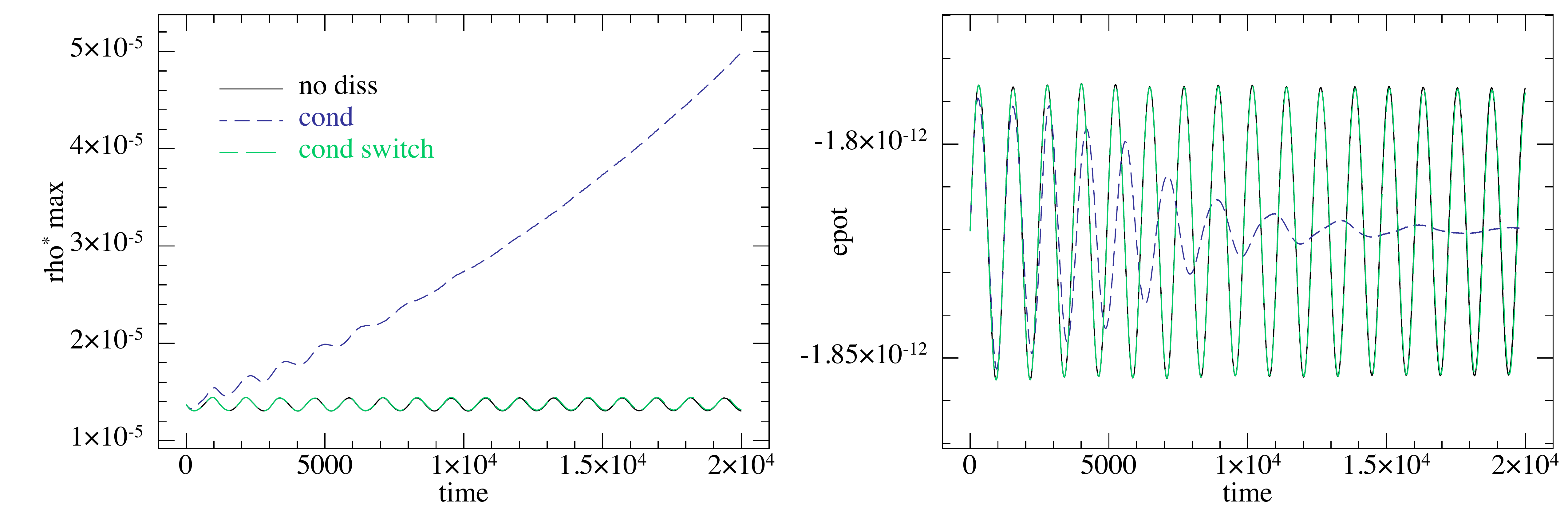}
      \caption{Maximum (central) density $\rho^*$ (left) and total gravitational potential energy (right) for the oscillating polytrope. With no dissipation the polytrope oscillates indefinitely about its equilibrium state (black solid line). Using artificial conductivity with the same signal speed as artificial viscosity results in a disturbed stellar profile (blue dot-dashed line). This is remedied by using the signal speed in Eq.~\ref{eq:vsigu2} which switches conductivity off where it is not needed (green dashed line).}
      \label{fig:polytrope}
   \end{center}
\end{figure*}

A concern with artificial conductivity (expressed by the referee of this paper) is that it may introduce unwanted side effects. For example, spuriously applied conductivity may seriously disrupt a stellar equilibrium state. To address this concern, Figure~\ref{fig:polytrope} shows the results of a 3D oscillating polytropic star with Newtonian self-gravity in the Minkowski metric. The initial conditions are a relaxed star consisting of $65752$ $\left(\approx 4\pi25^3/3\right)$ particles and polytropic equation of state $P(r)=K\rho(r)^{(n+1)/n}$ with $K=10^{-10}$ and $n=1.5$ (i.e. $\adgamma=5/3$).

To relax the star we apply an additional force to the particles in the momentum equation which depends on the particle velocity \citep{gingoldmonaghan77}
\begin{equation}
   f^i_\mathrm{damp} = -0.03 \, v^i,
\end{equation}
and turn off all dissipative terms. That is, we set $\alpha_\mathrm{AV}=0$ and $\alpha_u=0$. We integrate until $t=20 000$ when all particle motions have been sufficiently damped. 

To induce radial oscillations in the star, we perturb the relaxed star profile with a small velocity perturbation
\begin{equation}
   v^i(r) = 10^{-4} x^i (r).
\end{equation}
Figure~\ref{fig:polytrope} compares the results of three different simulations, one with no dissipation, a second with conductivity turned on using $v^u_{\mathrm{sig}} = v_{\mathrm{sig}}$, and a third using conductivity with $v^u_{\mathrm{sig}}$ as in Eq.~\ref{eq:vsigu2}.

With no dissipation the amplitude of the stellar oscillation does not decay with time since energy is conserved exactly (black line in both panels). However with $v^u_{\mathrm{sig}} = v_{\mathrm{sig}}$, the stellar profile slowly becomes disturbed, and the central density increases. The results of the third simulation demonstrate a way to remedy this issue. The signal speed in Eq.~\ref{eq:vsigu2} gives conductivity of $\mathcal{O}(h^2)$, and which vanishes when the star is in hydrostatic equilibrium. This allows us to reproduce the same undamped oscillations as in the dissipation-free simulation.

\begin{figure*}
   \begin{center}
      \includegraphics[width=0.33\textwidth]{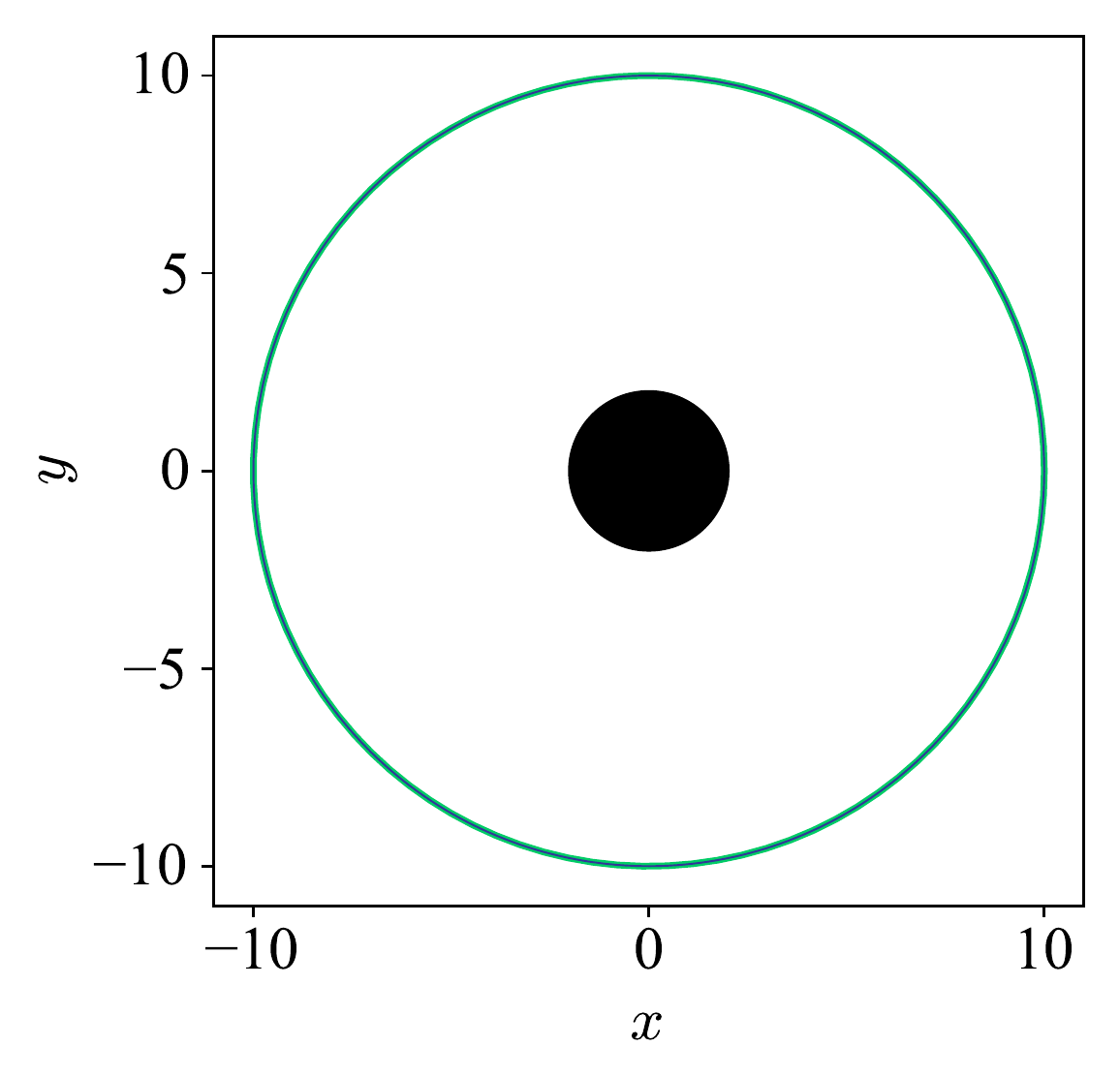}
      \hspace{1cm}
      \includegraphics[width=0.33\textwidth]{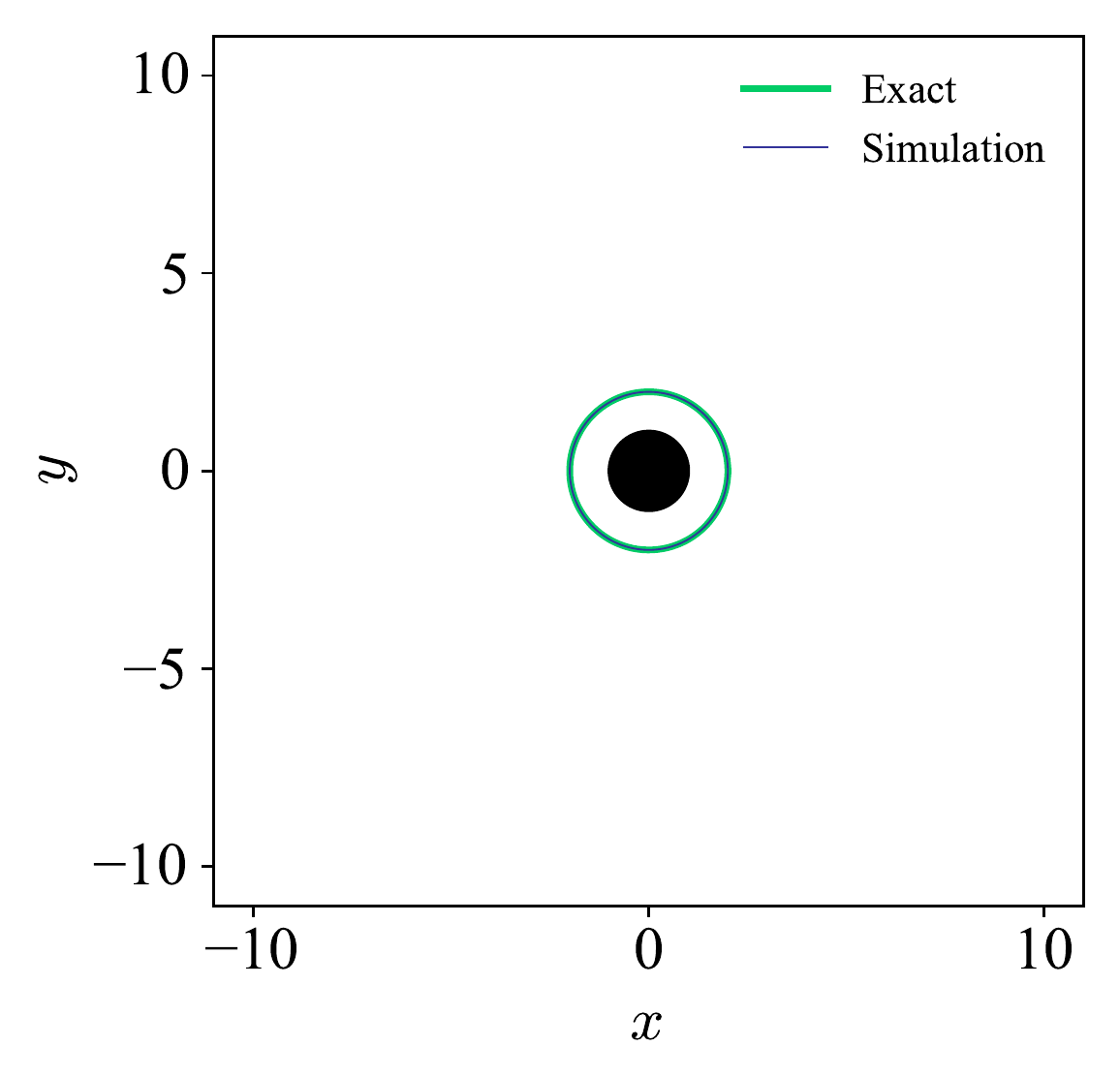}
      \caption{Circular orbits computed in the Schwarzschild (left) and Kerr (right) metrics at $r=10M$ and $r=2M$, respectively, with central mass $M=1$. The orbit in the Kerr metric was computed with maximal spin, $a/M=1$. Orbits are indicated by the blue line, while green lines show the exact orbit. The black holes are represented by the solid black circles i.e. the region interior to the event horizon $r_\mathrm{h}$. For the Schwarzschild metric, $r_\mathrm{h} = 2M$, while for the Kerr metric $r_\mathrm{h} = 1M$ when $a/M=1$.}
      \label{fig:circular-orbits}
   \end{center}
\end{figure*}

\begin{figure*}
   \begin{center}
      \includegraphics[width=0.44\textwidth]{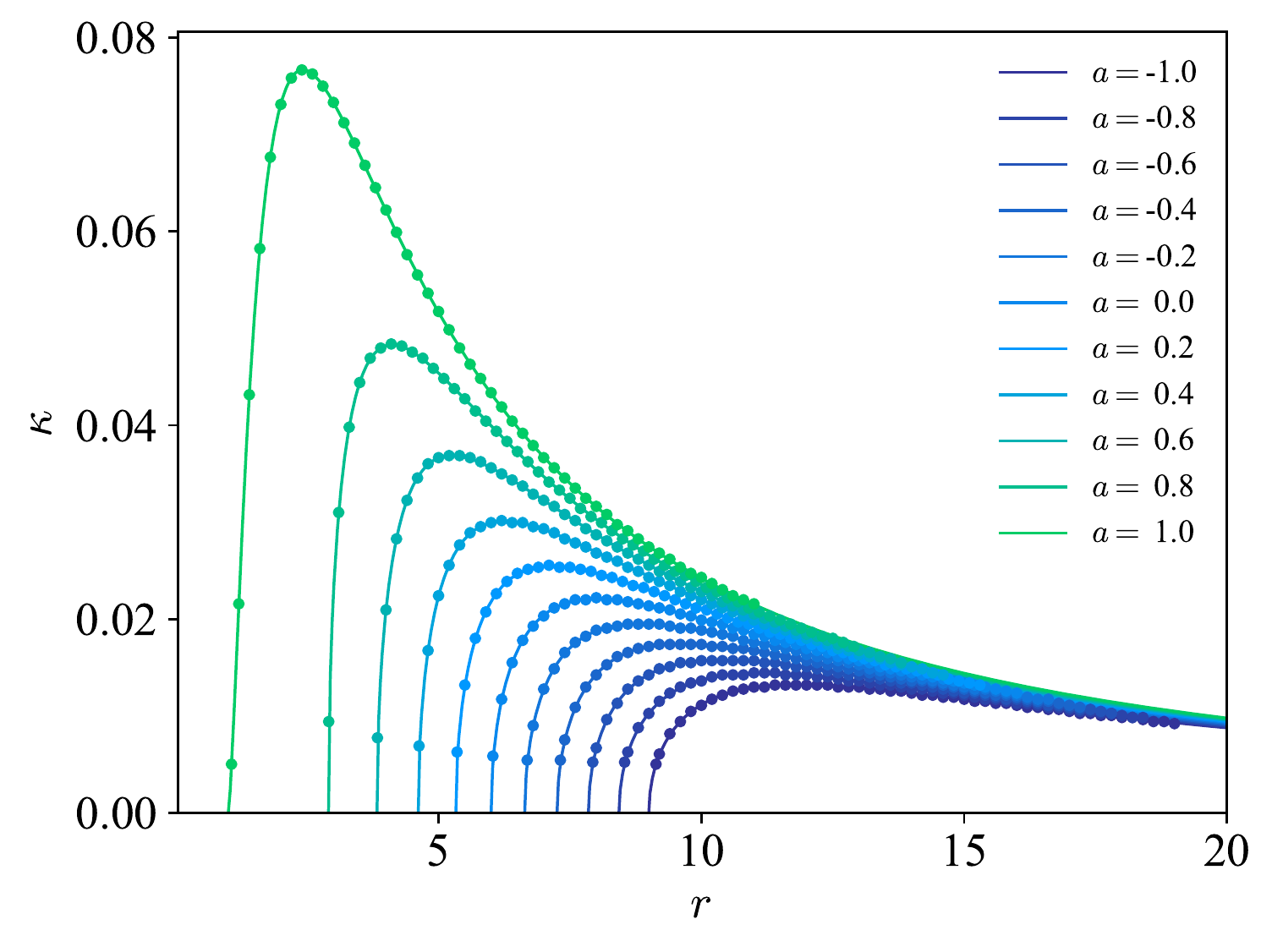}
      \hspace{0.5cm}
      \includegraphics[width=0.44\textwidth]{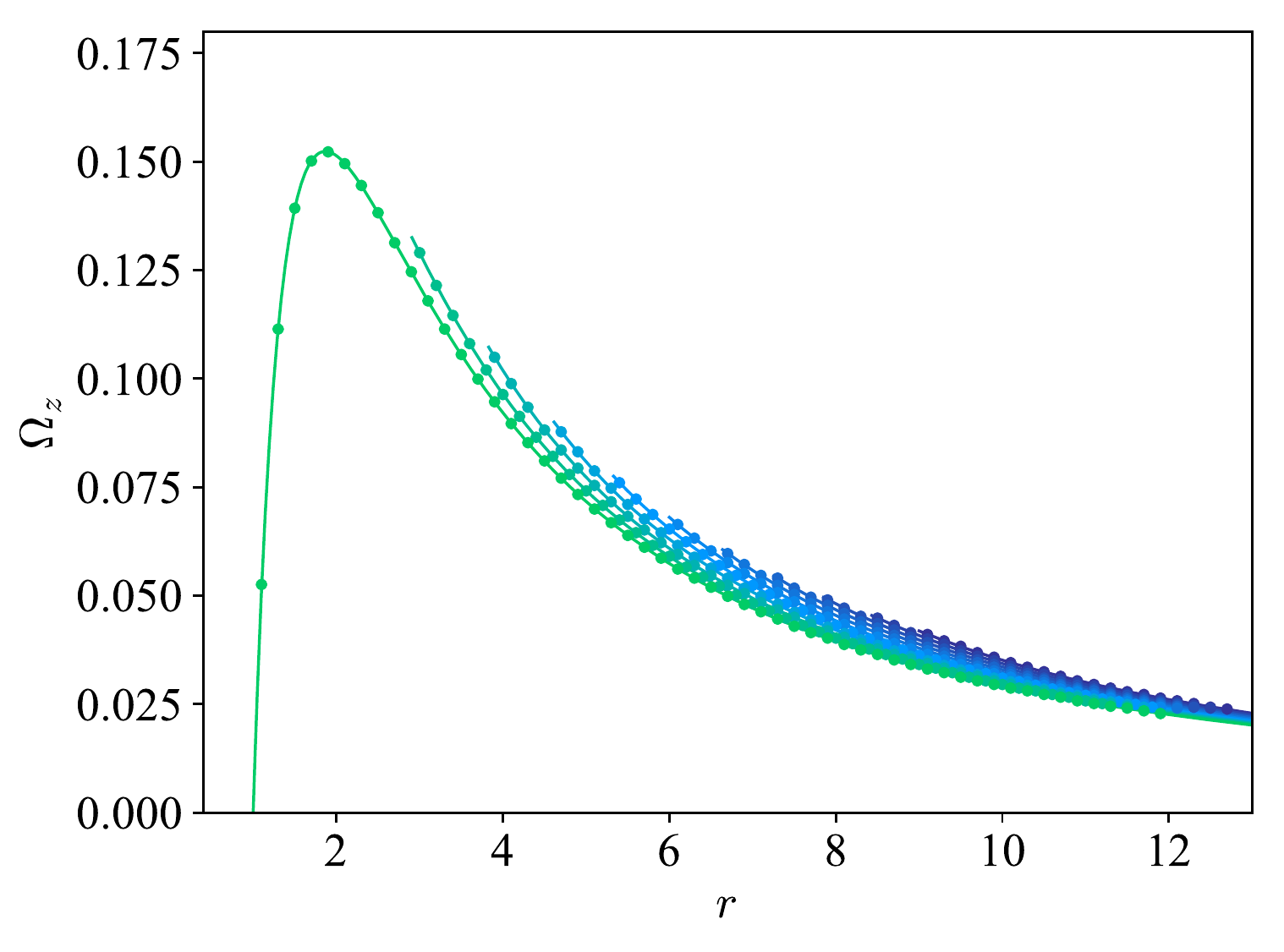}
      \caption{Epicyclic frequency (left) and vertical oscillation frequency (right) as a function of radius for $a/M \in [-1,1]$ shown in steps of $0.2$ (spin increasing right to left), showing numerical solution from test particle orbits (points), compared to the exact solution (lines). Central mass is $M=1$.}
      \label{fig:oscillations}
   \end{center}
\end{figure*}

\begin{figure}
   \begin{center}
      \includegraphics[width=0.87\columnwidth]{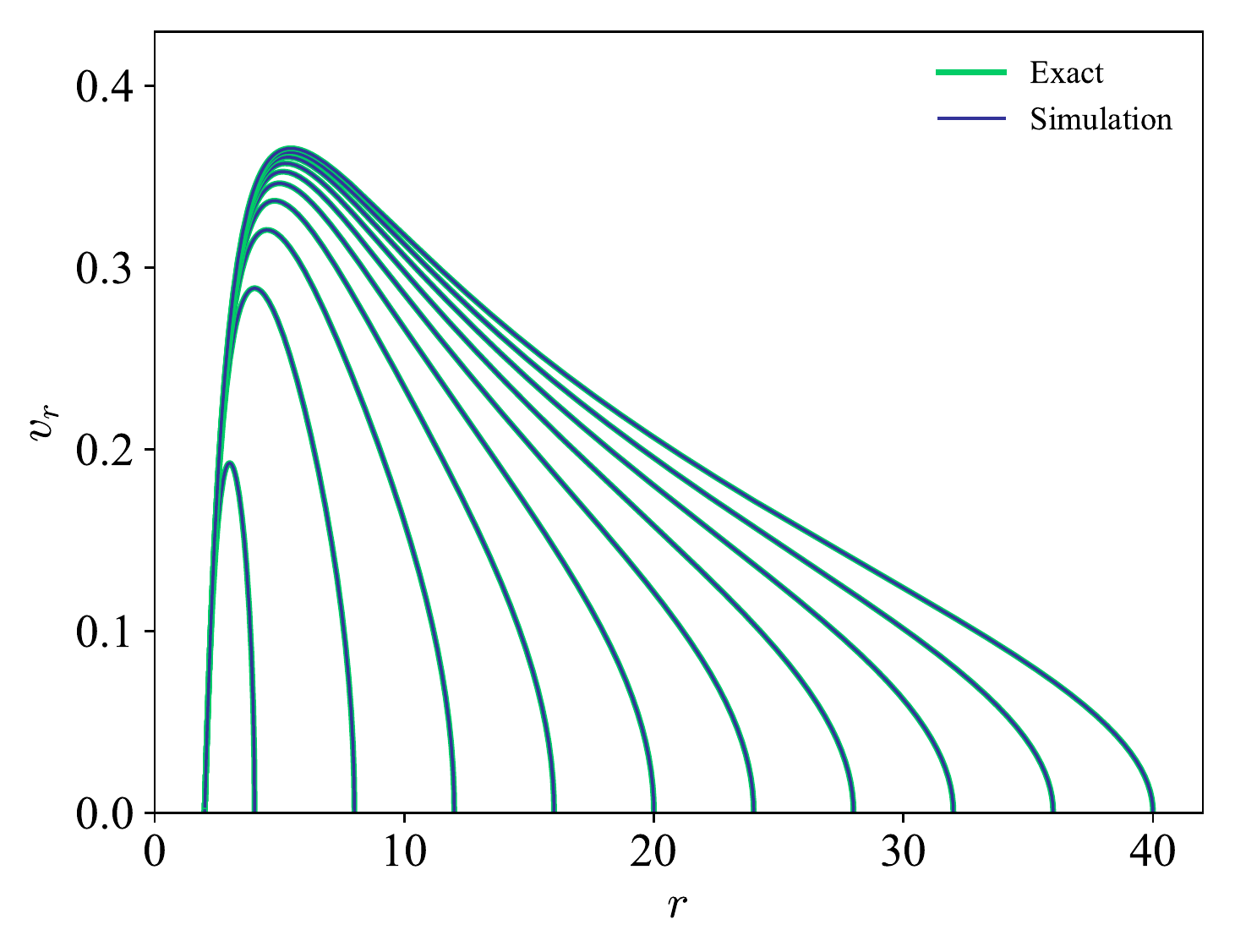}
      \caption{Velocity as a function of radius for radial geodesics computed from the infall of test particles (blue) starting at rest. The initial radii are spaced evenly for $r_0/M \in [4,40]$, and the corresponding exact solutions are shown in green. Central mass is $M=1$.}
      \label{fig:radial-multi}
   \end{center}
\end{figure}

\begin{figure*}
   \begin{center}
      \includegraphics[width=0.29\textwidth]{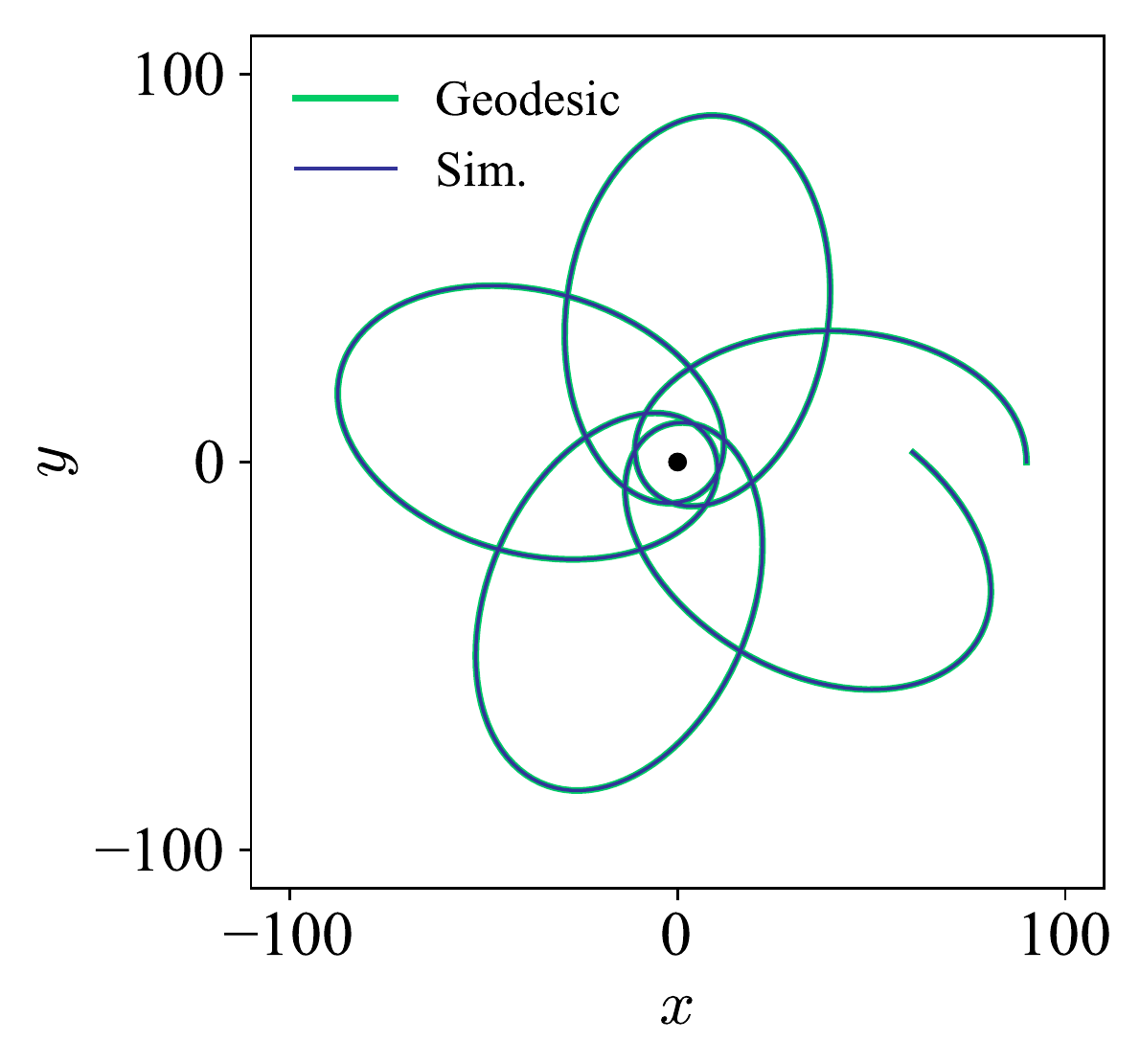}
      \includegraphics[width=0.29\textwidth]{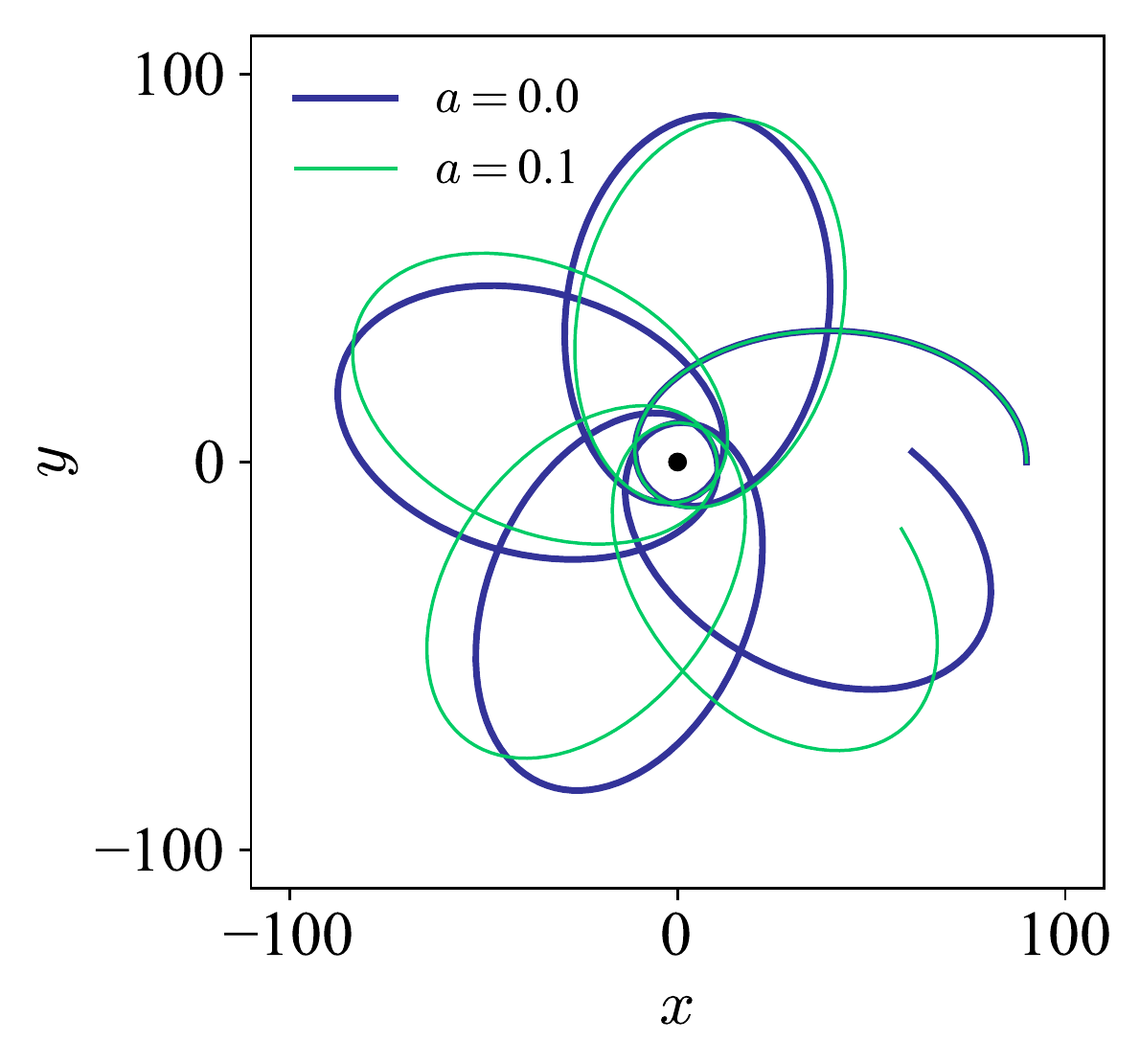}
      \includegraphics[width=0.29\textwidth]{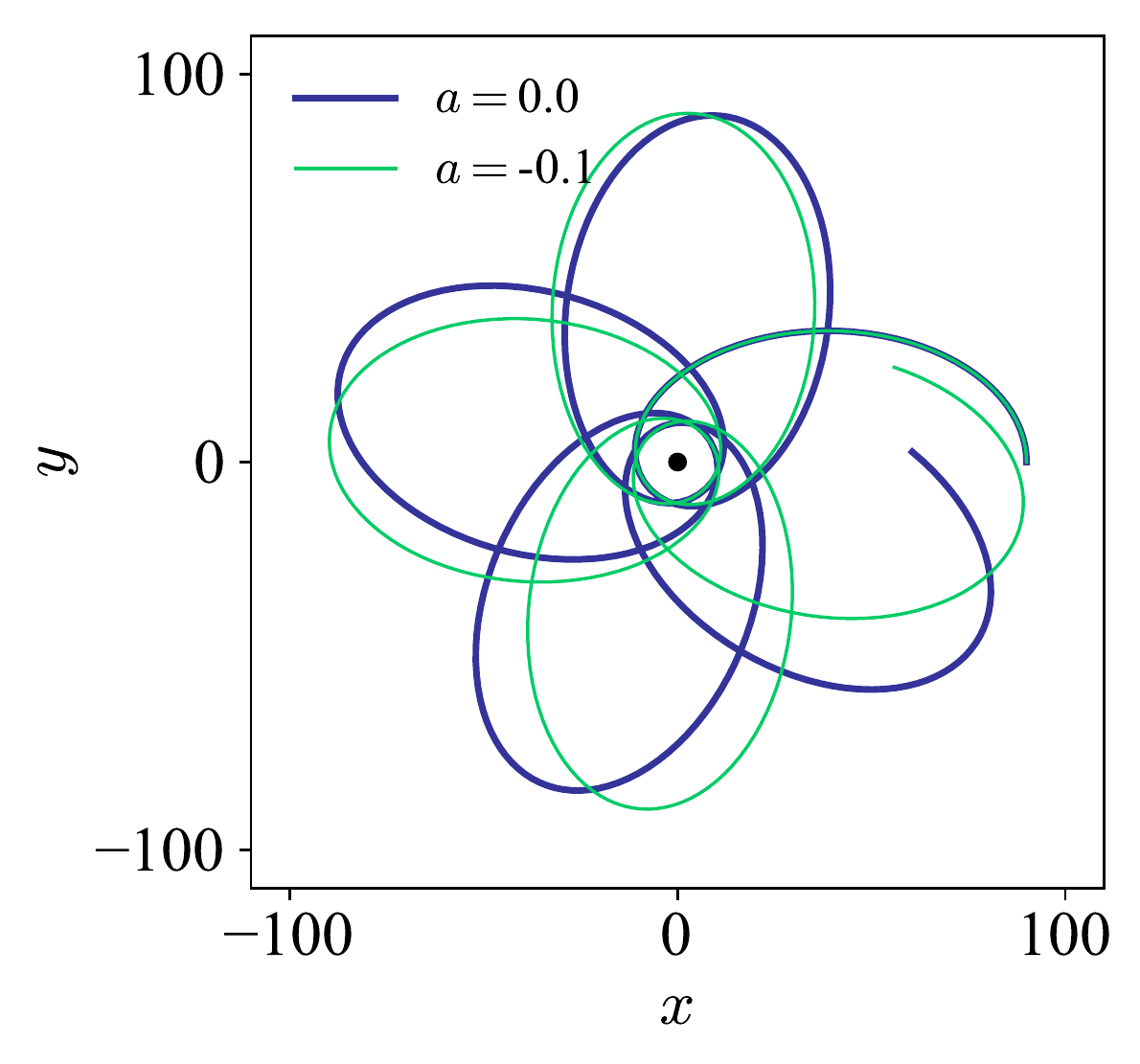}
      \caption{Precession in the Schwarzschild (left) and Kerr metrics (middle and right) with $M=1$. The orbit in the Schwarzschild metric matches the same orbit as computed from a direct integration of the geodesic equation. With $a/M=0$ in the Kerr metric we reproduce the Schwarzschild orbit as expected. With non-zero $a/M$ in the Kerr metric, the angle of apsidal advance is decreased for the prograde orbit ($a/M=0.1$) and increased for the retrograde orbit ($a/M=-0.1$). Large values of spin are also possible, however in those cases the direction of orbital advance becomes difficult to distinguish in a figure.}
      \label{fig:precession}
   \end{center}
\end{figure*}

\begin{figure}
   \begin{center}
      \includegraphics[width=0.8\columnwidth]{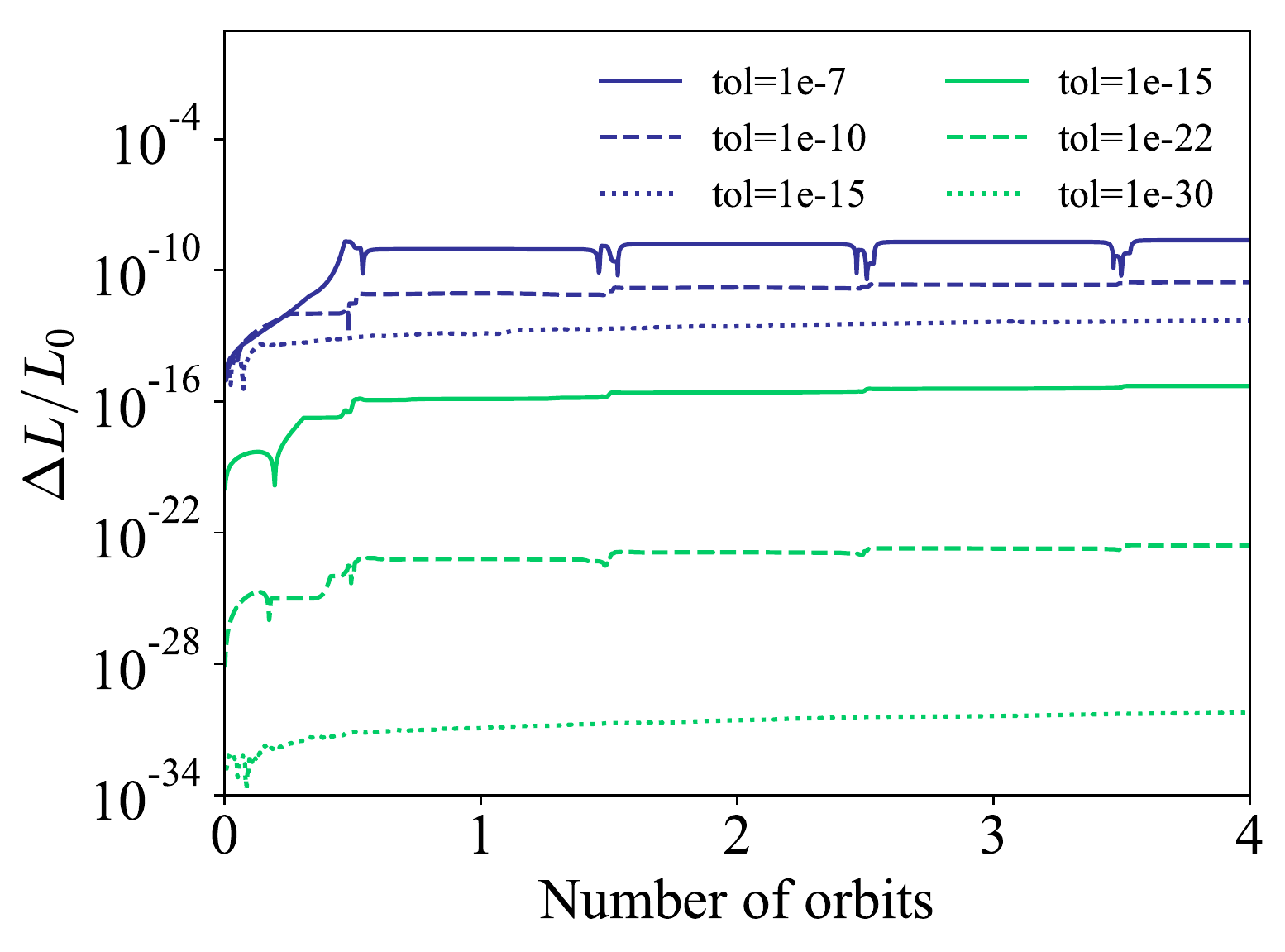}
      \caption{Fractional error in angular momentum ($L$) throughout the precessing orbit in the Schwarzschild metric, described in Section \ref{sec:precession-schwarzschild}. Calculations are computed in double precision (blue) and quad precision (green) with $10^4$ steps per orbit and varying values of $\epsilon_x = \epsilon_p = \mathrm{tol}$. Angular momentum is conserved to round off error with $\mathrm{tol}=10^{-15}$ and $\mathrm{tol}=10^{-30}$ for double and quad precision calculations, respectively.}
      \label{fig:timestepping}
   \end{center}
\end{figure}

\begin{figure*}
   \begin{center}
      \includegraphics[width=0.95\textwidth]{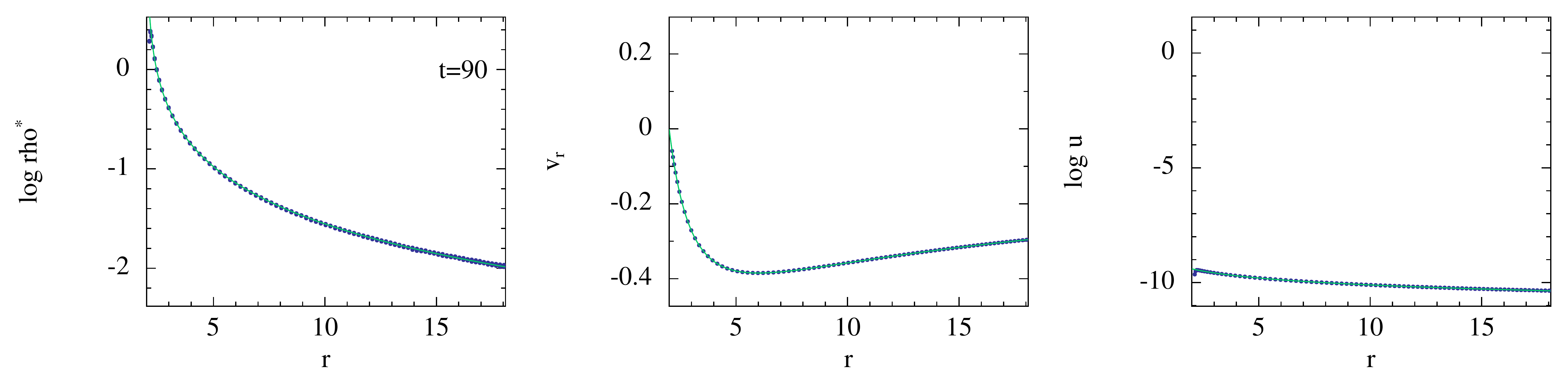}
      \caption{Results of spherically symmetric `geodesic' Bondi accretion in the Schwarzschild metric with $M=1$ and $\alpha_\mathrm{AV}=0$. Blue dots show SPH particles while green lines show the exact solution. This demonstrates that the time evolution of the energy equation is correct.}
      \label{fig:bondi-geodesic-nodiss}
   \end{center}
   \begin{center}
      \includegraphics[width=0.95\textwidth]{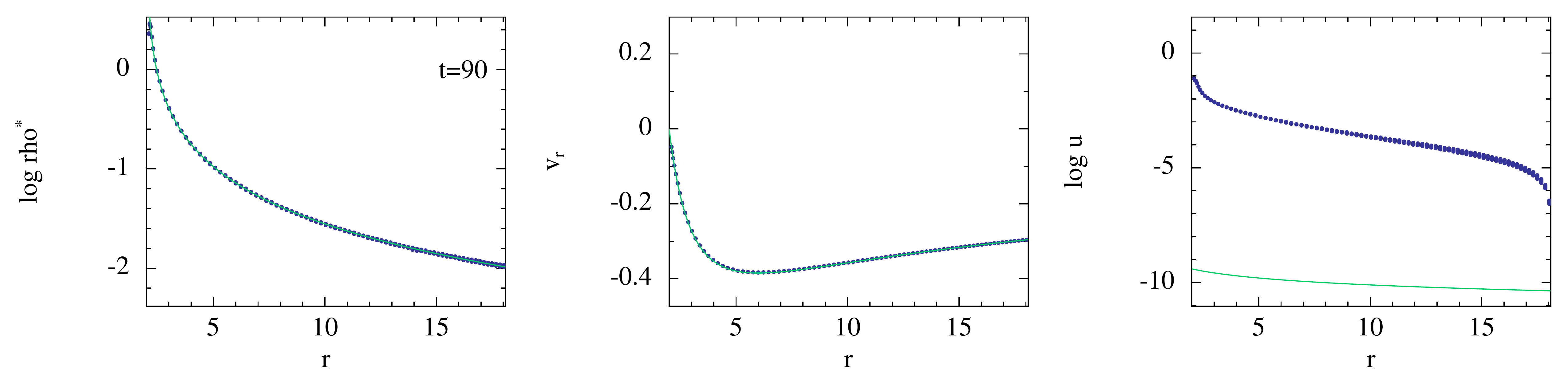}
      \caption{As in Fig.~\ref{fig:bondi-geodesic-nodiss} but artificial viscosity turned on as normal i.e. $\alpha_\mathrm{AV}=1$. This produces excessive viscous heating, evident in the thermal energy profile (right).}
      \label{fig:bondi-geodesic}
   \end{center}
\end{figure*}

\begin{figure*}
   \begin{center}
      \includegraphics[width=0.95\textwidth]{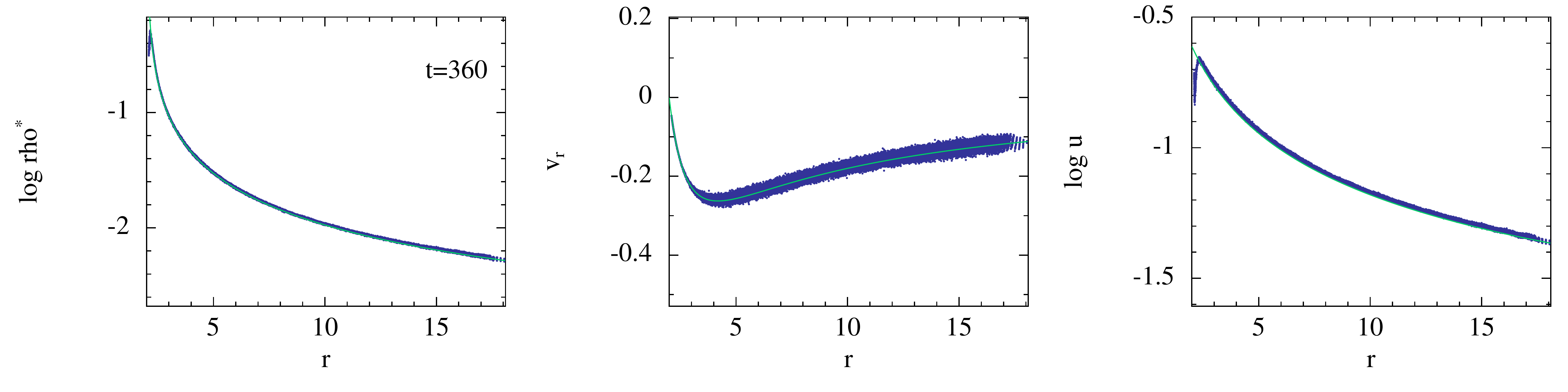}
      \caption{Results of spherically symmetric `sonic point' Bondi accretion in the Schwarzschild metric with $M=1$ and $\alpha_\mathrm{AV}=0.1$. Blue dots show SPH particles while green lines show the exact solution. The match between the numerical and analytic solutions demonstrates that our implementation of hydrodynamics in strong-field gravity is correct.}
      \label{fig:bondi-sonicpoint-smallalpha}
   \end{center}
   \begin{center}
      \includegraphics[width=0.95\textwidth]{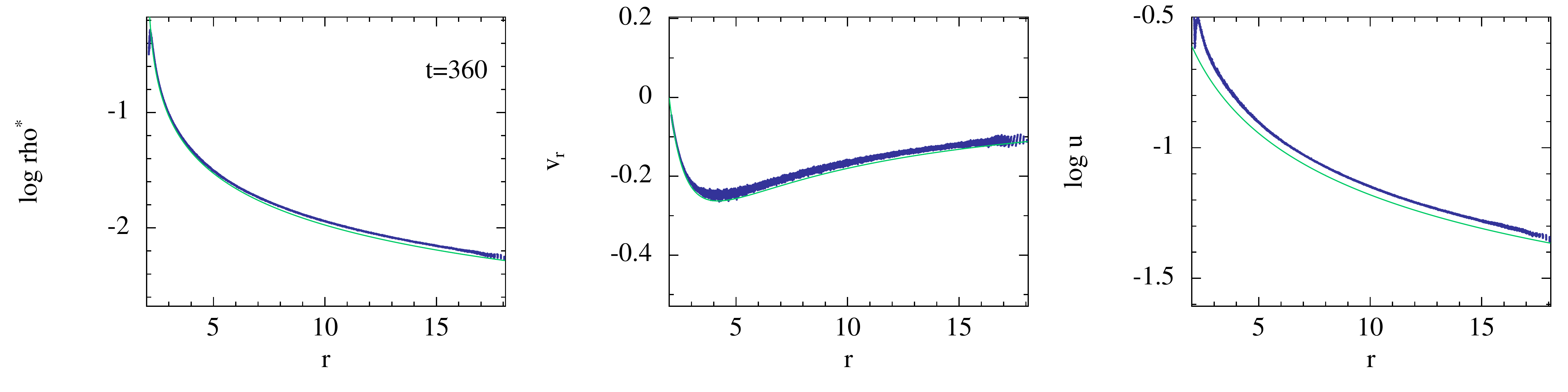}
      \caption{As in Fig.~\ref{fig:bondi-sonicpoint-smallalpha} but with $\alpha_\mathrm{AV}=1$. This produces excessive viscous heating (right), as in the geodesic case, however it does reduce particle noise in the velocity profile (middle).}
      \label{fig:bondi-sonicpoint}
   \end{center}
\end{figure*}

\begin{figure*}
   \begin{center}
      \includegraphics[width=\textwidth]{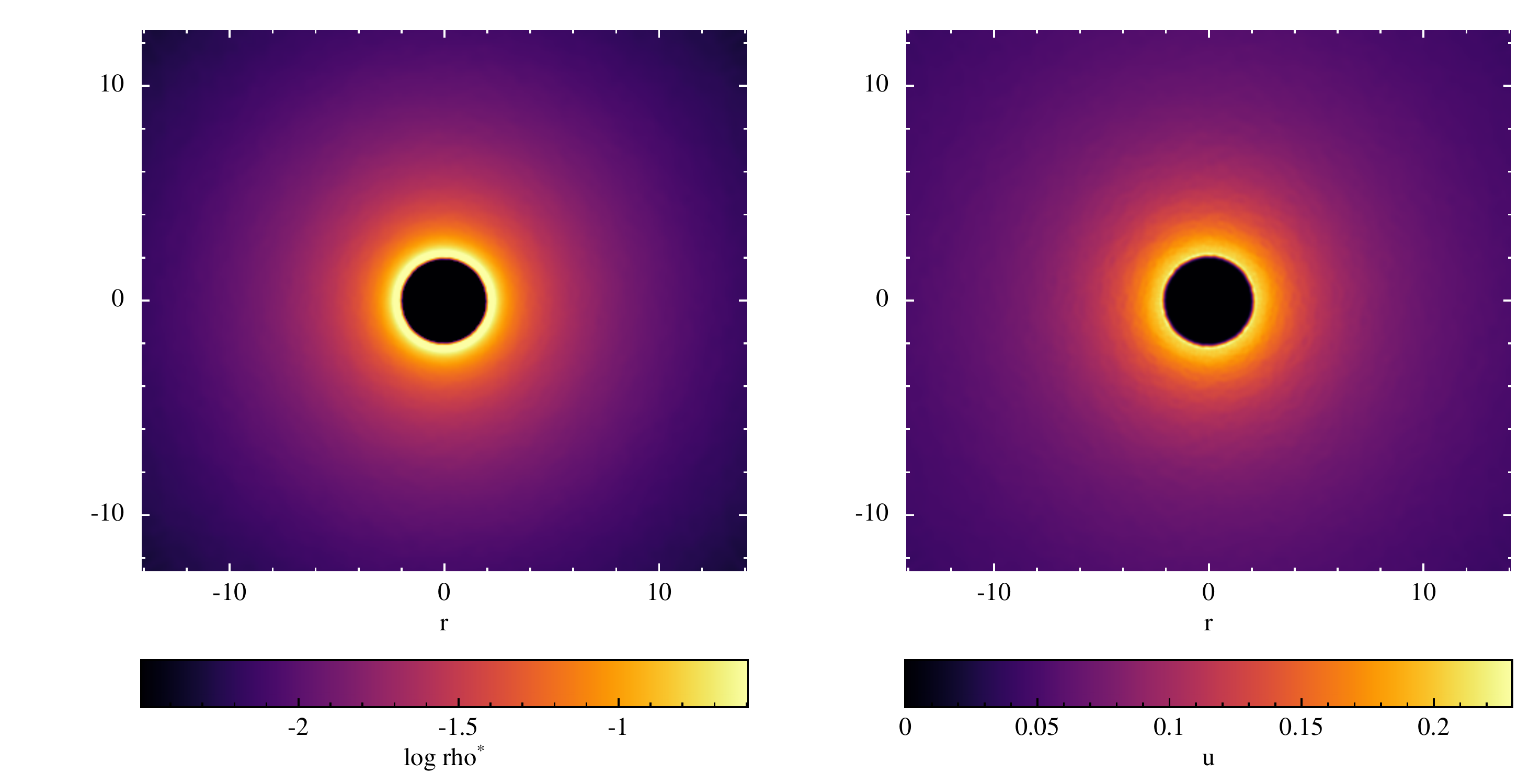}
      \caption{Rendered slices of spherically symmetric `sonic point' Bondi accretion as in Fig.~\ref{fig:bondi-sonicpoint-smallalpha}, showing conserved density $\rho^*$ (left) and specific thermal energy $u$ (right). Slice is taken through the $z=0$ plane. The density and thermal energy gradients are correctly captured at the inner boundary so long as the event horizon is well resolved.}
      \label{fig:bondi-render}
    \end{center}
\end{figure*}

\subsection{Orbital dynamics --- Schwarzschild Metric}
In order to test our ability to perform orbital dynamics accurately, we use the Schwarzschild metric in spherical Schwarzschild coordinates. We confirmed that the same results can be reproduced using the equivalent Cartesian representation (see Appendix \ref{sec:appendix-schwarzschild}). We use a black hole mass of $M=1$ in all simulations.

\subsubsection{Circular orbits}
The orbital frequency $\Omega$ of a circular geodesic in the Schwarzschild metric is identical to the frequency of a Newtonian orbit
\begin{equation}
   \Omega^2 = \brkt{\deriv{\phi}{t}}^2 = \frac{M}{r^{3}}.
\end{equation}
Figure \ref{fig:circular-orbits} (left) shows a test particle orbit with initial position $(r,\theta,\phi) = (10 M,\pi/2,0)$ and initial velocity $(v^r,v^\theta , v^\phi ) = (0,0,\Omega)$, simulated for $15$ orbital periods with constant time step $\Delta t=0.01 M$ throughout. Energy and angular momentum $(e,p_\phi)$ are conserved to machine precision ($\Delta E/E \approx 10^{-15}$), and the radius also remains constant to machine precision.

\subsubsection{Radial geodesics}
The velocity of a radially in-falling particle beginning at rest in the Schwarzschild metric is given by
\begin{equation}
   v^r(r) = \frac{1 - \frac{2M}{r}}{\sqrt{1 - \frac{2M}{r_0}}} \sqrt{2M\brkt{ \frac{1}{r} - \frac{1}{r_0}}},
\end{equation}
where $r_0$ is the starting radius. Figure \ref{fig:radial-multi} shows our calculations of radially in-falling test particles from several radii, compared to the exact solution. We integrate with constant time step $\Delta t=0.01 M$ until the particles reach $r=2.001 M$. The largest $L_2$ error out of the 10 simulations occurs for the $r_0=4 M$ case, with $L_2=6.7\times 10^{-8}$ when using data points every $31$ time steps. This confirms that radial infall is accurately captured by our code.

% Errors and final velocity when integrating up to r=2.001
%r0 =  4.0    L2 =  6.71140324087e-08  vmin =  0.000517065069444
%r0 =  8.0    L2 =  4.8393507444e-08    vmin =  0.000555603360157
%r0 =  12.0  L2 =  3.738544684e-08      vmin =  0.000566326647771
%r0 =  16.0  L2 =  3.08364393544e-08  vmin =  0.000558355398093
%r0 =  20.0  L2 =  2.64524868978e-08  vmin =  0.000536684600695
%r0 =  24.0  L2 =  2.32899811024e-08  vmin =  0.000516502107291
%r0 =  28.0  L2 =  2.0892241807e-08    vmin =  0.000523619061718
%r0 =  32.0  L2 =  1.89962738232e-08  vmin =  0.000513591505573
%r0 =  36.0  L2 =  1.74604296855e-08  vmin =  0.000530101654622
%r0 =  40.0  L2 =  1.62811853983e-08  vmin =  0.00345254447354

\subsubsection{Apsidal advance} \label{sec:precession-schwarzschild}
Figure~\ref{fig:precession} (left) shows the trajectory of a precessing orbit. The orbit begins at $r_0=90 M$ with tangential velocity $v_\mathrm{t}=0.0521157$ such that the pericentre distance is $\approx 10 M$, i.e.
\begin{align}
     (r,\theta,\phi)      &=\brkt{r_0,\frac\pi2,0},\\
     (v^r,v^\theta,v^\phi)&=\brkt{0,0,\frac{v_t}{r_0}}.
\end{align}
The trajectory of our simulation follows the trajectory as computed from a direct integration of the geodesic equation with coordinate time derivatives (see e.g. Equation 4.2 of \citealt{tejedagaftonrosswog17}).
We find a change in azimuthal angle of approximately $82.3^\circ$ between the first apocentre and the starting position. This is consistent with the exact value of $82.4^\circ$ (see e.g. Equation 2 of \citealt{wegg12}) to within measurement error. A more accurate value can be obtained with higher temporal resolution at the apocentre. This confirms that the apsidal advance is as expected from theory. 
%\begin{align}
%   \frac{\d^2 x^i}{d t^2} = \brkt{ \deriv{x^i}{t} g^{0 \nu} - g^{i \nu} } \brkt{ \pder{g_{\mu\nu}}{x^\alpha} -\frac12 \pder{g_{\mu\alpha}}{x^\nu} } \deriv{x^\mu}{t} \deriv{x^\alpha}{t}
%\end{align}

\subsection{Orbital dynamics --- Kerr Metric}
We perform further tests of orbital dynamics using the Kerr metric in spherical Boyer-Lindquist coordinates, however the same results can be reproduced using the equivalent Cartesian representation (see Appendix \ref{sec:appendix-kerr}). We again use a black hole mass of $M=1$ in all simulations.

\subsubsection{Circular orbits}
Circular geodesics are allowed in the Kerr metric when the plane of orbit is perpendicular to the spin axis of the black hole (i.e. $\theta = \pi/2$ in Boyer-Lindquist coordinates).
The orbital frequency for a circular orbit in the Kerr metric is given by \citep{abramowiczjaroszynskisikora78}
\begin{equation}
   \Omega = \frac{M^{1/2}}{r^{3/2}+aM^{1/2}},
\end{equation}
for orbits going in the positive $\phi$ direction, where $|a| \leq M$ is the angular momentum parameter of the black hole. Prograde orbits correspond to $a>0$, whilst retrograde orbits are when $a<0$. Figure \ref{fig:circular-orbits} (right) shows a circular orbit at a radius of $2M$ for a maximally rotating black hole ($a/M=1$), evolved for 15 orbital periods with constant $\Delta t=0.01 M$ and compared to the exact orbit. Again, energy and angular momentum $(e,p_\phi)$ are conserved to machine precision, and the radius also remains constant to machine precision.

\subsubsection{Epicyclic motion} \label{sec:epicycle}
An otherwise circular orbit given a small, linear, radial perturbation, will undergo epicyclic motion in the Kerr metric. The epicyclic frequency $\kappa$ is given by \citep{kato90,lubowogilviepringle02}
\begin{equation}
   \kappa^2 = \Omega^2 \brkt{ 1 - \left[\frac{6M}{r} - \frac{8aM^{1/2}}{r^{3/2}} + \frac{3a^2}{r^2}\right] }.
\end{equation}
We compute $\kappa$ for a range of values of $a$, by setting a series of test particles on epicyclic orbits at different radii. To induce the epicyclic motion we set the initial angular velocity to be $v_\phi = 1.00001\Omega$ (i.e. a 0.001\% increase from the circular velocity). For simplicity and because test particles do not interact, simulations with the same $a$ were computed simultaneously. Each simulation was run for $t=30000M$ to ensure that epicycles with the lowest frequency completed at least $20$ cycles. We then measure the frequency, as described in Appendix~\ref{sec:appendix-fft}. Figure \ref{fig:oscillations} (left) shows the results of all simulations for $a \in [-1,1]$ and $M=1$. The radii at which $\kappa=0$ correspond to the innermost stable circular orbit (ISCO). The typical errors are of order $\Delta \kappa/\kappa \approx 10^{-3}$. However, the errors can be made arbitrarily small by taking a longer time series, since the smallest frequency measurement possible with a Fourier transform is $\delta \Omega = 2\pi/t_\mathrm{max}$.

\subsubsection{Vertical-oscillation frequency} \label{sec:vertical-oscillations}
A circular orbit given a small, linear, vertical perturbation, will undergo vertical oscillations in the Kerr metric. The vertical oscillation frequency $\Omega_z$ is given by \citep{kato90,lubowogilviepringle02}
\begin{equation}
   \label{eq:vertical-oscillation}
   \Omega_z^2 =  \Omega^2 \brkt{ 1 - \left[ \frac{4aM^{1/2}}{r^{3/2}} + \frac{3a^2}{r^2} \right] }.
\end{equation}
Figure \ref{fig:oscillations} (right) shows our calculations of the vertical oscillation frequency, following a similar setup as above (see Section \ref{sec:epicycle}). To induce vertical oscillations we perturb test particles up out of the $\theta=\pi/2$ plane by 0.001\% (i.e. $\theta = \brkt{1-0.00001}\pi/2$). The exact solutions for each value of $a$ are truncated at the radius corresponding to the ISCO. Typical errors are similar to those for the epicyclic frequency ($\Delta \Omega_z/\Omega_z \approx 10^{-3}$), and can again be made smaller by taking a longer time series.

\subsubsection{Apsidal advance}
Figure \ref{fig:precession} (middle \& right) demonstrates the effect of black hole spin on apsidal advance. We use similar initial conditions as in Section \ref{sec:precession-schwarzschild} however with $a/M=\{-0.1,0,0.1\}$ in the Kerr metric and performed these calculations in Cartesian-like coordinates.
We chose relatively small values of spin in order to distinguish the direction of orbital advance, however there is no limit on the values of spin that the code can handle.
The initial conditions for this orbit are such that in the x--y plane the initial starting position and velocity is the same for all three orbits since $r^2 \neq x^2+y^2+z^2$ in Boyer-Lindquist coordinates, i.e.
\begin{align}
   (x,y,z)       &= (r_0,0,0), \\
   (v^x,v^y,v^z) &= (0,v_\mathrm{tan},0).
\end{align}
where again we use $r_0=90 M$ and $v_\mathrm{tan}=0.0521157$. The angle of precession decreases for prograde orbits, and increases for retrograde orbits, as expected. Combined with the high accuracy of our epicyclic frequency measurements, this confirms that we capture Kerr orbital dynamics accurately. 

\subsubsection{Conservation of angular momentum}
Figure \ref{fig:timestepping} shows the conservation of angular momentum in our timestepping algorithm described in Section \ref{sec:hybrid-leapfrog}. We compute the precessing orbit in the Schwarzschild metric in Cartesian coordinates (Section \ref{sec:precession-schwarzschild}) using both double (64 bit) and quad (128 bit) precision, whilst varying the number of steps per orbit (step size), and the tolerance ($\epsilon_x$, $\epsilon_p$) of the implicit solve. The conserved angular momentum in the Schwarzschild metric is
\begin{align}
   L &= r^2 \deriv{\phi}{t} U^0.
   %   &= \brkt{ x \deriv{y}{t} - y \deriv{x}{t} } U^0
\end{align}
We define the orbital period for this trajectory to be the time between consecutive apocentres ($t_{\mathrm{orbit}} \approx 2390 M$ with our setup). With $10^4$ steps per orbit, we find that we conserve angular momentum to round off error provided $\epsilon_p$ and $\epsilon_x$ are sufficiently small. Using tolerances of $10^{-15}$ (double precision) and $10^{-30}$ (quad precision) angular momentum is conserved to round-off error. Therefore in our 3D code, which is double precision, we set $\epsilon_p=\epsilon_x=10^{-15}$ by default.

\subsection{General relativity - Bondi accretion} \label{sec:bondi}
To test our ability to perform hydrodynamics in a curved spacetime and not just special relativity, we consider spherically symmetric flow in the Schwarzschild metric. \citet{hawleysmarrwilson84} give the analytic solution, which is a generalisation of non-relativistic Parker-Bondi flow. 
To simulate the accretion solution, we inject shells (geodesic spheres) of SPH particles from a particular radius at discrete, regular intervals. The particles are given density, velocity and thermal energy corresponding to the exact solution at the radius which they are injected. For details of the algorithm we refer the reader to the publicly available subroutine in \textsc{Phantom} used in applications of stellar winds (\citealt{toupinbraunsiess15,toupinbraunsiess15a}; Siess et al. in prep). 

\subsubsection{Geodesic flow} \label{sec:bondi-geodesic}
The solution for marginally bound, spherically symmetric, radial flow of `pressure-less' gas is given by \citep{hawleysmarrwilson84}\footnote{Note that equation (67b) of \citet{hawleysmarrwilson84} is written incorrectly}
\begin{align}
   v^r(r)     &= \sqrt{\frac{2M}{r}} \, \brkt{1-\frac{2M}{r}},\\
   \rho       &= \frac{\rho_0}{r^2} \sqrt{\frac{r}{2M}}, \\
   u(r)        &= u_0 \brkt{r^2 \sqrt{\frac{2M}{r}}}^{1-\adgamma},
\end{align}
where $\rho_0$ and $u_0$ are normalisation constants. The solution assumes that $u \ll 1$ everywhere, hence we must use $u_0 \ll 1$ to be consistent. The solution is also the same for inflow and outflow, aside from the sign of $v^r$. For our simulations we use normalisation constants $u_0=1\times10^{-9}$ and $\rho_0=1$. 

Following \citet{hawleysmarrwilson84a}, we inject fluid radially in toward the black hole from a radius of $r_\mathrm{in}=18.1M$. To have a smooth boundary condition at $r_\mathrm{in}$, we keep 10 shells of particles, uniformly spaced in time, outside of the injection radius at all times. The hydrodynamic quantities $v^r$, $\rho$ and $u$ are fixed to the exact solution while the shells remain outside the injection radius. As each shell moves inward and crosses $r_\mathrm{in}$, it is `released' and left to evolve on its own. Once a shell is released, a new boundary shell is added to the outside of the remaining boundary shells in order to maintain a constant number of boundary shells. For the inner boundary, we simply remove particles from the simulation when they cross $r=2.1M$. 

Figures \ref{fig:bondi-geodesic-nodiss} \& \ref{fig:bondi-geodesic} show our results for geodesic Bondi accretion with $8412$ particles per shell, a time spacing of $\delta t = 0.602M$ between consecutive shells, and individual particle masses of $m_a=9\times10^{-4}M$. We find that we achieve best results with artificial viscosity turned off i.e. $\alpha_\mathrm{AV}=0$ (Fig.~\ref{fig:bondi-geodesic-nodiss}). After evolving the simulation to $t=90M$ the $L_2$ errors are $5.4\times10^{-2}$, $1.8\times10^{-4}$ and $4.4\times10^{-2}$ for $\rho^*$, $v^r$ and $u$, respectively.

Figure~\ref{fig:bondi-geodesic} shows the same simulation but with the standard dissipation parameter $\alpha_\mathrm{AV}=1$ for comparison. Although there are no shocks present, viscous heating leads to excess thermal energy (third panel of Figure~\ref{fig:bondi-geodesic}). The excess heating is small but easily overwhelms the negligible thermal energy assumed in the test problem. To reduce viscous heating away from shocks, the standard approach in non-relativistic SPH is to implement shock detection switches. The most widely used is the switch proposed by \citep{cullendehnen10}, however this is non-trivial to generalise to relativity. We experimented with a modified form of the older \citet{morrismonaghan97} switch but found it ineffective for the Bondi problem since it relies on $(-\partial v^i / \partial x^i)$ as the shock detector, which does not switch off for adiabatic contraction.

There is a small turnover in density at the inner boundary in both cases. This is merely a consequence of our inner boundary condition and does not reflect the accuracy of our density calculation in a strong field regime. The drop off occurs because particles close to the accretion radius have fewer neighbours and thus a lower density. This effect is not unique to GRSPH and would occur wherever the inner boundary was chosen to be.

\subsubsection{Sonic point flow}
Dropping the assumption $u \ll 1$ means that the pressure between fluid elements is no longer negligible. The solution for this flow \citep{michel72,hawleysmarrwilson84} can be written in terms of a temperature variable $T \equiv P/\rho=(\adgamma-1) u$, not to be confused with the gas temperature as defined in equation~\ref{eq:idealgaslaw},
\begin{align}
   U^r(r) &= \frac{C_1}{r^2 T^n}, \\
   \rho(r) &= K_0 T^n, \\
   u(r)     &= n \, T,
\end{align}
where $n\equiv 1/(\adgamma-1)$ is the polytropic index and $K_0$ is an entropy normalisation constant. $T(r)$ is determined by the implicit equation
\begin{align}
   C_2   &= (1 + (n+1) T)^2 \brkt{1 -\frac{2M}{r} + \frac{{C_1}^2}{r^4 T^{2 n}}}. \label{eq:Tsolve}
\end{align}
The constants $C_1$ and $C_2$ can be both uniquely determined through the choice of a critical point $r_c$,
\begin{align}
   C_1   &= U_c \, {r_c}^2 \, {T_c}^n, \\
   C_2   &= \brkt{1 + (1+n)T_c}^2 \brkt{1 - \frac{2M}{r_c} + \frac{{C_1}^2}{{r_c}^4 {T_c}^{2n}}},
\end{align}
where
\begin{align}
   U_c   & \equiv U^r(r_c) = \sqrt{\frac{M}{2 r_c}}, \\
   T_c    & \equiv T(r_c)   = \frac{v_c^2 n}{(1+n)(1-n v_c^2)}, \\
   v_c    & \equiv v^r(r_c) = \frac{U_c}{\sqrt{1-3U_c^2}}.
\end{align}
To write the solution in terms of $v^r(r)$ and $\rho^*(r)$, we use equations~\ref{eq:coordinate-velocity} \& \ref{eq:rhostar} along with
\begin{align}
   U^0 = \frac{\sqrt{1- 2M/r + (U^r)^2}}{1- 2M/r},
\end{align}
as determined by equation~\ref{eq:u0} for the Schwarzschild metric. For $\adgamma=5/3$, equation~\ref{eq:Tsolve} yields two real solutions, one for inflow and the other for outflow \citep{michel72}. For our simulation we take care to choose the correct branch, corresponding to an inflow. We choose a critical value $r_c=8M$ and normalise the entropy to $K_0=1$.

Figures~\ref{fig:bondi-sonicpoint-smallalpha} \& \ref{fig:bondi-sonicpoint} show the results of our simulation. We follow the same procedure as for the geodesic flow simulation, injecting shells of particles from $r_\mathrm{in}=18.1M$, however this time we fill the domain all the way to the accretion radius at $r=2.1M$ before begining the simulation. This reduces the time taken to reach a steady state. Each shell consists of 8412 particles, with time intervals of $\delta t=1.404M$ between each consecutive shell, and individual particle masses of $m_a=4\times10^{-4}M$.

Figure~\ref{fig:bondi-sonicpoint-smallalpha} shows our results with a reduced artificial viscosity parameter of $\alpha_\mathrm{AV}=0.1$. After evolving the simulation to $t=360M$ in order to reach a steady state, the $L_2$ errors are $4.2\times10^{-2}$, $2.5\times10^{-2}$ and $3.2\times10^{-2}$ for $\rho^*$, $v^r$ and $u$, respectively.

In Figure~\ref{fig:bondi-sonicpoint} we show the same simulation but with $\alpha_\mathrm{AV}=1$ for comparison. As with the geodesic simulations above, we find excessive viscous heating when the standard artificial viscosity parameter is used. A small drop off in density can again be seen at the inner boundary, for the same reasons as discussed in the geodesic simulation. Figure~\ref{fig:bondi-render} shows a rendered cross section of the conserved density $\rho^*$ and the specific thermal energy $u$ close to the inner boundary for the same simulation as Fig.~\ref{fig:bondi-sonicpoint-smallalpha}. This demonstrates that the boundary condition is not a serious concern, so long as the event horizon is well resolved.

\section{Conclusions}
We have developed and benchmarked a 3D GRSPH code for hydrodynamics in strong field gravity. The novel aspects are:
\begin{enumerate}
\item Shock capturing: We demonstrated how to split viscous and thermal conduction components of the shock dissipation terms in relativistic SPH. This is important for accurate treatment of contact discontinuities \citep{price08}.
\item Entropy: We show how to formulate the equations with entropy as the conserved variable. This enables us to robustly simulate strong blast waves with no smoothing of the initial conditions without the code failing due to negative pressures.
\item Timestepping: We showed how to generalise the Leapfrog time integration scheme to relativistic SPH. We combined a symplectic and reversible method for non-separable Hamiltonians from \citet{leimkuhlerreich05} with a modified algorithm that is efficient for SPH. This allows for symplectic and reversible integration of the orbital dynamics and reversible integration of the SPH equations in the absence of dissipation. Our algorithm generalises the RESPA scheme \citep{tuckermanbernemartyna92} employed by \citet{pricewurstertricco18}.
\item Benchmarking: We benchmark both our algorithm and codes against a series of standardised tests in 1D and 3D. We showed results for mildly and ultra-relativistic shocks in 1D and 3D, circular and eccentric orbits, radial geodesics, vertical and epicyclic oscillation about circular orbits in Schwarzschild and Kerr metrics, 3D relativistic spherical blast, and generalised Bondi flow.
\item Public code: We implemented our algorithms into the public SPH code \textsc{Phantom} \citep{pricewurstertricco18}. Our intention is to make GR \textsc{Phantom} publicly available in due course.
\end{enumerate}
We showed how to use SPH, for flows in curved metrics of space. We captured the shocks, and adjusted the clocks. Our tests, they all worked in each case.

%----------------------------------------------------------------------------------------------------------------
\section*{Acknowledgments}
We thank Paul Lasky, Yuri Levin, Lionel Siess, Kimitake Hayasaki and Clement Bonnerot for useful discussions and feedback. DP thanks Walter Dehnen and Stephan Rosswog for useful discussions. We also thank the anonymous referee for comments which have improved the paper. We acknowledge CPU time on OzSTAR, funded by Swinburne University and the Australian Government. DL is funded through the Australian government RTP (Research Training Program) stipend. DP is grateful for funding from the Australian Research Council via DP130102078 and FT130100034.

\bibliography{dave}

\begin{thebibliography}{}
\makeatletter
\relax
\def\mn@urlcharsother{\let\do\@makeother \do\$\do\&\do\#\do\^\do\_\do\%\do\~}
\def\mn@doi{\begingroup\mn@urlcharsother \@ifnextchar [ {\mn@doi@}
  {\mn@doi@[]}}
\def\mn@doi@[#1]#2{\def\@tempa{#1}\ifx\@tempa\@empty \href
  {http://dx.doi.org/#2} {doi:#2}\else \href {http://dx.doi.org/#2} {#1}\fi
  \endgroup}
\def\mn@eprint#1#2{\mn@eprint@#1:#2::\@nil}
\def\mn@eprint@arXiv#1{\href {http://arxiv.org/abs/#1} {{\tt arXiv:#1}}}
\def\mn@eprint@dblp#1{\href {http://dblp.uni-trier.de/rec/bibtex/#1.xml}
  {dblp:#1}}
\def\mn@eprint@#1:#2:#3:#4\@nil{\def\@tempa {#1}\def\@tempb {#2}\def\@tempc
  {#3}\ifx \@tempc \@empty \let \@tempc \@tempb \let \@tempb \@tempa \fi \ifx
  \@tempb \@empty \def\@tempb {arXiv}\fi \@ifundefined
  {mn@eprint@\@tempb}{\@tempb:\@tempc}{\expandafter \expandafter \csname
  mn@eprint@\@tempb\endcsname \expandafter{\@tempc}}}

\bibitem[\protect\citeauthoryear{{Abbott} et~al.,}{{Abbott}
  et~al.}{2017a}]{abbottabbottabbott17}
{Abbott} B.~P.,  et~al., 2017a, Physical Review Letters, \href
  {http://adsabs.harvard.edu/abs/2017PhRvL.119p1101A} {119, 161101}

\bibitem[\protect\citeauthoryear{{Abbott} et~al.,}{{Abbott}
  et~al.}{2017b}]{abbottabbottabbott17a}
{Abbott} B.~P.,  et~al., 2017b, \apjl, \href
  {http://adsabs.harvard.edu/abs/2017ApJ...848L..12A} {848, L12}

\bibitem[\protect\citeauthoryear{{Abramowicz}, {Jaroszynski}  \&
  {Sikora}}{{Abramowicz} et~al.}{1978}]{abramowiczjaroszynskisikora78}
{Abramowicz} M.,  {Jaroszynski} M.,   {Sikora} M.,  1978, \aap, \href
  {http://adsabs.harvard.edu/abs/1978A%26A....63..221A} {63, 221}

\bibitem[\protect\citeauthoryear{{Akiyama} et~al.,}{{Akiyama}
  et~al.}{2017}]{akiyamakuramochiikeda17}
{Akiyama} K.,  et~al., 2017, \apj, \href
  {http://adsabs.harvard.edu/abs/2017ApJ...838....1A} {838, 1}

\bibitem[\protect\citeauthoryear{{Andersson} \& {Comer}}{{Andersson} \&
  {Comer}}{2007}]{anderssoncomer07}
{Andersson} N.,  {Comer} G.~L.,  2007, Living Reviews in Relativity, \href
  {http://adsabs.harvard.edu/abs/2007LRR....10....1A} {10, 1}

\bibitem[\protect\citeauthoryear{{Armitage} \& {Natarajan}}{{Armitage} \&
  {Natarajan}}{2002}]{armitagenatarajan02}
{Armitage} P.~J.,  {Natarajan} P.,  2002, \apjl, \href
  {http://adsabs.harvard.edu/abs/2002ApJ...567L...9A} {567, L9}

\bibitem[\protect\citeauthoryear{{Arnowitt}, {Deser}  \& {Misner}}{{Arnowitt}
  et~al.}{2008}]{arnowittdesermisner08}
{Arnowitt} R.,  {Deser} S.,   {Misner} C.~W.,  2008, General Relativity and
  Gravitation, \href {http://adsabs.harvard.edu/abs/2008GReGr..40.1997A} {40,
  1997}

\bibitem[\protect\citeauthoryear{{Bauswein}, {Oechslin}  \& {Janka}}{{Bauswein}
  et~al.}{2010}]{bausweinoechslinjanka10}
{Bauswein} A.,  {Oechslin} R.,   {Janka} H.-T.,  2010, \prd, \href
  {http://adsabs.harvard.edu/abs/2010PhRvD..81b4012B} {81, 024012}

\bibitem[\protect\citeauthoryear{Bonnerot, Rossi, Lodato  \& Price}{Bonnerot
  et~al.}{2016}]{bonnerotrossilodato16}
Bonnerot C.,  Rossi E.~M.,  Lodato G.,   Price D.~J.,  2016, Monthly Notices of
  the Royal Astronomical Society, 455, 2253

\bibitem[\protect\citeauthoryear{{Cerioli}, {Lodato}  \& {Price}}{{Cerioli}
  et~al.}{2016}]{ceriolilodatoprice16}
{Cerioli} A.,  {Lodato} G.,   {Price} D.~J.,  2016, \mnras, \href
  {http://adsabs.harvard.edu/abs/2016MNRAS.457..939C} {457, 939}

\bibitem[\protect\citeauthoryear{Chow \& Monaghan}{Chow \&
  Monaghan}{1997}]{chowmonaghan97}
Chow E.,  Monaghan J.~J.,  1997, J. Comp. Phys., 134, 296

\bibitem[\protect\citeauthoryear{{Cullen} \& {Dehnen}}{{Cullen} \&
  {Dehnen}}{2010}]{cullendehnen10}
{Cullen} L.,  {Dehnen} W.,  2010, \mnras, \href
  {http://adsabs.harvard.edu/abs/2010MNRAS.408..669C} {408, 669}

\bibitem[\protect\citeauthoryear{{Del Zanna} \& {Bucciantini}}{{Del Zanna} \&
  {Bucciantini}}{2002}]{del-zannabucciantini02}
{Del Zanna} L.,  {Bucciantini} N.,  2002, \aap, \href
  {http://adsabs.harvard.edu/abs/2002A%26A...390.1177D} {390, 1177}

\bibitem[\protect\citeauthoryear{{Faber}, {Grandcl{\'e}ment}  \&
  {Rasio}}{{Faber} et~al.}{2004}]{fabergrandclementrasio04}
{Faber} J.~A.,  {Grandcl{\'e}ment} P.,   {Rasio} F.~A.,  2004, \prd, \href
  {http://adsabs.harvard.edu/abs/2004PhRvD..69l4036F} {69, 124036}

\bibitem[\protect\citeauthoryear{{Gammie}, {McKinney}  \& {T{\'o}th}}{{Gammie}
  et~al.}{2003}]{gammiemckinneytoth03}
{Gammie} C.~F.,  {McKinney} J.~C.,   {T{\'o}th} G.,  2003, \apj, \href
  {http://adsabs.harvard.edu/abs/2003ApJ...589..444G} {589, 444}

\bibitem[\protect\citeauthoryear{{Giacomazzo} \& {Rezzolla}}{{Giacomazzo} \&
  {Rezzolla}}{2006}]{giacomazzorezzolla06}
{Giacomazzo} B.,  {Rezzolla} L.,  2006, Journal of Fluid Mechanics, \href
  {http://adsabs.harvard.edu/abs/2006JFM...562..223G} {562, 223}

\bibitem[\protect\citeauthoryear{{Gingold} \& {Monaghan}}{{Gingold} \&
  {Monaghan}}{1977}]{gingoldmonaghan77}
{Gingold} R.~A.,  {Monaghan} J.~J.,  1977, \mnras, \href
  {http://adsabs.harvard.edu/abs/1977MNRAS.181..375G} {181, 375}

\bibitem[\protect\citeauthoryear{{Hawley}, {Smarr}  \& {Wilson}}{{Hawley}
  et~al.}{1984a}]{hawleysmarrwilson84a}
{Hawley} J.~F.,  {Smarr} L.~L.,   {Wilson} J.~R.,  1984a, \apjs, \href
  {http://adsabs.harvard.edu/abs/1984ApJS...55..211H} {55, 211}

\bibitem[\protect\citeauthoryear{{Hawley}, {Smarr}  \& {Wilson}}{{Hawley}
  et~al.}{1984b}]{hawleysmarrwilson84}
{Hawley} J.~F.,  {Smarr} L.~L.,   {Wilson} J.~R.,  1984b, \apj, \href
  {http://adsabs.harvard.edu/abs/1984ApJ...277..296H} {277, 296}

\bibitem[\protect\citeauthoryear{{Hayasaki}, {Stone}  \& {Loeb}}{{Hayasaki}
  et~al.}{2013}]{hayasakistoneloeb13}
{Hayasaki} K.,  {Stone} N.,   {Loeb} A.,  2013, \mnras, \href
  {http://adsabs.harvard.edu/abs/2013MNRAS.434..909H} {434, 909}

\bibitem[\protect\citeauthoryear{{Hayasaki}, {Stone}  \& {Loeb}}{{Hayasaki}
  et~al.}{2016}]{hayasakistoneloeb16}
{Hayasaki} K.,  {Stone} N.,   {Loeb} A.,  2016, \mnras, \href
  {http://adsabs.harvard.edu/abs/2016MNRAS.461.3760H} {461, 3760}

\bibitem[\protect\citeauthoryear{{Kato}}{{Kato}}{1990}]{kato90}
{Kato} S.,  1990, \pasj, \href
  {http://adsabs.harvard.edu/abs/1990PASJ...42...99K} {42, 99}

\bibitem[\protect\citeauthoryear{{Kheyfets}, {Miller}  \& {Zurek}}{{Kheyfets}
  et~al.}{1990}]{kheyfetsmillerzurek90}
{Kheyfets} A.,  {Miller} W.~A.,   {Zurek} W.~H.,  1990, \prd, \href
  {http://adsabs.harvard.edu/abs/1990PhRvD..41..451K} {41, 451}

\bibitem[\protect\citeauthoryear{{Laguna}, {Miller}  \& {Zurek}}{{Laguna}
  et~al.}{1993}]{lagunamillerzurek93}
{Laguna} P.,  {Miller} W.~A.,   {Zurek} W.~H.,  1993, \apj, \href
  {http://adsabs.harvard.edu/abs/1993ApJ...404..678L} {404, 678}

\bibitem[\protect\citeauthoryear{{Leimkuhler} \& {Reich}}{{Leimkuhler} \&
  {Reich}}{2005}]{leimkuhlerreich05}
{Leimkuhler} B.,  {Reich} S.,  2005, {Simulating Hamiltonian Dynamics}.
Cambridge Monographs on Applied and Computational Mathematics, Cambridge
  University Press

\bibitem[\protect\citeauthoryear{{Lubow}, {Ogilvie}  \& {Pringle}}{{Lubow}
  et~al.}{2002}]{lubowogilviepringle02}
{Lubow} S.~H.,  {Ogilvie} G.~I.,   {Pringle} J.~E.,  2002, \mnras, \href
  {http://adsabs.harvard.edu/abs/2002MNRAS.337..706L} {337, 706}

\bibitem[\protect\citeauthoryear{{Lucy}}{{Lucy}}{1977}]{lucy77}
{Lucy} L.~B.,  1977, \aj, \href
  {http://adsabs.harvard.edu/abs/1977AJ.....82.1013L} {82, 1013}

\bibitem[\protect\citeauthoryear{{Mann}}{{Mann}}{1991}]{mann91}
{Mann} P.~J.,  1991, Computer Physics Communications, \href
  {http://adsabs.harvard.edu/abs/1991CoPhC..67..245M} {67, 245}

\bibitem[\protect\citeauthoryear{{Mart{\'\i}} \& {M{\"u}ller}}{{Mart{\'\i}} \&
  {M{\"u}ller}}{1994}]{martimuller94}
{Mart{\'\i}} J.~M.,  {M{\"u}ller} E.,  1994, Journal of Fluid Mechanics, \href
  {http://adsabs.harvard.edu/abs/1994JFM...258..317M} {258, 317}

\bibitem[\protect\citeauthoryear{{Mart{\'\i}} \& {M{\"u}ller}}{{Mart{\'\i}} \&
  {M{\"u}ller}}{2003}]{martimuller03}
{Mart{\'\i}} J.~M.,  {M{\"u}ller} E.,  2003, Living Reviews in Relativity,
  \href {http://adsabs.harvard.edu/abs/2003LRR.....6....7M} {6, 7}

\bibitem[\protect\citeauthoryear{{Mart{\'\i}} \& {M{\"u}ller}}{{Mart{\'\i}} \&
  {M{\"u}ller}}{2015}]{martimuller15}
{Mart{\'\i}} J.~M.,  {M{\"u}ller} E.,  2015, Living Reviews in Computational
  Astrophysics, \href {http://adsabs.harvard.edu/abs/2015LRCA....1....3M} {1,
  3}

\bibitem[\protect\citeauthoryear{{Mart{\'\i}}, {Ib{\'a}ez}  \&
  {Miralles}}{{Mart{\'\i}} et~al.}{1991}]{martibaezmiralles91}
{Mart{\'\i}} J.~M.,  {Ib{\'a}ez} J.~M.,   {Miralles} J.~A.,  1991, \prd, \href
  {http://adsabs.harvard.edu/abs/1991PhRvD..43.3794M} {43, 3794}

\bibitem[\protect\citeauthoryear{{Metzger}}{{Metzger}}{2017}]{metzger17}
{Metzger} B.~D.,  2017, Living Reviews in Relativity, \href
  {http://adsabs.harvard.edu/abs/2017LRR....20....3M} {20, 3}

\bibitem[\protect\citeauthoryear{{Michel}}{{Michel}}{1972}]{michel72}
{Michel} F.~C.,  1972, \apss, \href
  {http://adsabs.harvard.edu/abs/1972Ap%26SS..15..153M} {15, 153}

\bibitem[\protect\citeauthoryear{{Milosavljevi{\'c}} \&
  {Phinney}}{{Milosavljevi{\'c}} \& {Phinney}}{2005}]{milosavljevicphinney05}
{Milosavljevi{\'c}} M.,  {Phinney} E.~S.,  2005, \apjl, \href
  {http://adsabs.harvard.edu/abs/2005ApJ...622L..93M} {622, L93}

\bibitem[\protect\citeauthoryear{{Monaghan}}{{Monaghan}}{1997}]{monaghan97}
{Monaghan} J.~J.,  1997, Journal of Computational Physics, \href
  {http://adsabs.harvard.edu/abs/1997JCoPh.136..298M} {136, 298}

\bibitem[\protect\citeauthoryear{{Monaghan}}{{Monaghan}}{2002}]{monaghan02}
{Monaghan} J.~J.,  2002, \mnras, \href
  {http://adsabs.harvard.edu/abs/2002MNRAS.335..843M} {335, 843}

\bibitem[\protect\citeauthoryear{Monaghan \& Price}{Monaghan \&
  Price}{2001}]{monaghanprice01}
Monaghan J.~J.,  Price D.~J.,  2001, Monthly Notices of the Royal Astronomical
  Society, 328, 381

\bibitem[\protect\citeauthoryear{{Morris} \& {Monaghan}}{{Morris} \&
  {Monaghan}}{1997}]{morrismonaghan97}
{Morris} J.~P.,  {Monaghan} J.~J.,  1997, Journal of Computational Physics,
  \href {http://adsabs.harvard.edu/abs/1997JCoPh.136...41M} {136, 41}

\bibitem[\protect\citeauthoryear{Nealon, Price  \& Nixon}{Nealon
  et~al.}{2015}]{nealonpricenixon15}
Nealon R.,  Price D.~J.,   Nixon C.~J.,  2015, Monthly Notices of the Royal
  Astronomical Society, 448, 1526

\bibitem[\protect\citeauthoryear{{Nixon} \& {Salvesen}}{{Nixon} \&
  {Salvesen}}{2014}]{nixonsalvesen14}
{Nixon} C.,  {Salvesen} G.,  2014, \mnras, \href
  {http://adsabs.harvard.edu/abs/2014MNRAS.437.3994N} {437, 3994}

\bibitem[\protect\citeauthoryear{{Nixon}, {King}, {Price}  \& {Frank}}{{Nixon}
  et~al.}{2012}]{nixonkingprice12}
{Nixon} C.,  {King} A.,  {Price} D.,   {Frank} J.,  2012, \apjl, \href
  {http://adsabs.harvard.edu/abs/2012ApJ...757L..24N} {757, L24}

\bibitem[\protect\citeauthoryear{Oechslin, Rosswog  \& Thielemann}{Oechslin
  et~al.}{2002}]{oechslinrosswogthielemann02}
Oechslin R.,  Rosswog S.,   Thielemann F.-K.,  2002, Physical Review D, 65,
  103005

\bibitem[\protect\citeauthoryear{{Okazaki}, {Nagataki}, {Naito}, {Kawachi},
  {Hayasaki}, {Owocki}  \& {Takata}}{{Okazaki}
  et~al.}{2011}]{okazakinagatakinaito11}
{Okazaki} A.~T.,  {Nagataki} S.,  {Naito} T.,  {Kawachi} A.,  {Hayasaki} K.,
  {Owocki} S.~P.,   {Takata} J.,  2011, \pasj, \href
  {http://adsabs.harvard.edu/abs/2011PASJ...63..893O} {63, 893}

\bibitem[\protect\citeauthoryear{{Price}}{{Price}}{2008}]{price08}
{Price} D.~J.,  2008, Journal of Computational Physics, \href
  {http://adsabs.harvard.edu/abs/2008JCoPh.22710040P} {227, 10040}

\bibitem[\protect\citeauthoryear{{Price}}{{Price}}{2012}]{price12}
{Price} D.~J.,  2012, Journal of Computational Physics, \href
  {http://adsabs.harvard.edu/abs/2012JCoPh.231..759P} {231, 759}

\bibitem[\protect\citeauthoryear{{Price} \& {Federrath}}{{Price} \&
  {Federrath}}{2010}]{pricefederrath10}
{Price} D.~J.,  {Federrath} C.,  2010, \mnras, \href
  {http://adsabs.harvard.edu/abs/2010MNRAS.406.1659P} {406, 1659}

\bibitem[\protect\citeauthoryear{{Price} \& {Monaghan}}{{Price} \&
  {Monaghan}}{2007}]{pricemonaghan07}
{Price} D.~J.,  {Monaghan} J.~J.,  2007, \mnras, \href
  {http://adsabs.harvard.edu/abs/2007MNRAS.374.1347P} {374, 1347}

\bibitem[\protect\citeauthoryear{{Price} et~al.,}{{Price}
  et~al.}{2018}]{pricewurstertricco18}
{Price} D.~J.,  et~al., 2018, \pasa, \href
  {http://adsabs.harvard.edu/abs/2018PASA...35...31P} {35, e031}

\bibitem[\protect\citeauthoryear{Rosswog}{Rosswog}{2009}]{rosswog09}
Rosswog S.,  2009, New Astronomy Reviews, 53, 78

\bibitem[\protect\citeauthoryear{Rosswog}{Rosswog}{2010a}]{rosswog10a}
Rosswog S.,  2010a, Classical and Quantum Gravity, 27, 114108

\bibitem[\protect\citeauthoryear{{Rosswog}}{{Rosswog}}{2010b}]{rosswog10}
{Rosswog} S.,  2010b, Journal of Computational Physics, \href
  {http://adsabs.harvard.edu/abs/2010JCoPh.229.8591R} {229, 8591}

\bibitem[\protect\citeauthoryear{{Siegler} \& {Riffert}}{{Siegler} \&
  {Riffert}}{2000}]{sieglerriffert00}
{Siegler} S.,  {Riffert} H.,  2000, \apj, \href
  {http://adsabs.harvard.edu/abs/2000ApJ...531.1053S} {531, 1053}

\bibitem[\protect\citeauthoryear{{Springel} \& {Hernquist}}{{Springel} \&
  {Hernquist}}{2002}]{springelhernquist02}
{Springel} V.,  {Hernquist} L.,  2002, \mnras, \href
  {http://adsabs.harvard.edu/abs/2002MNRAS.333..649S} {333, 649}

\bibitem[\protect\citeauthoryear{Tejeda}{Tejeda}{2012}]{tejeda12}
Tejeda E.,  2012, PhD thesis, International School for Advanced Studies Via
  Bonomea 265, 34136 Trieste, Italy

\bibitem[\protect\citeauthoryear{Tejeda \& Rosswog}{Tejeda \&
  Rosswog}{2013}]{tejedarosswog13}
Tejeda E.,  Rosswog S.,  2013, Monthly Notices of the Royal Astronomical
  Society, 433, 1930

\bibitem[\protect\citeauthoryear{{Tejeda}, {Gafton}, {Rosswog}  \&
  {Miller}}{{Tejeda} et~al.}{2017}]{tejedagaftonrosswog17}
{Tejeda} E.,  {Gafton} E.,  {Rosswog} S.,   {Miller} J.~C.,  2017, \mnras,
  \href {http://adsabs.harvard.edu/abs/2017MNRAS.469.4483T} {469, 4483}

\bibitem[\protect\citeauthoryear{{Toupin}, {Braun}, {Siess}, {Jorissen}, {Gail}
   \& {Price}}{{Toupin} et~al.}{2015a}]{toupinbraunsiess15}
{Toupin} S.,  {Braun} K.,  {Siess} L.,  {Jorissen} A.,  {Gail} H.-P.,   {Price}
  D.,  2015a, in EAS Publications Series. pp 173--174

\bibitem[\protect\citeauthoryear{{Toupin}, {Braun}, {Siess}, {Jorissen}  \&
  {Price}}{{Toupin} et~al.}{2015b}]{toupinbraunsiess15a}
{Toupin} S.,  {Braun} K.,  {Siess} L.,  {Jorissen} A.,   {Price} D.,  2015b, in
  {Kerschbaum} F.,  {Wing} R.~F.,   {Hron} J.,  eds,  Astronomical Society of
  the Pacific Conference Series Vol. 497, Why Galaxies Care about AGB Stars
  III: A Closer Look in Space and Time. p.~225

\bibitem[\protect\citeauthoryear{{Tuckerman}, {Berne}  \&
  {Martyna}}{{Tuckerman} et~al.}{1992}]{tuckermanbernemartyna92}
{Tuckerman} M.,  {Berne} B.~J.,   {Martyna} G.~J.,  1992, \jcp, \href
  {http://adsabs.harvard.edu/abs/1992JChPh..97.1990T} {97, 1990}

\bibitem[\protect\citeauthoryear{{Wadsley}, {Veeravalli}  \&
  {Couchman}}{{Wadsley} et~al.}{2008}]{wadsleyveeravallicouchman08}
{Wadsley} J.~W.,  {Veeravalli} G.,   {Couchman} H.~M.~P.,  2008, \mnras, \href
  {http://adsabs.harvard.edu/abs/2008MNRAS.387..427W} {387, 427}

\bibitem[\protect\citeauthoryear{{Wegg}}{{Wegg}}{2012}]{wegg12}
{Wegg} C.,  2012, \apj, \href
  {http://adsabs.harvard.edu/abs/2012ApJ...749..183W} {749, 183}

\bibitem[\protect\citeauthoryear{{Yoshida}}{{Yoshida}}{1993}]{yoshida93}
{Yoshida} H.,  1993, Celestial Mechanics and Dynamical Astronomy, \href
  {http://adsabs.harvard.edu/abs/1993CeMDA..56...27Y} {56, 27}

\bibitem[\protect\citeauthoryear{{Zhang} \& {MacFadyen}}{{Zhang} \&
  {MacFadyen}}{2006}]{zhangmacfadyen06}
{Zhang} W.,  {MacFadyen} A.~I.,  2006, \apjs, \href
  {http://adsabs.harvard.edu/abs/2006ApJS..164..255Z} {164, 255}

\makeatother
\end{thebibliography}
\appendix
\section{Positivity of the entropy change} \label{sec:appendix-pos_def}
\subsection{Conductivity} \label{sec:appendix-conductivity}
First, we will prove that the artificial conductivity contributes to a positive definite change to the total entropy.
Taking Equation~\ref{eq:entropy-sph} and relating it back to the specific entropy
\begin{align}
   T_a \deriv{s_a}{t} = U^0_a \brkt{\Pi^a_\mathrm{cond} + \smb \frac{q_a \brkt{v_a^i - v_b^i}}{{\rho_a^*}^2} D^a_i }.
\end{align}
The change in total entropy is
\begin{align}
   \deriv{S}{t} &= \sma \deriv{s_a}{t},
\end{align}
and thus the contribution from artificial conductivity is
\begin{align}
   \evalat{\deriv{S}{t}}{\mathrm{cond}} &= \sum_a \frac{m_a U^0_a}{T_a} \Pi^a_{\mathrm{cond}}, \\
                                                            &= \sum_a \sum_b \zeta_{ab} \frac{U^0_a}{T_a} \brkt{\frac{u_a}{U^0_a} - \frac{u_b}{U^0_b}}, \label{eq:tot-entropy-cond}
\end{align}
where we have used the relation $U^0=\Gamma/\alpha$, and defined the negative, symmetric term
\begin{equation}
   \zeta_{ab} \equiv \frac12 \alpha_u m_a m_b \brkt{\frac{v^u_{\mathrm{sig},a}G_a}{\rho_a^*} + \frac{v^u_{\mathrm{sig},b}G_b}{\rho_b^*}} \leq 0.
\end{equation}
Splitting Eq.~\ref{eq:tot-entropy-cond} into two equal sums, interchanging the labels on one, and then recombining into a single sum, we get
\begin{align}
   \evalat{\deriv{S}{t}}{\mathrm{cond}} &= \sum_a \sum_b \zeta_{ab} \brkt{\frac{U^0_a}{T_a} - \frac{U^0_b}{T_b}}\brkt{\frac{u_a}{U^0_a} - \frac{u_b}{U^0_b}}.
\end{align}
We note that $U^0$ is a strictly positive quantity, and $u \propto T$. Hence, if $u_a > u_b$, we must have $T_a > T_b$, and
\begin{align}
   \brkt{\frac{U^0_a}{T_a} - \frac{U^0_b}{T_b}}\brkt{\frac{u_a}{U^0_a} - \frac{u_b}{U^0_b}} < 0.
\end{align}
Thus the contribution to the total entropy, due to artificial conductivity, is always positive in both special and general relativity.

%Proof requires Cartesian like coordinates
\subsection{Viscosity}
Next we consider the change in entropy per particle due to artificial viscosity in Cartesian coordinates
\begin{align}
   \evalat{\deriv{s_a}{t}}{\mathrm{visc}} &= \frac{U^0_a}{T_a} \smb \frac{q_a \brkt{v_a^i - v_b^i}}{{\rho_a^*}^2} D^a_i, \\
   									     &= \sum_b  Q_{ab} \brkt{\Gamma^*_a V^*_a - \Gamma^*_b V^*_b} \brkt{v_a^* - v_b^*}, \label{eq:entropy-visc}
\end{align}
where $v^*= N_i \, v^i$ is a line of sight velocity, and we defined the positive quantity
\begin{equation}
   Q_{ab} \equiv -\frac12 \frac{U^0_a}{T_a \rho_a^*} m_b \alpha_\mathrm{AV} v_{\mathrm{sig},a} w_a G_a \geq 0.
\end{equation}
In the case that $V^*$ is a monotonically increasing function of $v^*$ in a given spacetime, if $v_a^* > v_b^*$ we must have $V_a^* > V_b^*$ as well as $\Gamma_a^* > \Gamma_b^*$, and hence
\begin{equation}
   \brkt{\Gamma^*_a V^*_a - \Gamma^*_b V^*_b} \brkt{v_a^* - v_b^*} > 0.
\end{equation}
The condition that $V^*$ be monotonic is easily satisfied in the Minkowski metric, where $V^i=v^i$, since there is no shift, $\beta^i=0$, and the lapse $\alpha=1$. Thus in the case of special relativity, the artificial viscosity will only ever increase the entropy of a particle, however this cannot be guaranteed in all spacetimes.

\section{Recovery of primitive variables} \label{sec:appendix-cons2prim}
Step \ref{step:rootfind} of section~\ref{sec:cons2prim} involves solving the equation
\begin{equation}
   f(w) \equiv w(\rho,P) - w = 0
\end{equation}
for $w$, where $\rho$ and $P$ are functions of $w$ (Eq.~\ref{eq:cons2prim_rho} and~\ref{eq:cons2prim_pre}). For this we use the standard Newton-Raphson algorithm
\begin{equation}
   \label{eq:newtonraphson}
   w_{n+1} = w_n - \frac{f(w_n)}{f'(w_n)},
\end{equation}
which requires the derivative $f'(w)$. Here, the subscript $n$ corresponds to the iteration number. For the initial prediction, $w_0$, we use the enthalpy computed at the previous timestep. Iterations of Equation~\ref{eq:newtonraphson} are repeated until the following convergence criteria is satisfied
\begin{equation}
   \frac{w_{n+1} - w_n}{w_{n+1}} \leq \epsilon_w,
\end{equation}
for some sufficiently small number $\epsilon_w$. In this paper we use $\epsilon_w = 10^{-12}$.
Given the ideal gas equation of state (Eq.~\ref{eq:eos}), we can write
\begin{equation}
   f(w) = \left[ 1 + \frac{P}{\rho}\brkt{ \frac{\adgamma}{\adgamma - 1} } \right] - w
\end{equation}
and hence the derivative can be written as
\begin{equation}
   f'(w) = \frac{\adgamma }{\adgamma - 1} \brkt{ 1 - \frac{p^i p_i ~P}{ w^3 \rho ~\Gamma^2} } - 1,
\end{equation}
where $\rho$, $P$, and $\Gamma$ are all functions of $w$ (Eq.~\ref{eq:cons2prim_rho}, \ref{eq:cons2prim_pre} and \ref{eq:cons2prim_gamma}). Equivalently, when evolving the entropy variable $K$
\begin{equation}
   f'(w) = \frac{\adgamma ~p^i p_i P}{w^3 \rho ~\Gamma^2} - 1,
\end{equation}
where instead of Equation~\ref{eq:cons2prim_pre} for the pressure, we use
\begin{align}
   P(w) &= K \left[\rho(w)\right]^\adgamma, \\
           &= K \left[\frac{ \rho^*}{\sqrt{\gamma} ~\Gamma(w)}\right]^\adgamma.
\end{align}

\section{Metrics in Cartesian Coordinates}
\subsection{Schwarzschild metric} \label{sec:appendix-schwarzschild}
The Schwarzschild metric line element, in standard spherical coordinates $(t,r,\theta,\phi)$ is
\begin{equation}
   ds^2 = - \brkt{ 1 - \frac{2M}{r} } \d  t^2 + \frac{\d r^2}{1 -\frac{2M}{r}}  + r^2 \brkt{ \d  \theta^2 + \sin^2 \theta \, \d  \phi^2 }.
\end{equation}
If we make the coordinate transformation
\begin{align}
   x &= r \sin \theta \cos \phi, \nonumber \\
   y &= r \sin \theta \sin \phi, \\
   z &= r \cos \theta, \nonumber
\end{align}
we can write the components of the covariant metric in Cartesian coordinates $(t,x,y,z)$
\begin{align}
 g_{tt} &= -\brkt{1 - \frac{2M}{r}}, \nonumber \\
 g_{xx} &= \frac{1}{1 - \frac{2M}{r}}  \brkt{1- \frac{2M}{r^3}  (y^2+z^2)}, \nonumber \\
 g_{yy} &= \frac{1}{1 - \frac{2M}{r}}  \brkt{1- \frac{2M}{r^3}  (x^2+z^2)}, \nonumber \\
 g_{zz} &= \frac{1}{1 - \frac{2M}{r}}  \brkt{1- \frac{2M}{r^3}  (x^2+y^2)}, \\
 g_{yx} &= g_{xy} = \frac{x y}{1 - \frac{2M}{r}} \frac{2M}{r^3}, \nonumber \\
 g_{zx} &= g_{xz} = \frac{x z}{1 - \frac{2M}{r}}    \frac{2M}{r^3}, \nonumber \\
 g_{zy} &= g_{yz} = \frac{y z}{1 - \frac{2M}{r}}    \frac{2M}{r^3}, \nonumber \\
 g_{tx}  &= g_{ty} = g_{tz} = g_{xt} = g_{yt} = g_{zt} = 0, \nonumber
\end{align}
as well as the components of the contravariant metric
\begin{align}
 g^{tt} &= -\frac{1}{1 - \frac{2M}{r}}, \nonumber \\
 g^{xx} &= 1- \frac{2M}{r^3}  x^2, \nonumber \\
 g^{yy} &= 1- \frac{2M}{r^3}  y^2, \nonumber \\
 g^{zz} &= 1- \frac{2M}{r^3}  z^2, \\
 g^{yx} &= g_{xy} = - \frac{2M}{r^3}  x y, \nonumber \\
 g^{zx} &= g_{xz} = - \frac{2M}{r^3}  x z, \nonumber \\
 g^{zy} &= g_{yz} = - \frac{2M}{r^3}  y z, \nonumber \\
 g^{tx}  &= g^{ty} = g^{tz} = g^{xt} = g^{yt} = g^{zt} = 0, \nonumber
\end{align}
where $r = \sqrt{x^2 + y^2 + z^2}$. In these coordinates $\sqrt{-g} = 1$. The derivatives $\partial g_{\mu\nu}/ \partial x^i$ can be found analytically by differentiating $g_{\mu\nu}$, or through a finite difference approximation.
%The independent components are
%\begin{align}
%\pder{g_{xt}}{x} &= \pder{g_{yt}}{x} = \pder{g_{zt}}{x} = \pder{g_{tx}}{x} = 0 , \\
%\pder{g_{tt}}{x} &= -\frac{{r_\mathrm{s}} x}{r^3}, \\
%\pder{g_{xx}}{x} &= \frac{x (2 r^4 - 2 r^3 {r_\mathrm{s}} - 2 ({r_\mathrm{s}})^2 (y^2 + z^2) - 2 r^4 + r {r_\mathrm{s}} (x^2 + 4 (y^2 + z^2)))}{(r^4 (r - {r_\mathrm{s}})^2)}, \\
%\pder{g_{yx}}{x} &= ({r_\mathrm{s}} (r^3 - r^2 {r_\mathrm{s}} - 3 r x^2 + 2 {r_\mathrm{s}} x^2) y)/(r^4 (r - {r_\mathrm{s}})^2), \\
%\pder{g_{zx}}{x} &= ({r_\mathrm{s}} (r^3 - r^2 {r_\mathrm{s}} - 3 r x^2 + 2 {r_\mathrm{s}} x^2) z)/(r^4 (r - {r_\mathrm{s}})^2), \\
%\pder{g_{yy}}{x} &= (x (2 r^4 - 4 r^3 {r_\mathrm{s}} + 2 r^2 (({r_\mathrm{s}})^2 - r^2) - 2 ({r_\mathrm{s}})^2 (x^2 + z^2) + r {r_\mathrm{s}} (4 x^2 + y^2 + 4 z^2)))/(r^4 (r - {r_\mathrm{s}})^2), \\
%\pder{g_{zy}}{x} &= ({r_\mathrm{s}} (-3 r + 2 {r_\mathrm{s}}) x y z)/(r^4 (r - {r_\mathrm{s}})^2), \\
%\pder{g_{zz}}{x} &= (-2 (r - {r_\mathrm{s}})^2 x (-z^2) + r (-2 r + {r_\mathrm{s}}) x z^2)/(r^4 (r - {r_\mathrm{s}})^2),
%\end{align}

\subsection{Kerr metric} \label{sec:appendix-kerr}
The Kerr metric line element in standard Boyer-Lindquist coordinates $(t,r,\theta,\phi)$ is
\begin{align}
   ds^2 = -&\brkt{1 - \frac{2Mr}{\rho^2} } \d t^2  - \frac{4 M r a \sin^2 \theta}{\rho^2}\d t \, \d \phi + \frac{\rho^2}{\Delta} \d r^2 \nonumber \\
                &+ \rho^2 \d \theta^2 + \brkt{a^2 + r^2 + \frac{2Mr}{\rho^2} a^2 \sin^2 \theta} \sin^2 \theta \, \d \phi^2,
\end{align}
where
\begin{align}
   \rho^2    &= r^2 + a^2 \cos^2 \theta, \\
   \Delta &= r^2 - 2Mr + a^2,
\end{align}
and $|a| \leq M$ is the spin parameter. If we make the coordinate transformation
\begin{align}
   x &= \sqrt{r^2 + a^2} \sin \theta \cos \phi, \nonumber \\
   y &= \sqrt{r^2 + a^2} \sin \theta \sin \phi, \\
   z &= r \cos \theta, \nonumber
\end{align}
we can write the components of the covariant metric in Cartesian-like coordinates $(t,x,y,z)$
\begin{align}
g_{tt} &= - \brkt{1 - \frac{2Mr}{\rho^2}}, \nonumber \\
g_{xt} &= g_{tx} = \frac{-y}{x^2 + y^2} g_{t \phi}, \nonumber \\
g_{yt} &= g_{ty} = \frac{x}{x^2 + y^2} g_{t \phi}, \nonumber \\
g_{zt} &= g_{tz} = 0, \nonumber \\
g_{xx} &= \frac{r^2 x^2}{\rho^2 \Delta} + g_{\phi\phi} \brkt{\frac{y}{x^2 + y^2}}^2 + \frac{x^2 z^2}{\rho^2 (r^2 - z^2)}, \\
g_{yy} &= \frac{r^2 y^2}{\rho^2 \Delta} + g_{\phi\phi} \brkt{\frac{x}{x^2 + y^2}}^2 + \frac{y^2 z^2}{\rho^2 (r^2 - z^2)}, \nonumber \\
g_{zz} &= \frac{ z^2 (a^2 + r^2)^2 }{r^2 \rho^2 \Delta} + \frac{\rho^2}{r^2 - z^2} \brkt{1 - \frac{z^2 (a^2 + r^2)}{r^2 \rho^2}}^2, \nonumber \\
g_{xy} &= g_{yx} = r^2 \frac{x y}{\rho^2 \Delta} - g_{\phi\phi} \frac{x y}{(x^2 + y^2)^2} + \frac{x y z^2}{\rho^2 (r^2 - z^2)}, \nonumber \\
g_{xz} &= g_{zx} = (a^2 + r^2) \frac{x z}{\rho^2 \Delta} - \frac{x z}{(r^2 - z^2)} \brkt{1 - \frac{z^2 (a^2 + r^2)}{r^2 \rho^2}}, \nonumber \\
g_{yz} &= g_{zy} = (a^2 + r^2) \frac{y z}{\rho^2 \Delta} - \frac{y z}{(r^2 - z^2)} \brkt{1 - \frac{z^2 (a^2 + r^2)}{r^2 \rho^2}}, \nonumber
\end{align}
and the components of the contravariant metric
\begin{align}
g^{tt} &= \frac{-g_{\phi\phi}}{{g_{t \phi}}^2 - g_{\phi\phi} g_{tt}}, \nonumber \\
g^{xt} &= g^{tx} = \frac{-y g_{t \phi}}{{g_{t \phi}}^2 - g_{\phi\phi} g_{tt}}, \nonumber \\
g^{yt} &= g^{ty} = \frac{ x g_{t \phi}}{{g_{t \phi}}^2 - g_{\phi\phi} g_{tt}}, \nonumber \\
g^{zt} &= g^{tz} = 0, \nonumber \\
g^{xx} &= \frac{r^2 x^2 \Delta}{\rho^2 (a^2 + r^2)^2} - \frac{y^2 g_{tt}}{{g_{t \phi}}^2 - g_{\phi\phi} g_{tt}} + \frac{x^2 z^2}{\rho^2 (r^2 - z^2)}, \\
g^{yy} &= \frac{r^2 y^2 \Delta}{\rho^2 (a^2 + r^2)^2} - \frac{x^2 g_{tt}}{{g_{t \phi}}^2 - g_{\phi\phi} g_{tt}} + \frac{y^2 z^2}{\rho^2 (r^2 - z^2)}, \nonumber \\
g^{zz} &= \frac{r^2 - z^2}{\rho^2} + \frac{z^2 \Delta}{r^2 \rho^2}, \nonumber \\
g^{xy} &= g^{yx} = \frac{r^2 x y \Delta}{\rho^2 (a^2 + r^2)^2} +  \frac{x y g_{tt}}{{g_{t \phi}}^2 - g_{\phi\phi} g_{tt}} + \frac{x y z^2}{\rho^2 (r^2 - z^2)}, \nonumber \\
g^{xz} &= g^{zx} = -\frac{x z}{\rho^2} + \frac{x z \Delta}{\rho^2 (a^2 + r^2)}, \nonumber \\
g^{yz} &= g^{zy} = -\frac{y z}{\rho^2} + \frac{y z \Delta}{\rho^2 (a^2 + r^2)}, \nonumber
\end{align}
where
\begin{align}
   r                 &= \sqrt{\frac{ R^2-a^2+\sqrt{(R^2-a^2)^2 + 4 a^2 z^2} }{2}}, \\
   R^2            &= x^2+y^2+z^2, \\
   \rho^2        &= r^2 + a^2 \frac{z^2}{r^2}, \\
   g_{tt}          &= -\brkt{1 - \frac{2Mr}{\rho^2}}, \\
   g_{\phi\phi}&= \brkt{r^2+a^2 + \frac{2Mr}{\rho^2} a^2 \sin^2\theta} \sin^2\theta, \\
   g_{t \phi}    &= - \frac{2 M r a}{\rho^2} \sin^2\theta, \\
   \sin^2\theta&= 1 - \frac{z^2}{r^2}.
\end{align}
We again have $\sqrt{-g} = 1$ in the new coordinates and the derivatives $\partial g_{\mu\nu}/\partial x^i$ can be found analytically by differentiating $g_{\mu\nu}$, or through a finite difference approximation.
%\begin{equation}
%   g_{\mu \nu} =
%   \begin{pmatrix}
%      -\brkt{1 - \frac{2Mr}{\rho^2} } & 0 & 0 & - \frac{2Mar \sin^2 \theta}{\rho^2}\\
%      0 & \frac{\rho^2}{\Delta} & 0 & 0\\
%      0 & 0 & \rho^2 & 0\\
%      - \frac{2Mar \sin^2 \theta}{\rho^2} & 0 & 0 & \frac{\sin^2 \theta}{\rho^2} \left[\brkt{r^2+a^2}^2 - a^2 \Delta \sin^2 \theta \right]
%   \end{pmatrix}
%\end{equation}

\subsection{Derivatives of the metric} \label{sec:appendix-finite_difference}
We compute the derivatives of the Schwarzschild metric by analytically differentiating $g_{\mu\nu}$, however for the Kerr metric we choose to compute the derivatives through a 2nd order centred finite difference approximation
\begin{align}
\pder{g_{\mu\nu}(x^i)}{x^i} \approx \frac{g_{\mu\nu}(x^i+\epsilon) - g_{\mu\nu}(x^i-\epsilon)}{2\epsilon},
\end{align}
for $\epsilon \ll 1$. In our simulations we use $\epsilon=10^{-8}$.

\subsection{Inverse}
The contravariant metric $g^{\mu\nu}$ can also be found by inverting the matrix $g_{\mu\nu}$ at every point, since
\begin{align} \label{eq:metric-inverse}
   g_{\mu\alpha} g^{\nu\alpha} = \delta_\mu^\nu.
\end{align}
Given the determinant of the covariant metric $g$, we simply compute the adjugate matrix of $g_{\mu\nu}$, and then
\begin{align}
   g^{\mu\nu} = \frac{1}{g} \mathrm{adj}(g_{\mu\nu}).
\end{align}
In practice we find that this conserves Eq.\ref{eq:metric-inverse} better than the analytic contravariant terms near the event horizon. As such, we compute the contravariant metric in this manner by default.

\section{Computing oscillation frequencies} \label{sec:appendix-fft}
\begin{figure}
   \begin{center}
      \includegraphics[height=0.35\columnwidth]{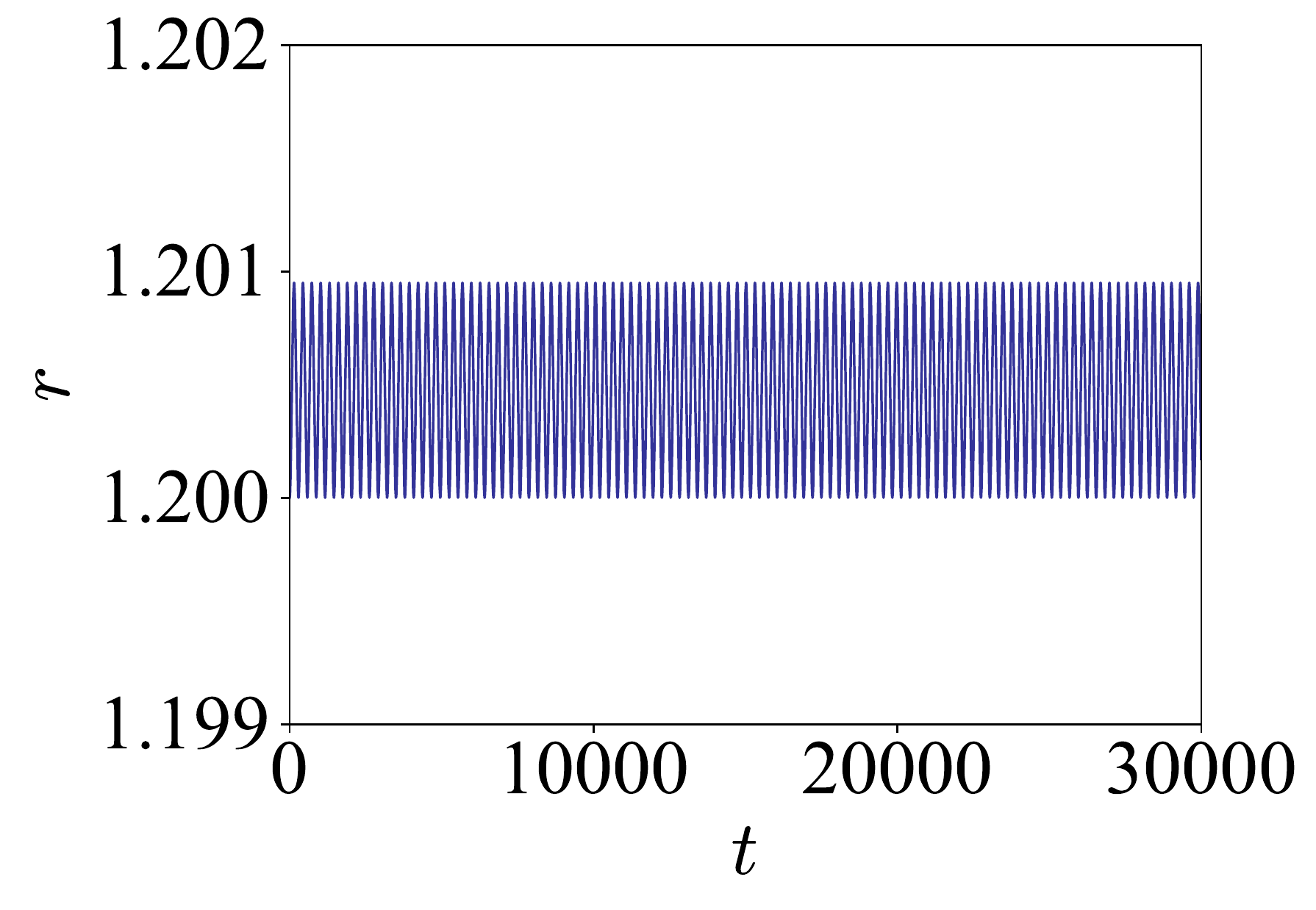}
      \includegraphics[height=0.35\columnwidth]{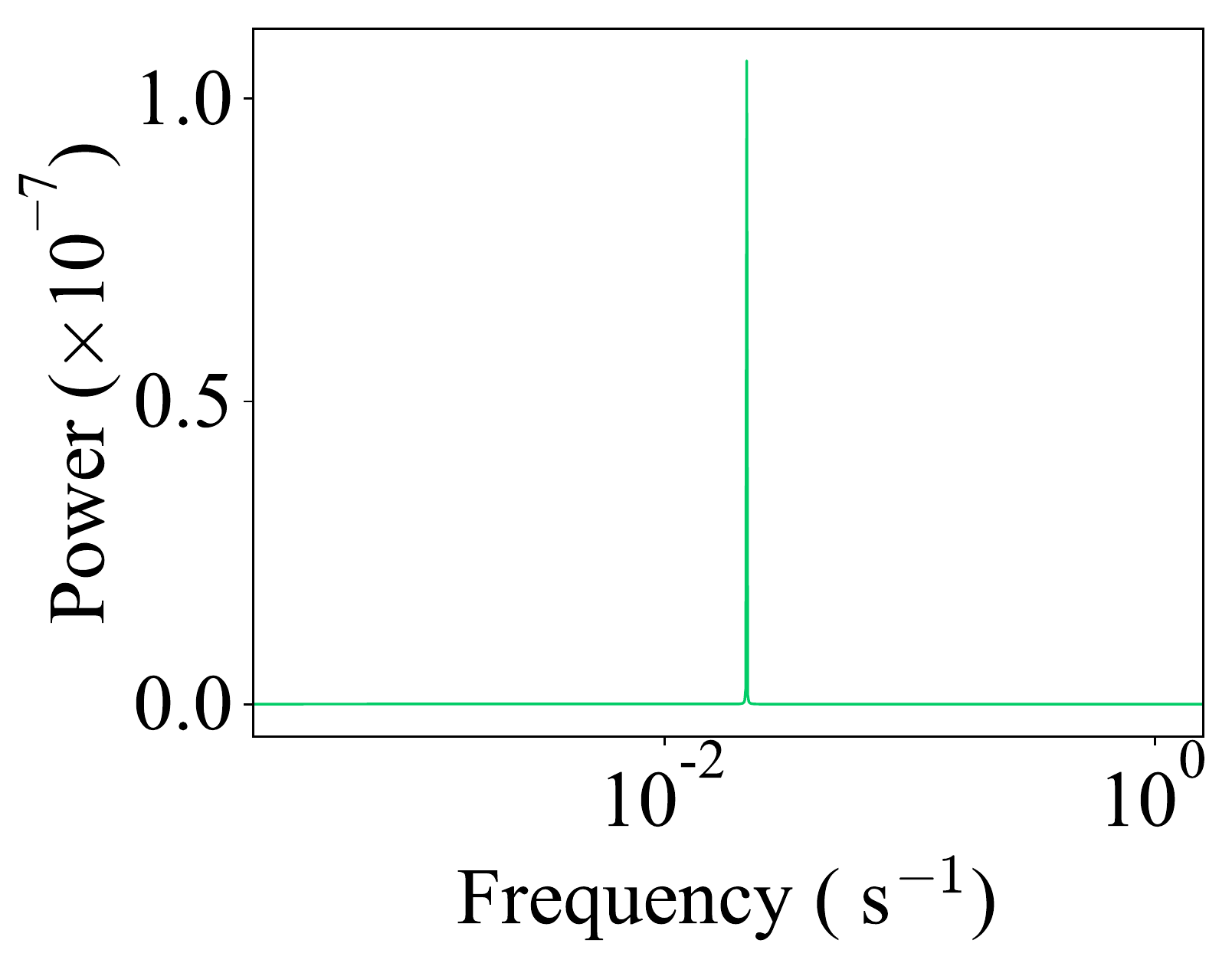}
      \caption{Radius as a function of time (left), and the corresponding 1D power spectrum (right), for the epicycle at $r=1.2$ with black hole spin $a=1$.}
      \label{fig:getting_frequency}
   \end{center}
\end{figure}

To measure the oscillation frequencies described in Sections~\ref{sec:epicycle} and \ref{sec:vertical-oscillations}, we take the maximum value of the 1-D power spectrum of $r(t)$, using \verb+numpy.fft.fft+ in \textsc{Python}.
Figure \ref{fig:getting_frequency} shows the time series (left) and corresponding power spectrum (right) for the $a=1$, $r=1.2$ case.

\label{lastpage}
\end{document}